\documentclass[onecolumn,12pt]{article}
\usepackage{jheppub}
\usepackage{ifpdf}
\usepackage[bb=boondox]{mathalfa}
\graphicspath{ {figures/} }

\usepackage{amsfonts}
\usepackage{latexsym}
\usepackage{color}

\usepackage{amsmath,braket}
\usepackage{amssymb,physics}
\usepackage{amsthm}
\usepackage{mathtools}

\usepackage{epsfig}

\usepackage{tensor}
\usepackage{pdfpages}
\usepackage{bbm}
\usepackage{graphicx,epstopdf}
\usepackage[numbers]{natbib} 
\usepackage[makeroom]{cancel}
\usepackage{hyperref}
\usepackage{array}
\usepackage[export]{adjustbox}

\usepackage[normalem]{ulem}
\usepackage{romannum}

\usepackage{ulem}

\usepackage{tikz}
\usetikzlibrary{intersections}
\usepackage{circuitikz}

\numberwithin{equation}{section}									

\usepackage{here}

\usepackage{comment}

\newcommand{\de}{\partial}
\newcommand{\be}{\begin{equation}}
	\newcommand{\ba}{\begin{eqnarray}}
		\newcommand{\ea}{\end{eqnarray}}
	\newcommand{\ee}{\end{equation}}

\newcommand{\ca}{\mathcal}

\newcommand{\f}{\frac}
\newcommand{\s}{\sqrt}
\newcommand{\vp}{\varphi}

\newcommand{\ti}{\tilde}
\newcommand{\ap}{\alpha}

\newcommand{\ddd}{\cdot\cdot\cdot}
\newcommand{\no}{\nonumber \\}
\newcommand{\la}{\langle}
\newcommand{\lb}{\rangle}
\newcommand{\bea}{\begin{eqnarray}}
	\newcommand{\eea}{\end{eqnarray}}
\newcommand{\bes}{\begin{equation*}}
	\newcommand{\beas}{\begin{eqnarray*}}
		\newcommand{\eeas}{\end{eqnarray*}}
	\newcommand{\bas}{\begin{array*}}
		\newcommand{\eas}{\end{array*}}
	\newcommand{\ees}{\end{equation*}}

\newcommand{\ep}{\epsilon}

          \let\m=\mu 
   \let\vp=\varphi \let\x=\xi

\newcommand{\ie}{{\it i.e.,}\ }

\newcommand{\mt}[1]{\textrm{\tiny #1}}

\newcommand{\eff}{\mathrm{eff}}

\newcommand{\GN}{G_\mt{N}}

\newcommand{\mL}{\mathcal{L}}

\newcommand{\mO}{\mathcal{O}}

\newcommand{\arcsinh}{\text{arcsinh}}

\newcommand{\const}{\text{const.}}
\newcommand{\OTT}{{\mO}_{T\overline{T}}}

\newcommand{\gz}{{g_{(0)}}}
\newcommand{\gt}{{g_{(2)}}}
\newcommand{\gf}{{g_{(4)}}}

\newcommand{\gaz}{{\gamma_{[0]}}}
\newcommand{\gam}{{\gamma_{[\mu]}}}
\newcommand{\Tmu}{{T_{[\mu]}}}

\newcommand{\LandauO}[1]{O\qty(#1)}

\newcommand{\mtL}[1]{\Tilde{\mathcal{L}}_{(#1)}}
\newcommand{\Pie}[1]{{\Pi_{(#1)}}}
\newcommand{\Pite}[1]{{\Tilde{\Pi}_{(#1)}}}

\newcommand{\EllipticK}[1]{K\left(#1\right)}

\newcommand{\TTbar}{T\overline{T}}

\subheader{\today}
\title{\boldmath Traversable AdS Wormhole via Non-local Double Trace or Janus Deformation}

\author[a]{Taishi Kawamoto,}
\author[a]{Ryota Maeda,}
\author[a]{Nanami Nakamura,}
\author[a,b]{Tadashi Takayanagi}

\affiliation[a]{Center\! for Gravitational\! Physics\! and Quantum \! Information, Yukawa\! Institute\! for\! Theoretical\! Physics, Kyoto\! University, Kitashirakawa\! Oiwakecho, Sakyo-ku, Kyoto 606-8502, Japan}
\affiliation[b]{Inamori\! Research\! Institute\! for\! Science,\! 620\! Suiginya-cho,\! Shimogyo-ku,\! Kyoto\! 600-8411, Japan}

\emailAdd{taishi.kawamoto@yukawa.kyoto-u.ac.jp}
\emailAdd{ryota.maeda@yukawa.kyoto-u.ac.jp}
\emailAdd{nanami.nakamura@yukawa.kyoto-u.ac.jp}
\emailAdd{takayana@yukawa.kyoto-u.ac.jp}

\abstract{
We study (i) Janus deformations and (ii) non-local double trace deformations of a pair of CFTs, as two different ways to construct CFT duals of traversable AdS wormholes. First we construct a simple model of traversable wormholes by gluing two Poincar\'e AdS geometries and BTZ black holes and compute holographic two point functions and (pseudo) entanglement entropy. We point out that a Janus gravity solution describes a traversable wormhole when the deformation parameter takes imaginary values. On the other hand, we show that double trace deformations between two decoupled CFTs can reproduce two point functions of traversable AdS wormholes. By considering the case where the double trace deformation is given by a non-local $T\overline{T}$ deformation, we analyze the dual gravity which implies emergence of wormholes. We present toy model of these deformed CFTs by using free scalars and obtain qualitative behaviors expected for them. We argue that the crucial difference between the two constructions is that a global time slice of wormhole is described by a pure state for Janus deformations, while it is a mixed state for the double trace deformations.}

\begin{document} 
	
\begin{flushright}
YITP-25-09
\\
\end{flushright}
\maketitle
\flushbottom

\section{Introduction}
\label{sec:intro}

One of the most fascinating and important aspects of quantum gravity is the emergence of diverse spacetime topologies. Among them, wormholes are quite fundamental and intriguing as they can connect different worlds. Holography \cite{Susskind:1994vu,tHooft:1993dmi} or more specifically the AdS/CFT \cite{Maldacena:1997re,Gubser:1998bc,Witten:1998qj} provides a very promising approach to quantum gravity as they rewrite it in terms of a microscopic and non-gravitational theory.

In AdS/CFT, if there is a lot of quantum entanglement between two CFTs, they are dual to an asymptotically AdS wormhole such as the eternal AdS black holes \cite{Maldacena:2001kr}. The amount of quantum entanglement is measured by entanglement entropy \cite{Bombelli:1986rw,Srednicki:1993im,Holzhey:1994we,Calabrese:2004eu} and this is computed as the area of extremal surface in AdS \cite{Ryu:2006bv,Ryu:2006ef,Hubeny:2007xt,Nishioka:2009un,Rangamani:2016dms}. This holographic entanglement entropy quantifies the size of wormhole and suggests the emergence of spacetime in gravity from quantum entanglement \cite{Swingle:2009bg,VanRaamsdonk:2010pw,Maldacena:2013xja}. Also the AdS wormhole appears as an important contribution to the entanglement entropy under black hole evaporation \cite{Penington:2019kki,Almheiri:2019qdq}, the late time spectral form factor \cite{Saad:2018bqo,Cotler:2021cqa,Marolf:2021kjc,DiUbaldo:2023qli}, the boundary thermal correlation functions \cite{Saad:2019pqd,Kawamoto:2024vzd}. This kind of wormhole also used for the regulating the UV divergence of the theories including AdS gravity \cite{Chen:2023hra,Kawamoto:2023ade}.\par

Even though these geometries dual to entangled CFTs are macroscopic wormholes, they are not traversable. As pioneered in \cite{Gao:2016bin}, we can make an eternal AdS black hole traversable by turning on interactions between two CFTs, which is called a double trace deformation \cite{Aharony:2001pa,Witten:2001ua}. The construction of a traversable AdS wormhole via the double trance deformation has been very successful and thoroughly analyzed in AdS$_2$ gravity \cite{Maldacena:2017axo,Maldacena:2018gjk,Maldacena:2018lmt}.
An eternal traversable wormhole in AdS$_3$ was also found by deforming the BTZ black hole \cite{Harvey:2023oom}.
In all such examples, the traversable wormholes are produced by perturbing eternal black hole geometries, which are non-traversable wormholes. This raises the question of what will happen if we introduce double trace interactions between two CFTs which are originally decoupled and are not entangled at all. Also in \cite{Mori:2024gwe}, it is discussed that the (gravitatinal) LOCCs promotes the existence of the traversal wormholes in AdS. Another honest question is whether there are other ways to create traversable wormhole in AdS/CFT.

Motivated by these questions, we would like to study general setups of static traversable wormholes in AdS and analyze their CFT duals. First we will work out the properties of traversable AdS wormholes by considering a simple model of traversable AdS wormhole by gluing two AdS geometries in any dimensions along interior surfaces as depicted in Fig.\ref{fig:setup}. We will find characteristic behaviors of holographic two point functions. For example, the traversable property leads to a divergent two point function between two points connected by a bulk null geodesic. We will argue that there are at least two different ways to realize such wormholes in AdS/CFT. 

One obvious construction of the CFT dual is to introduce the double trace interactions between two CFTs, which we call the model B. An important point we emphasize is that we need to consider non-local interactions with a UV cut off so that they do not affect the high energy physics, which guarantees the presence of two asymptotically AdS boundaries. In this sense, this setup makes a sharp contrast with earlier works \cite{Aharony:2006hz,Kiritsis:2006hy}, where the double trace interactions between two CFTs without UV cut off were analyzed and were argued to be dual to gluing two AdS geometries along the AdS boundaries. Our argument can be applicable when we turn on the interactions between two decoupled CFTs, dual to gluing two disconnected AdS geometries. Moreover, a version of this model B can be obtained by considering a double trance deformation by the energy stress tensors in the two CFTs. This can be regarded as an extension of the $\TTbar$ deformation \cite{Zamolodchikov:2004ce,Cavaglia:2016oda,McGough:2016lol}
(see also \cite{Bzowski:2020umc,Ferko:2022dpg} for setups related to ours). This allows us to directly study the gravitational dynamics which implies the wormhole structure under this deformation as we will see later in this paper.

We will also point out that there is the second approach, called the model A, to construct traversable AdS wormholes. This is to employ Janus deformations in AdS/CFT \cite{Bak:2003jk,Freedman:2003ax,DHoker:2007zhm,Bak:2007jm}. 
This introduces a localized deformation on a codimension one interface in a given CFT. A coupling constant in the CFT lagrangian jumps across the interface. In particular, the Janus deformation of BTZ black holes in \cite{Bak:2007jm,Bak:2007qw,Nakaguchi:2014eiu}  gives an one parameter family of AdS wormhole which is not traversable. This is dual to thermofield double (TFD) states of two CFTs, which are parameterized by the Janus deformation parameter. However if we continue the parameter to imaginary values, we find that the wormhole becomes traversable. This setup is now dual to not a single state but a post selection process where the initial state and final state are different. We will argue that post-selections instead of the double trace deformations can also give rise to the traversable wormhole in AdS/CFT. 

Interestingly, these two different constructions have different properties in terms of a global nature of the corresponding quantum states. In the Janus deformation case (model A), even though the initial state and final state are different, there are no interactions between the two CFTs under the time evolutions. Accordingly if we consider the time slice A at $t=t_1$ in the CFT$_{(1)}$ and the one B at  $t=t_2$ in the CFT$_{(2)}$, the quantum state realized on $A\cup B$ is a pure state. However, in the case of the double trace deformation (model B), even though the initial and final state are the same, due to the interactions between the two CFTs, the quantum state on the time slice $A\cup B$ becomes a mixed state as we will explain in detail later in this paper. The common feature in both cases is that the quantum state is not described by a hermitian density matrix but given by a transition matrix which is non-hermitian. The entropy for this state should be interpreted as pseudo entropy \cite{Nakata:2021ubr,Mollabashi:2020yie,Mollabashi:2021xsd}, instead of entanglement entropy. In earlier works \cite{Kanda:2023jyi,Kanda:2023zse}, an imaginary valued Janus deformation of a boundary CFT was holographically studied and a novel type of phase transition in the pseudo entropy was found. In the present paper, we will show that the imaginary Janus deformation of a bulk CFT enhances the pseudo entropy and this makes the non-traversable wormhole into traversable one.

This paper is organized as follows. In section 2, we present the basic model of traversable AdS wormhole by gluing two Poincar\'e AdS geometries. We compute the two point functions as well as holographic pseudo entropy. We also analyze gluing of two BTZ geometries. In section 3, we explain two different setups of CFT duals: namely the Janus deformation (model A) and the double trace deformation (model B), with illustrations using toy examples. In section 4, we show that a Janus deformation with an imaginary valued parameter leads to a traversable AdS wormhole. We also examine a CFT example of Janus deformation by considering the free scalar CFT and compute the two point functions. In section 5, we study the non-local double trace deformations and show that this can reproduce the same two point functions as those in the traversable AdS wormholes. We also present a toy CFT example of the deformation defined by a free scalar CFT and calculate the pseudo entropy for the total system. In section 6, we analyze a non-local double trace deformation by the energy stress tensors in the two CFTs and present its gravity dual, which implies the presence of wormholes.  In section 7, we summarize our conclusions.

\section{Simple examples of traversable AdS wormholes and two point functions}
\label{sec:AdSWH}

\begin{figure}[ttt]
		\centering
		\includegraphics[width=10cm]{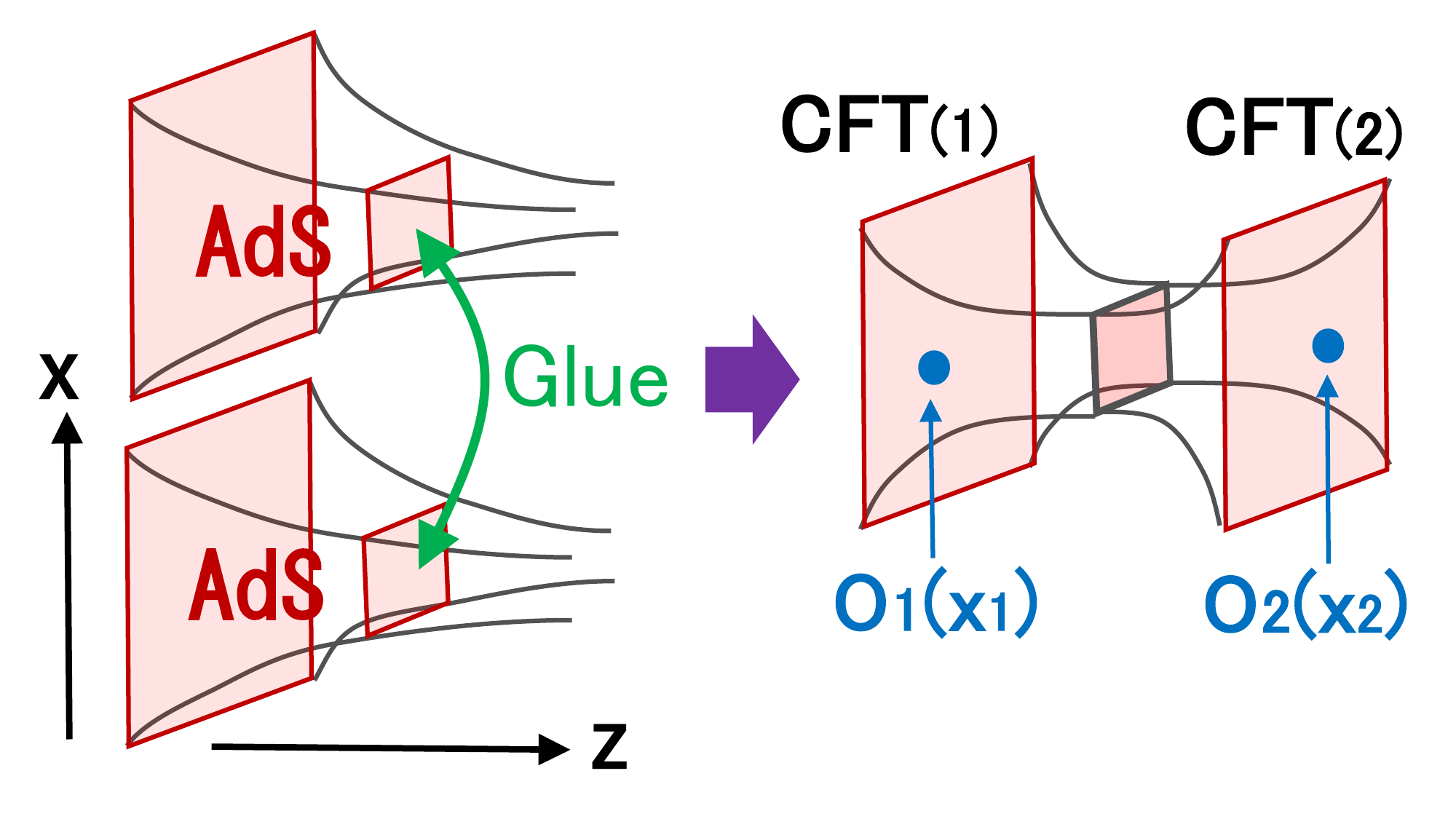}
		\caption{A simple construction of traversable AdS wormhole defined by (\ref{adSwm}) and (\ref{WHR}).} 
		\label{fig:setup}
\end{figure}

We would like to start with analyzing a simple class of traversable AdS wormholes to see what kinds of  properties we expect for the dual CFTs. We first prepare an asymptotically AdS region by removing an interior part (or IR region) from the whole AdS, along a surface. Then we glue a pair of such geometries on the surface as depicted in Fig.\ref{fig:setup}.

Gluing two Poincar\'e AdS geometries provides the simplest example and is explicitly given as follows.
Consider the Poincar\'e AdS$_{d+1}$:
\ba
ds^2=\frac{dz^2+\sum_{i=0}^{d-1}dx_i^2}{z^2}.
\ea
We would like to consider a traversable wormhole
where two AdS$_3$ are glued along $z=z_0$. This geometry is written as 
\ba
ds^2=R(z)\left(dz^2+\sum_{i=0}^{d-1}dx_i^2\right),\label{adSwm}
\ea
where
\ba
R(z)=\frac{1}{z^2} \ \ (0<z<z_0),\ \ \ \ \ \ R(z)=\frac{1}{(2z_0-z)^2} \ \ (z_0<z<2z_0).  \label{WHR}
\ea
We put the UV cut off at $z=\ep$ and $z=2z_0-\ep$ for the two AdS boundaries.

We expect that the gravity in this wormhole geometry is dual to a pair of $d$ dimensional CFTs, denoted by CFT$_{(1)}$ and CFT$_{(2)}$, which live on the two AdS boundaries $z=0$ and $z=2z_0$. To regulate the UV divergence of the CFTs we introduce the geometrical cut off $\ep$ such that they live at $z=\ep$ and $z=2z_0-\ep$. We will study the properties of these CFTs by analyzing two point functions of scalar operator and also by computing the holographic (pseudo) entanglement entropy below.
In addition we will also examine gluing two AdS black hole geometries.

\subsection{Two point functions}

To probe basic properties of the CFTs dual to the wormhole, we would like to calculate holographic two point functions. Below we calculate the holographic two point functions by extending the standard prescription \cite{Gubser:1998bc,Witten:1998qj,Klebanov:1999tb} to our wormhole model. For simplicity, we consider that the two CFTs have the same central charge and shares some light spectrum. We consider a scalar field $\Phi(z,x)$ in the $d+1$ dimensional bulk geometry with a mass $m$. The action of the scalar field reads
\ba
I_{\mathrm{scalar}}=\int dzd^dx\left[\frac{1}{z^{d-1}}\left((\de_z\Phi)^2+
(\de_x\Phi)^2\right)+\frac{m^2}{z^{d+1}}\Phi^2\right]. \label{saction}
\ea
By taking the Fourier transformation $\Phi(x,z)=e^{ikx}\Phi_k(z)$, the equation of motion reads 
\ba
\Phi''-\frac{d-1}{z}\Phi'-\left(k^2+\frac{m^2}{z^2}\right)\Phi=0. \label{EOMs}
\ea
Now we introduce the coordinate $w=2z_0-z$ and another representation $\Psi(w)$ of the same scalar field such that $\Psi(w)=\Phi(z)$, to make the second boundary clearer. 

The general solutions to (\ref{EOMs}) can be written as follows (we set  $\nu = \sqrt{m^2 - \frac{d^2}{4}}$):
\begin{align}
    \Phi(z) &= A  z^{d/2} K_\nu(k z)+B  z^{d/2} I_\nu(k z) \\
    \Psi(w = 2z_0 - z) &= a w^{d/2} K_\nu(k w) + b w^{d/2} I_\nu(k w).  \label{gbc}
\end{align}

In the boundary $z,w \rightarrow 0$ limits,  we obtain
\begin{align}
    \Phi(z) &= \alpha_1 z^{d-\Delta} + \beta_1 z^{\Delta} + ...\\
    \alpha_1 &= A \frac{\pi e^{ikx}}{2 \sin \nu\pi} \frac{1}{\Gamma(1-\nu)} \qty(\frac{k}{2})^{-\nu} \\
    \beta_1 &= -A \frac{\pi e^{ikx}}{2 \sin \nu\pi} \frac{1}{\Gamma(1+\nu)} \qty(\frac{k}{2})^\nu + B \frac{e^{ikx}}{\Gamma(1+\nu)} \qty(\frac{k}{2})^\nu \\
    \Psi(w) &= \alpha_2 w^{d-\Delta} + \beta_2 w^{\Delta} + ...\\ \label{psi}
    \alpha_2 &= a \frac{\pi e^{ikx}}{2 \sin \nu\pi} \frac{1}{\Gamma(1-\nu)} \qty(\frac{k}{2})^{-\nu} \\
    \beta_2 &= -a \frac{\pi e^{ikx}}{2 \sin \nu\pi} \frac{1}{\Gamma(1+\nu)} \qty(\frac{k}{2})^{\nu} + b \frac{e^{ikx}}{\Gamma(1+\nu)} \qty(\frac{k}{2})^{\nu},
\end{align}
where we introduced the conformal dimension $\Delta=\frac{d}{2}+\nu$ of the operators $\mO_1$ and $\mO_2$.

Note that $\ap_i$ and $\beta_i$ are dual to the sources $J_i$ and the expectation values 
$\la O_i\lb$, respectively, for the dual scalar operators $\mO_i$ 
in CFT$_{(i)}$ with $i=1,2$, which is a straightforward extension of those in the standard AdS/CFT \cite{Klebanov:1999tb}. We are interested in the two point functions $\la \mO_1(k)\mO_1(-k)\lb(=\la \mO_2(k)\mO_2(-k)\lb)$ in the same CFT and $\la \mO_1(k)\mO_2(-k)\lb$ in the two different CFTs. They can be computed from the ratio $\frac{\beta_1}{\ap_1}$ and  $\frac{\beta_2}{\ap_1}$ when the source of CFT$_{(2)}$ is vanishing $\ap_2=0$, following the standard prescription.

At the gluing point $z=z_0$, we require that the scalar field and its derivative is continuous: 
\begin{align}
   \Phi(z=z_0)&=\Psi(w=z_0),  \label{gluingcnfda} \\ 
   \partial_z\Phi(z=z_0)&+\partial_w\Psi(w=z_0)=0.  \label{gluingcnfdb}
\end{align}

We assume that there are no source in the CFT$_{(2)}$ i.e. $a = 0$. Then, by imposing the condition (\ref{gbc}) we can express $\beta$ and $b$ in terms of $\ap$. In this way we obtain the two point functions:
\begin{align}
  &P(\nu,k,z=z_0,d) \coloneqq \ev{\mO_1(k)\mO_1(-k)} = -\frac{\beta_1}{\alpha_1} \no
&=\frac{\Gamma(1-\nu)}{\Gamma(1+\nu)}\left(\frac{k}{2}\right)^{2\nu} \frac{kz_0I_{\nu-1}(kz_0)I_{-\nu}(kz_0)+
\left(kz_0 I_{1-\nu}(kz_0)+(d-2\nu)I_{-\nu}(kz_0)\right)I_{\nu}(kz_0)}
{(d-2 \nu) I_\nu(k z_0){}^2+2 k z_0 I_{\nu-1}(k z_0) I_\nu(k z_0)}.\\
    &Q(\nu,k,z=z_0,d) \coloneqq \ev{\mO_1(k)\mO_2(-k)} = \frac{\beta_2}{\alpha_1} \no
  &= \frac{\Gamma(1-\nu)}{\Gamma(1+\nu)}\qty(\frac{k}{2})^{2\nu}\frac{2 \sin \nu\pi}{\pi}\frac{1}{(d-2 \nu) I_\nu(k z_0){}^2+2 k z_0 I_{\nu-1}(k z_0) I_\nu(k z_0)}.
  \label{Qtpwh}
\end{align}

\begin{figure}[h]
  \centering
  \begin{minipage}{0.3\columnwidth}
    \centering
    \includegraphics[width=4.5cm]{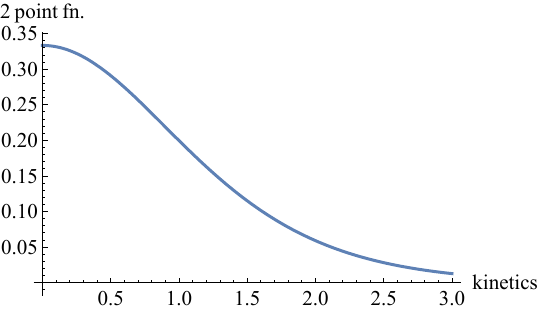}
  \end{minipage}
  \hspace{5mm}
  \begin{minipage}{0.3\columnwidth}
    \centering
    \includegraphics[width=4.5cm]{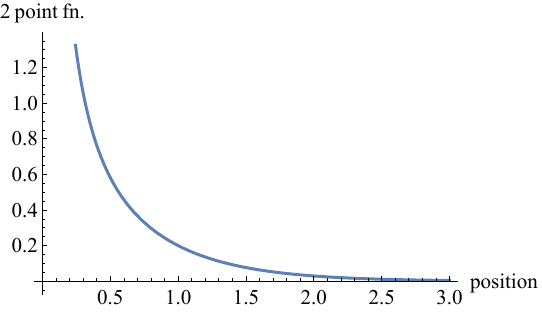}
  \end{minipage}
  \hspace{5mm}
  \begin{minipage}{0.3\columnwidth}
    \centering
   \includegraphics[width=6cm]{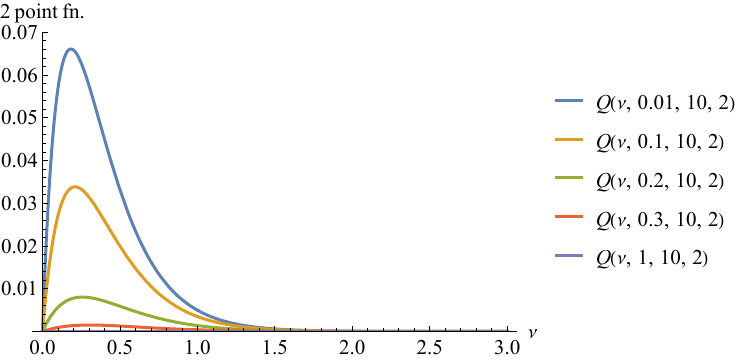}
  \end{minipage}
  \caption{Plots of $\la \mO_1\mO_2\lb$ for $d=2$. The left graph shows $\la \mO_1\mO_2\lb$ as a function of $k$ at $\nu=\frac{1}{2}$. The middle one does  $\la \mO_1\mO_2\lb$ as a function of $z_0$ at $\nu=\frac{1}{2}$. The right one does  $\la \mO_1\mO_2\lb$ as a function of $\nu$ for various values of $k$ at $z_0=10$.}
    \label{fig:twopt}
\end{figure}

In the UV limit ($k z_0 \gg 1$), we obtain
\begin{align}
  P(\nu,k,z=z_0,d)   &\simeq \frac{\Gamma(1-\nu)}{\Gamma(1+\nu)}\qty(\frac{k}{2})^{2\nu} \label{UVTPP}\\
  Q(\nu,k,z=z_0,d) &\simeq \frac{2 \sin \nu\pi\Gamma(1-\nu)}{\Gamma(1+\nu)}\qty(\frac{k}{2})^{2\nu}e^{-2k z_0}.  \label{qqcow}
\end{align}
Since the two AdS geometries are connected at $z=z_0$, we expect that only low energy modes 
$k\sim 1/z_0$ can detect that CFT$_{(1)}$ is connected to CFT$_{(2)}$. Indeed, the two point function $\la \mO_1(k)\mO_2(-k)\lb$ shows the exponential decay for high energy modes in
(\ref{qqcow}), while $\la \mO_1(k)\mO_1(-k)\lb$ on the same CFT does not decay. We draw graphs about two point function of $\ev{\mO_1(k)\mO_2(-k)}$ in Fig.\ref{fig:twopt}.

In the IR limit ($k z_0 \ll 1$), we obtain 
\begin{align}
  P(\nu,k,z=z_0,d) &\simeq \frac{d}{d+2\nu} \frac{1}{z_0^{2\nu}} + \LandauO{kz_0} 
  \label{PIR}\\
  Q(\nu,k,z=z_0,d) &\simeq \frac{2\nu}{d+2\nu}\frac{1}{z_0^{2\nu}}+ \LandauO{kz_0}. \label{QIR}
\end{align}

It might also be useful to write an example which allows us to express the two point functions in terms of elementary functions by choose $d=2$ (i.e. AdS$_3$) and 
$m^2=-\frac{3}{4}$ or $\nu=\frac{1}{2}$. They are explicitly given by
\begin{align*}
  P(1/2, k, z, 2) &= \frac{k(-1+2kz_0+e^{4kz_0}(1+2kz_0))}{1-2e^{2kz_0}-2kz_0+e^{4kz_0}(1+2kz_0)},\\
  Q(1/2, k, z, 2) &= 
  \frac{4 k^2 z_0 e^{2kz_0}}{1-2e^{2kz_0}-2kz_0+e^{4kz_0}(1+2kz_0)}.
\end{align*}

\begin{figure}[t]
		\centering
		\includegraphics[width=9cm]{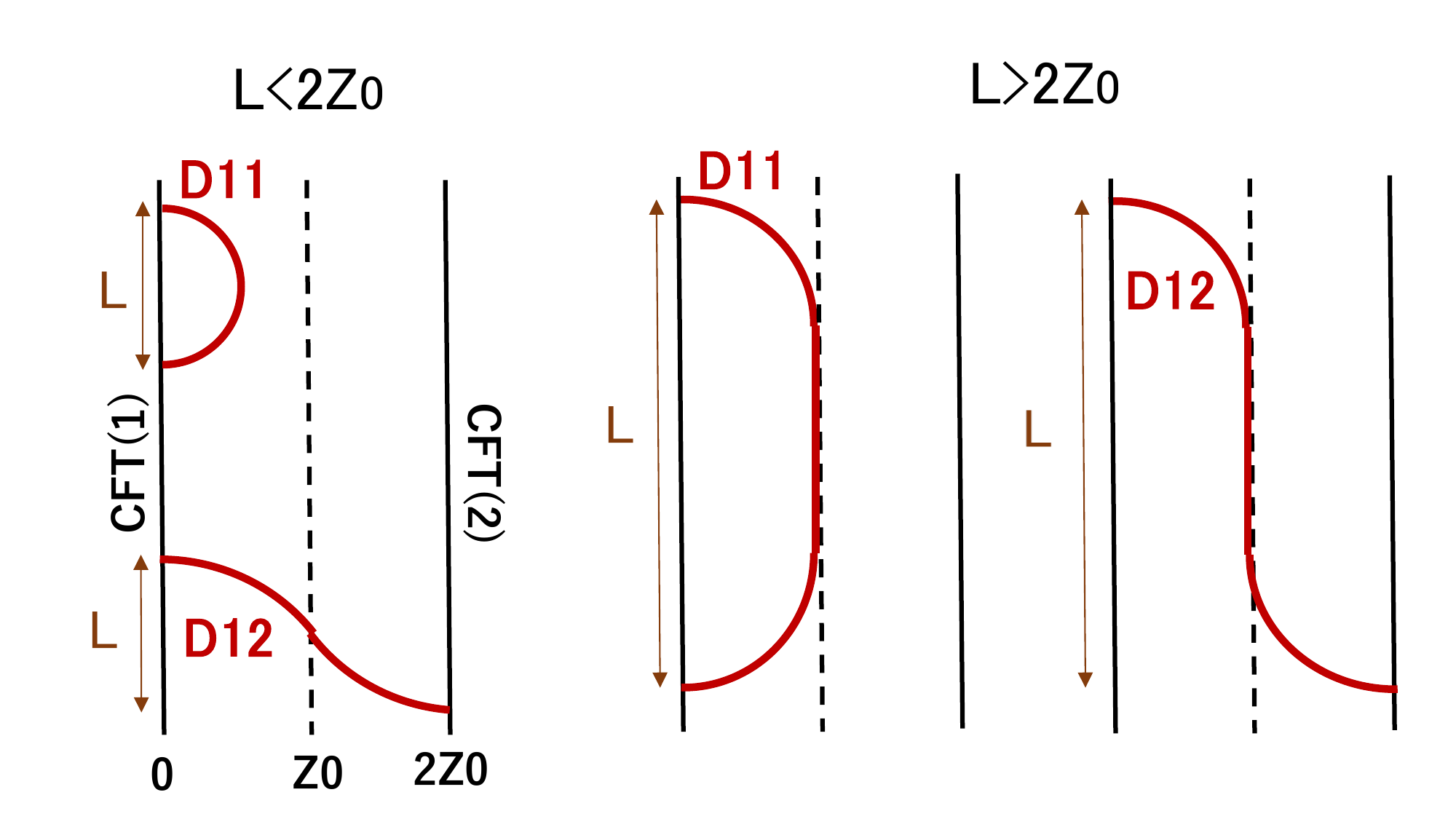}
		\caption{Geodesics in the traversable AdS wormhole. The geodesic curves show the phase transition at $l=2z_0$, where $l$ is the size of the interval $l=|x_a-x_b|$.} 
		\label{fig:geodesiclength}
\end{figure}

For $\Delta$ large, we can approximate the 
two point function by the bulk geodesic length \cite{Balasubramanian:1999zv}.  The geodesic distance $D$ between $(z_a,x_a)$ and $(z_b,x_b)$ in a Poincar\'e AdS$_{d+1}$ reads
\ba
\cosh D=\frac{(z_a)^2+(z_b)^2+(x_a-x_b)^2}{2z_az_b}.  \label{geodesicL}
\ea
The geodesic length $D_{11}$ between $(\ep,x_a)$ and $(\ep,x_b)$ on the same boundary, dual to CFT$_{(1)}$ is given by (refer to the left panel of Fig.\ref{fig:geodesiclength})
\ba
D_{11}= \left\{
\begin{array}{ll}
\log \frac{(x_a-x_b)^2}{\ep^2} &\ \ \  (|x_a-x_b|\leq 2z_0)\\
\log \frac{4z_0^2}{\ep^2}+\frac{|x_a-x_b|-2z_0}{z_0}  &\ \ \   (|x_a-x_b|> 2z_0).
\end{array}
\right.\label{geodesicA}
\ea
On the other hand, the geodesic length $D_{12}$ between $(\ep,x_a)$ and $(2z_0-\ep,x_b)$ which bridges the two different CFTs reads (refer to the middle and right panel of Fig.\ref{fig:geodesiclength}):
\ba
D_{12}= \left\{
\begin{array}{ll}
2\log \frac{\frac{(x_a-x_b)^2}{4}+z_0^2}{\ep z_0} &\ \ \  (|x_a-x_b|\leq 2z_0)\\
\log \frac{4z_0^2}{\ep^2}+\frac{|x_a-x_b|-2z_0}{z_0}  &\ \ \   (|x_a-x_b|>2z_0).
\end{array}
\right.\label{geodesicB}
\ea

The geodesics approximations of $\la \mO_1(x_a)\mO_1(x_b)\lb$ and $\la \mO_1(x_a)\mO_2(x_b)\lb$ are given by 
\ba
&& \la \mO_1(x_a)\mO_1(x_b)\lb \simeq  e^{-\Delta D_{11}},\ \ \   \la \mO_1(x_a)\mO_2(x_b)\lb \simeq  e^{-\Delta D_{12}},\label{geoapw}
\ea
It is clear that when $kz_0\gg 1$, $\la \mO_1(x_a)\mO_1(x_b)\lb$ is the same as the standard two point function in CFT $\la \mO_1(x_a)\mO_1(x_b)\lb\propto |x_a-x_b|^{-2\Delta}$, matching with the Fourier transformed result (\ref{UVTPP}). The other two point functon $\la \mO_1(x_a)\mO_2(x_b)\lb$ can be written after the Fourier transformation (we set $d=2$ for simplicity):
\ba
\la \mO_1(k)\mO_2(-k)\lb&=&\int^{2\pi}_0  d\theta \int^\infty_0 rdr \left(\frac{4\ep z_0}{4z_0^2+r^2}\right)^{2\Delta}
e^{ik\cos\theta r}\no
&=&2\pi \int^\infty_0 rdr J_0(kr)\left(\frac{4\ep z_0}{4z_0^2+r^2}\right)^{2\Delta}\propto k^{2\Delta-1}K_{1-2\Delta}(2kz_0).
\ea
This behaves as $\propto k^{2\Delta-\frac{3}{2}} e^{-2kz_0}$ at $kz_0\to\infty$, which agrees with (\ref{qqcow})  when  $\Delta\gg 1$.

In the low momenta limit $kz_0\ll 1$, we apply the geodesic length formula for $|x_a-x_b|>z_0$ and find that both $\la \mO_1(x_a)\mO_1(x_b)\lb$ and $\la \mO_1(x_a)\mO_2(x_b)\lb$ decays exponentially as a function of $|x_a-x_b|$. Therefore, Fourier transformed two point functions approach to constants, being consistent with 
the behaviors (\ref{PIR}) and (\ref{QIR}).

It is also helpful to consider the Lorentzian continuation $x^0\to it$ of $\la \mO_1\mO_2\lb$ as it reflects the property of the traversable wormhole.\footnote{Here we define the Lorentzian two point functions in the path-integral formalism by simply Wick rotating the Euclidean ones. Note that since it dual CFT may include interactions between the two CFTs, we cannot simply define the orderings of the operators and employ the time-ordered products.} From the UV expression 
(\ref{qqcow}) we obtain the following behavior near the singularity 
at $-t^2+x^2+4z_0^2=0$:
\ba
\la \mO_1(t,x)\mO_2(0,0)\lb\sim \frac{1}{\left(-t^2+x^2+4z_0^2\right)^{d+2\nu-\frac{1}{2}}}.\label{shortq}
\ea
This also agrees with the geodesic approximation by applying (\ref{geoapw}) for $\nu\gg 1$. This singularity at $-t^2+x^2+4z_0^2=0$ is clearly explained in the gravity dual because the two points are connected by a null geodesic in our wormhole geometry.

\subsection{Holographic entanglement (pseudo) entropy}

For a quantum state described by a density matrix $\rho$, which is hermitian, the entanglement entropy is defined as follows. First we decompose the total system into $A$ and $B$, which is conveniently done by dividing a time slice into two subregions $A$ and $B$. This bipartite decomposition allows us to define the reduced density matrix  $\rho_A=\mbox{Tr}_B\rho$,
by tracing out the subsystem $B$. Finally the entanglement entropy is defined by the von-Neumann entropy
\ba
S_A=-\mbox{Tr}[\rho_A\log\rho_A].  \label{vN}
\ea
In the AdS/CFT, we can calculate $S_A$ from the area of extremal surface 
$\Gamma_A$, which ends on the boundary of $A$ by the area formula \cite{Ryu:2006bv,Ryu:2006ef,Hubeny:2007xt}:
\ba
S_A=\frac{\mbox{Area}(\Gamma_A)}{4\GN }.
\ea
In the present setup of the traversable wormholes, however, as we will explain in detail next section, generically we expect that the dual state is described not by a hermitian density matrix but by a non-hermitian transition matrix, which is still denoted by $\rho$. In this case, the quantity defined by the same formula (\ref{vN}) is called the pseudo entropy \cite{Nakata:2021ubr}, which takes complex values in general. Below we present results of the holographic pseudo entropy, focusing on the three dimensional (i.e. $d=2$) case for simplicity.

If we choose $A$ and $B$ to be the entire CFT$_{(1)}$ and CFT$_{(2)}$, respectively, we find
\ba
S_A=\frac{L}{4\GN z_0}=\frac{c}{6}\cdot \frac{L}{z_0},
\ea
where $L$ is the infinite length in $x$ direction and $c$ is the central charge in the dual CFT, given in terms of the Newton constant by $c=\frac{3}{2G_N}$ \cite{Brown1986}.

Next we consider the simple setup where the subsystem $A$ is given by an interval length $l=|x_a-x_b|$ in the CFT$_{(1)}$. We already computed the geodesic length in (\ref{geodesicA}) and the entropy is given by $S_A=\frac{c}{6}D_{11}$, which shows the phase transition at $l=2z_0$. Note that when $L\gg 2z_0$, $S_A$ grows linearly with respect to $L$. Similarly, when $A$ and $B$ are given by semi-infinite lines $x\leq x_a$ and $x\leq x_b$ in the CFT$_{(1)}$ and CFT$_{(2)}$, respectively, the entropy is $S_{AB}=\frac{c}{6}D_{12}$, which shows a similar phase transition.

\subsubsection{Two intervals (different sides)}
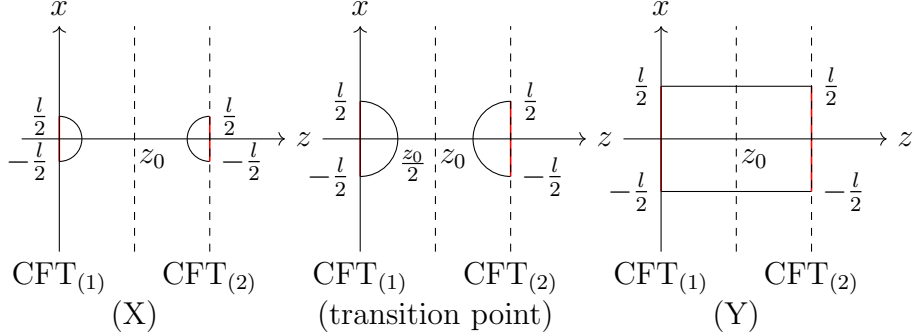
\begin{figure}[ht]
  \centering
  \begin{tikzpicture}
    \draw(90:0.3) arc (90:-90:0.3cm);
    \draw[red,semithick](0,-0.3) -- (0,0.3);
    \draw(0,0.3)node[anchor=east]{$\frac{l}{2}$};
    \draw(0,-0.3)node[anchor=east]{$-\frac{l}{2}$};
    \draw[xshift=2cm](90:0.3) arc (90:270:0.3cm);
    \draw[red,semithick](2,-0.3) -- (2,0.3);
    \draw[dashed] (1,-1.5) -- (1,1.5);
    \draw(1,-2.0)node[anchor=north]{(X)};
    \draw(1.1,0)node[anchor=north]{$\ \ z_0$};
    \draw[dashed] (2,-1.5)node[anchor=north]{CFT$_{(2)}$} -- (2,1.5);
    \draw(2,0.3)node[anchor=west]{$\frac{l}{2}$};
    \draw(2,-0.3)node[anchor=west]{$-\frac{l}{2}$};
    \draw[->] (-0.5,0) -- (3, 0)node[anchor=west]{$z$};
    \draw[->] (0,-1.5)node[anchor=north]{CFT$_{(1)}$} -- (0,1.5)node[anchor=south]{$x$};
  \begin{scope}[xshift=4cm]
    \draw(90:0.5) arc (90:-90:0.5cm);
    \draw[red,semithick](0,-0.5) -- (0,0.5);
    \draw(0,0.5)node[anchor=east]{$\frac{l}{2}$};
    \draw(0,-0.5)node[anchor=east]{$-\frac{l}{2}$};
    \draw[xshift=2cm](90:0.5) arc (90:270:0.5cm);
    \draw[red,semithick](2,-0.5) -- (2,0.5);
    \draw[dashed] (1,-1.5) -- (1,1.5);
    \draw(1,-2.0)node[anchor=north]{(transition point)};
    \draw(0.7,0)node[anchor=north]{{\small $\frac{z_0}{2}$}};
    \draw(1.1,0)node[anchor=north]{$\ \ z_0$};
    \draw[dashed] (2,-1.5)node[anchor=north]{CFT$_{(2)}$} -- (2,1.5);
    \draw(2,0.5)node[anchor=west]{$\frac{l}{2}$};
    \draw(2,-0.5)node[anchor=west]{$-\frac{l}{2}$};
    \draw[->] (-0.5,0) -- (3, 0)node[anchor=west]{$z$};
    \draw[->] (0,-1.5)node[anchor=north]{CFT$_{(1)}$} -- (0,1.5)node[anchor=south]{$x$};
  \end{scope}  
  \begin{scope}[xshift=8cm]
    \draw(0,0.7) -- (2,0.7);
    \draw[red,semithick](0,-0.7) -- (0,0.7);
    \draw(0,0.7)node[anchor=east]{$\frac{l}{2}$};
    \draw(0,-0.7)node[anchor=east]{$-\frac{l}{2}$};
    \draw(0,-0.7) -- (2,-0.7);
    \draw[red,semithick](2,-0.7) -- (2,0.7);
    \draw[dashed] (1,-1.5) -- (1,1.5);
    \draw(1,-2.0)node[anchor=north]{(Y)};
    \draw(1.1,0)node[anchor=north]{$\ \ z_0$};
    \draw[dashed] (2,-1.5)node[anchor=north]{CFT$_{(2)}$} -- (2,1.5);
    \draw(2,0.7)node[anchor=west]{$\frac{l}{2}$};
    \draw(2,-0.7)node[anchor=west]{$-\frac{l}{2}$};
    \draw[->] (-0.5,0) -- (3, 0)node[anchor=west]{$z$};
    \draw[->] (0,-1.5)node[anchor=north]{CFT$_{(1)}$} -- (0,1.5)node[anchor=south]{$x$};
  \end{scope}
\end{tikzpicture}
\caption{EE with two symmetric regions in different side. Phase $X$ and $Y$ corresponds to $l<z_0$ and $l>z_0$.}\label{tatA1}
\end{figure}

Now we consider  an interval $A$ in CFT$_{(1)}$ and another interval $B$ in CFT$_{(2)}$. We assume that the length of each interval is $l$. In the symmetric case: $A=B=\left[-\frac{l}{2},\frac{l}{2}\right]$, we find
\ba
S_{AB}= \left\{
\begin{array}{ll}
\frac{2c}{3}\log \frac{l}{\ep} &\ \ \  (l\leq z_0)\\
\frac{2c}{3}\log \frac{z_0}{\ep}  &\ \ \   (l> z_0).
\end{array}
\right.\label{doubleS}
\ea
Refer to Fig.\ref{tatA1} for the profile of the geodesic $\Gamma_{AB}$.

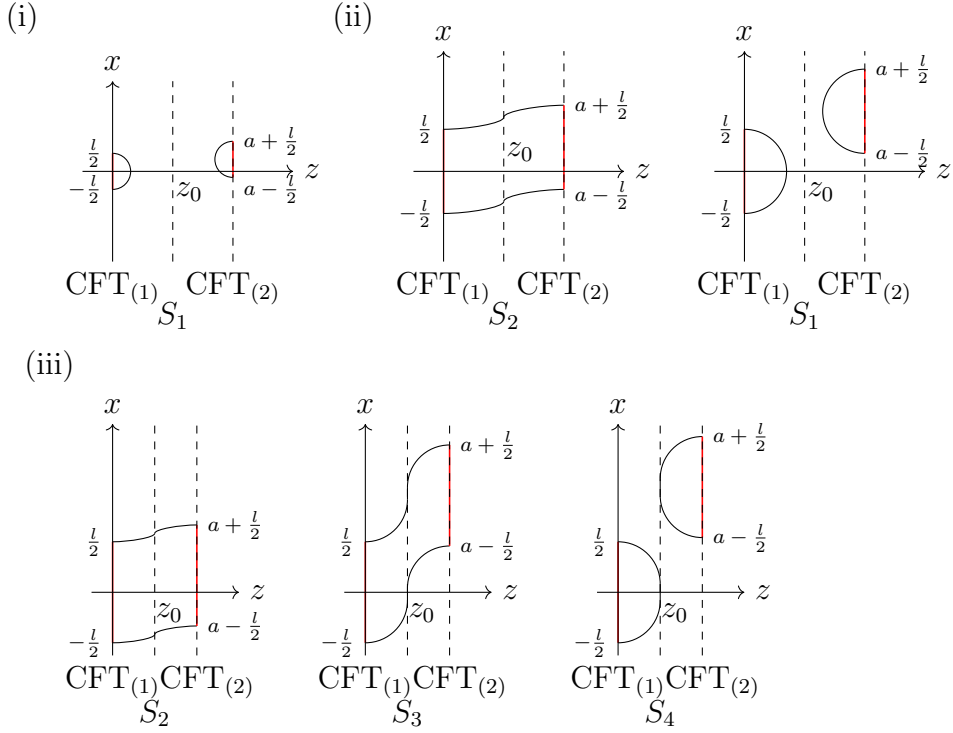
\begin{figure}[ht]
  \label{Fig:btdl}
  \centering
  \begin{tikzpicture}
  \begin{scope}[scale=0.8]
  \draw(-1.5,3)node[anchor=north]{(i)};
    \draw(90:0.3) arc (90:-90:0.3cm);
    \draw[red,semithick](0,-0.3) -- (0,0.3);
    \draw(0,0.3)node[anchor=east,font=\scriptsize]{$\frac{l}{2}$};
    \draw(0,-0.3)node[anchor=east,font=\scriptsize]{$-\frac{l}{2}$};
    \draw[xshift=2cm](90:0.5) arc (90:270:0.3cm);
    \draw[red,semithick](2,-0.1) -- (2,0.5);
    \draw[dashed] (1,-1.5) -- (1,1.5);
    \draw(1,-2.0)node[anchor=north]{$S_1$};
    \draw(1.1,0)node[anchor=north]{$\ \ z_0$};
    \draw[dashed] (2,-1.5)node[anchor=north]{CFT$_{(2)}$} -- (2,1.5);
    \draw(2,0.5)node[anchor=west,font=\scriptsize]{$a+\frac{l}{2}$};
    \draw(2,-0.3)node[anchor=west,font=\scriptsize]{$a-\frac{l}{2}$};
    \draw[->] (-0.5,0) -- (3, 0)node[anchor=west]{$z$};
    \draw[->] (0,-1.5)node[anchor=north]{CFT$_{(1)}$} -- (0,1.5)node[anchor=south]{$x$};
  \begin{scope}[xshift=5.5cm]
  \draw(-1.5,3)node[anchor=north]{(ii)};
    \draw[red,semithick](0,-0.7) -- (0,0.7);
    \draw(0,0.7)node[anchor=east,font=\scriptsize]{$\frac{l}{2}$};
    \draw(0,-0.7)node[anchor=east,font=\scriptsize]{$-\frac{l}{2}$};
    \draw(0, 0.7) .. controls (0.4,0.7) and (1,0.8) ..(1,0.9);
    \draw(1, 0.9) .. controls (1,1) and (1.6,1.1) ..(2,1.1);
    \draw(0, -0.7) .. controls (0.4,-0.7) and (1,-0.6) ..(1,-0.5);
    \draw(1, -0.5) .. controls (1,-0.4) and (1.6,-0.3) ..(2,-0.3);
    \draw[red,semithick](2,-0.3) -- (2,1.1);
    \draw[dashed] (1,-1.5) -- (1,2);
    \draw(1,-2.0)node[anchor=north]{$S_2$};
    \draw(1.1,0)node[anchor=south]{$\ \ z_0$};
    \draw[dashed] (2,-1.5)node[anchor=north]{CFT$_{(2)}$} -- (2,2);
    \draw(2,1.1)node[anchor=west,font=\scriptsize]{$a+\frac{l}{2}$};
    \draw(2,-0.4)node[anchor=west,font=\scriptsize]{$a-\frac{l}{2}$};
    \draw[->] (-0.5,0) -- (3, 0)node[anchor=west]{$z$};
    \draw[->] (0,-1.5)node[anchor=north]{CFT$_{(1)}$} -- (0,2)node[anchor=south]{$x$};
  \end{scope}
  \begin{scope}[xshift=10.5cm]
    \draw(90:0.7) arc (90:-90:0.7cm);
    \draw[red,semithick](0,-0.7) -- (0,0.7);
    \draw(0,0.7)node[anchor=east,font=\scriptsize]{$\frac{l}{2}$};
    \draw(0,-0.7)node[anchor=east,font=\scriptsize]{$-\frac{l}{2}$};
    \draw[xshift=2cm](90:1.7) arc (90:270:0.7cm);
    \draw[red,semithick](2,0.3) -- (2,1.7);
    \draw[dashed] (1,-1.5) -- (1,2);
    \draw(1,-2.0)node[anchor=north]{$S_1$};
    \draw(1.1,0)node[anchor=north]{$\ \ z_0$};
    \draw[dashed] (2,-1.5)node[anchor=north]{CFT$_{(2)}$} -- (2,2);
    \draw(2,1.7)node[anchor=west,font=\scriptsize]{$a+\frac{l}{2}$};
    \draw(2,0.3)node[anchor=west,font=\scriptsize]{$a-\frac{l}{2}$};
    \draw[->] (-0.5,0) -- (3, 0)node[anchor=west]{$z$};
    \draw[->] (0,-1.5)node[anchor=north]{CFT$_{(1)}$} -- (0,2)node[anchor=south]{$x$};
  \end{scope}  
  \begin{scope}[yshift=-7cm,scale=0.7]
  \draw(-1.5,6)node[anchor=north]{(iii)};
    
  \begin{scope}[xshift=0cm]
    \draw[red,semithick](0,-1.2) -- (0,1.2);
    \draw(0,1.2)node[anchor=east,font=\scriptsize]{$\frac{l}{2}$};
    \draw(0,-1.2)node[anchor=east,font=\scriptsize]{$-\frac{l}{2}$};
    \draw(0, 1.2) .. controls (0.4,1.2) and (1,1.3) ..(1,1.4);
    \draw(1, 1.4) .. controls (1,1.5) and (1.6,1.6) ..(2,1.6);
    \draw(0, -1.2) .. controls (0.4,-1.2) and (1,-1.1) ..(1,-1);
    \draw(1, -1) .. controls (1,-0.9) and (1.6,-0.8) ..(2,-0.8);
    \draw[red,semithick](2,-0.8) -- (2,1.6);
    \draw[dashed] (1,-1.5) -- (1,4);
    \draw(1,-2.3)node[anchor=north]{$S_2$};
    \draw(1.1,0)node[anchor=north]{$\ \ z_0$};
    \draw[dashed] (2,-1.5)node[anchor=north]{$\ \ $CFT$_{(2)}$} -- (2,4);
    \draw(2,1.6)node[anchor=west,font=\scriptsize]{$a+\frac{l}{2}$};
    \draw(2,-0.8)node[anchor=west,font=\scriptsize]{$a-\frac{l}{2}$};
    \draw[->] (-0.5,0) -- (3, 0)node[anchor=west]{$z$};
    \draw[->] (0,-1.5)node[anchor=north]{CFT$_{(1)}$} -- (0,4)node[anchor=south]{$x$};
  \end{scope}
  \begin{scope}[xshift=6cm]
    \draw(90:1.2) arc (-90:0:1.0cm);
    \draw(2,3.5) arc (90:180:1.0cm);
    \draw[red,semithick](0,-1.2) -- (0,1.2);
    \draw(0,-1.2)node[anchor=east,font=\scriptsize]{$-\frac{l}{2}$};
    \draw(0,1.2)node[anchor=east,font=\scriptsize]{$\frac{l}{2}$};
    \draw(1,2.2) -- (1,2.5);
    \draw[yshift=-2.4cm](90:1.2) arc (-90:0:1.0cm);
    \draw[yshift=-2.4cm](2,3.5) arc (90:180:1.0cm);
    \draw[red,semithick](2,1.1) -- (2,3.5);
    \draw(2,1.1)node[anchor=west,font=\scriptsize]{$a-\frac{l}{2}$};
    \draw(2,3.5)node[anchor=west,font=\scriptsize]{$a+\frac{l}{2}$};
    \draw(1,-0.2) -- (1,0.1);
    \draw[dashed] (1,-1.5) -- (1,4);
    \draw(1,-2.3)node[anchor=north]{$S_3$};
    \draw(1.1,0)node[anchor=north]{$\ \ z_0$};
    \draw[dashed] (2,-1.5)node[anchor=north]{$\ \ $CFT$_{(2)}$} -- (2,4);
    \draw[->] (-0.5,0) -- (3, 0)node[anchor=west]{$z$};
    \draw[->] (0,-1.5)node[anchor=north]{CFT$_{(1)}$} -- (0,4)node[anchor=south]{$x$};
  \end{scope}  
  \begin{scope}[xshift=12cm]
    \draw(90:1.2) arc (90:0:1.0cm);
    \draw(2,3.7) arc (90:180:1.0cm);
    \draw(1,0.2) -- (1,-0.2);
    \draw[red,semithick](0,-1.2) -- (0,1.2);
    \draw(0,-1.2)node[anchor=east,font=\scriptsize]{$-\frac{l}{2}$};
    \draw(0,1.2)node[anchor=east,font=\scriptsize]{$\frac{l}{2}$};
    \draw[yshift=-2.4cm](90:1.2) arc (-90:0:1.0cm);
    \draw[yshift=-2.4cm](2,3.7) arc (-90:-180:1.0cm);
    \draw(1,2.3) -- (1,2.7);
    \draw[red,semithick](2,1.3) -- (2,3.7);
    \draw(2,1.3)node[anchor=west,font=\scriptsize]{$a-\frac{l}{2}$};
    \draw(2,3.7)node[anchor=west,font=\scriptsize]{$a+\frac{l}{2}$};
    \draw[dashed] (1,-1.5) -- (1,4);
    \draw(1,-2.3)node[anchor=north]{$S_4$};
    \draw(1.1,0)node[anchor=north]{$\ \ z_0$};
    \draw[dashed] (2,-1.5)node[anchor=north]{$\ \ $CFT$_{(2)}$} -- (2,4);
    \draw[->] (-0.5,0) -- (3, 0)node[anchor=west]{$z$};
    \draw[->] (0,-1.5)node[anchor=north]{CFT$_{(1)}$} -- (0,4)node[anchor=south]{$x$};
  \end{scope}  
  \end{scope}
  \end{scope}
\end{tikzpicture}
\caption{EE with two regions in different side}\label{figasym}
\end{figure}

More generally, let us consider the case when the location of the two intervals is shifted i.e. $A=\left[-\frac{l}{2},\frac{l}{2}\right]$ and $B=\left[-\frac{l}{2}+a,\frac{l}{2}+a\right]$. Eventually, the resulting entropy is found as follows (refer to Fig.\ref{figasym} for a sketch of each phase $S_1,S_2,S_3$ and $S_4$). When (i) $0\leq l<z_0$, the disconnected phase is always favored and we have $S_{AB}=S_1=\frac{2c}{3}\log\frac{l}{\ep}$.  When (ii) $z_0\leq l<2z_0$, we have a phase transition between the connected and disconnected phase:
\ba
S_{AB}= \left\{
\begin{array}{ll}
S_2=\frac{2c}{3}\log\frac{z_0^2+\frac{a^2}{4}}{\ep z_0} &\ \ \  (0\leq a<2\s{(l-z_0)z_0})\\
S_1= \frac{2c}{3}\log\frac{l}{\ep} &\ \ \   (a>2\s{(l-z_0)z_0}),
\end{array}
\right.\label{geodesicBC}
\ea
where we apply the standard rule of phase transition \cite{Headrick:2007km} by choosing the extremal surface with the smallest area among multiple candidates. 

Finally, when (iii) $l\geq 2z_0$, there are three phases:
\ba
S_{AB}= \left\{
\begin{array}{ll}
S_2=\frac{2c}{3}\log\frac{z_0^2+\frac{a^2}{4}}{\ep z_0} &\ \ \  (0\leq a<2z_0)\\
S_3= \frac{2c}{3}\log\frac{2z_0}{\ep}+\frac{c}{3 }\cdot\frac{a-2z_0}{z_0} &\ \ \   (2z_0\leq a<l)\\
S_4=\frac{2c}{3}\log\frac{2z_0}{\ep}+\frac{c}{3}\cdot\frac{l-2z_0}{z_0} &\ \ \   (a\geq l)
\end{array}
\right.\label{geodesicDF}
\ea

\subsubsection{Two intervals (same side)}

\begin{figure}[ht]
  \centering
  \begin{tikzpicture}
  \begin{scope}[scale=0.8]
  \begin{scope}[xshift=0cm]
    \draw(-1.5,2.0)node[anchor=south]{(i)};
    \draw[yshift=0.7cm](90:0.4) arc (90:-90:0.4cm);
    \draw[yshift=-0.7cm](90:0.4) arc (90:-90:0.4cm);
    \draw[red,semithick](0,-1.1) -- (0,-0.3);
    \draw(0,1.1)node[anchor=east,font=\scriptsize]{$a+\frac{l}{2}$};
    \draw(0,0.4)node[anchor=east,font=\scriptsize]{$a-\frac{l}{2}$};
    \draw[red,semithick](0,0.3) -- (0,1.1);
    \draw(1,-2.0)node[anchor=north]{$S_1$};
    \draw(0,-0.4)node[anchor=east,font=\scriptsize]{$-a+\frac{l}{2}$};
    \draw(0,-1.1)node[anchor=east,font=\scriptsize]{$-a-\frac{l}{2}$};
    \draw[dashed] (1.2,-1.5) -- (1.2,1.5);
    \draw(1.3,0)node[anchor=north]{$\ \ \ z_0$};
    \draw[->] (-0.5,0) -- (1.8, 0)node[anchor=west]{$\ $};
    \draw[->] (0,-1.5)node[anchor=north]{CFT$_{(1)}$} -- (0,1.5)node[anchor=south]{$x$};
  \end{scope}
  \begin{scope}[xshift=3.5cm]
    \draw(90:0.9) arc (90:-90:0.9cm);
    \draw(90:0.1) arc (90:-90:0.1cm);
    \draw[red,semithick](0,-0.9) -- (0,-0.1);
    \draw(0,1)node[anchor=east,font=\scriptsize]{$a+\frac{l}{2}$};
    \draw(0,0.4)node[anchor=east,font=\scriptsize]{$a-\frac{l}{2}$};
    \draw[red,semithick](0,0.1) -- (0,0.9);
    \draw(1,-2.0)node[anchor=north]{$S_2$};
    \draw(0,-0.4)node[anchor=east,font=\scriptsize]{$-a+\frac{l}{2}$};
    \draw(0,-1)node[anchor=east,font=\scriptsize]{$-a-\frac{l}{2}$};
    \draw[dashed] (1.5,-1.5) -- (1.5,1.5);
    \draw(1.5,0)node[anchor=north]{$\ \ \ z_0$};
    \draw[->] (-0.5,0) -- (2.0, 0)node[anchor=west]{$z$};
    \draw[->] (0,-1.5)node[anchor=north]{CFT$_{(1)}$} -- (0,1.5)node[anchor=south]{$x$};
  \end{scope}  
  \begin{scope}[xshift=9cm]]
    \draw(-1.5,2.0)node[anchor=south]{(ii)};
    \draw[yshift=0.7cm](90:0.45) arc (90:-90:0.45cm);
    \draw[yshift=-0.7cm](90:0.45) arc (90:-90:0.45cm);
    \draw[red,semithick](0,-1.15) -- (0,-0.25);
    \draw(0,1.1)node[anchor=east,font=\scriptsize]{$a+\frac{l}{2}$};
    \draw(0,0.4)node[anchor=east,font=\scriptsize]{$a-\frac{l}{2}$};
    \draw[red,semithick](0,0.25) -- (0,1.15);
    \draw[dashed] (1,-1.5) -- (1,1.5);
    \draw(1,-2.0)node[anchor=north]{$S_1$};
    \draw(1.1,0)node[anchor=north]{$\ \ z_0$};
    \draw(0,-0.4)node[anchor=east,font=\scriptsize]{$-a+\frac{l}{2}$};
    \draw(0,-1.1)node[anchor=east,font=\scriptsize]{$-a-\frac{l}{2}$};
    \draw[->] (-0.5,0) -- (2.0, 0)node[anchor=west]{$\ $};
    \draw[->] (0,-1.5)node[anchor=north]{CFT$_{(1)}$} -- (0,1.5)node[anchor=south]{$x$};
  \end{scope}
  \begin{scope}[xshift=13cm]
    \draw(90:1.1) arc (90:0:1cm);
    \draw(90:-1.1) arc (-90:0:1cm);
    \draw(90:0.15) arc (90:-90:0.15cm);
    \draw[red,semithick](0,-1.1) -- (0,-0.15);
    \draw(0,1.1)node[anchor=east,font=\scriptsize]{$a+\frac{l}{2}$};
    \draw(0,0.4)node[anchor=east,font=\scriptsize]{$a-\frac{l}{2}$};
    \draw[red,semithick](0,0.15) -- (0,1.1);
    \draw[dashed] (1,-1.5) -- (1,1.5);
    \draw(1,-0.1) -- (1,0.1);
    \draw(1,-2.0)node[anchor=north]{$S_3$};
    \draw(1.1,0)node[anchor=north]{$\ \ z_0$};
    \draw(0,-0.3)node[anchor=east,font=\scriptsize]{$-a+\frac{l}{2}$};
    \draw(0,-1.1)node[anchor=east,font=\scriptsize]{$-a-\frac{l}{2}$};
    \draw[->] (-0.5,0) -- (2.0, 0)node[anchor=west]{$\ $};
    \draw[->] (0,-1.5)node[anchor=north]{CFT$_{(1)}$} -- (0,1.5)node[anchor=south]{$x$};
  \end{scope}
  \begin{scope}[xshift=17cm]
    \draw(90:0.95) arc (90:-90:0.95cm);
    \draw(90:0.1) arc (90:-90:0.1cm);
    \draw[red,semithick](0,-0.95) -- (0,-0.1);
    \draw(0,1)node[anchor=east,font=\scriptsize]{$a+\frac{l}{2}$};
    \draw(0,0.4)node[anchor=east,font=\scriptsize]{$a-\frac{l}{2}$};
    \draw[red,semithick](0,0.1) -- (0,0.95);
    \draw(1,-2.0)node[anchor=north]{$S_2$};
    \draw(0,-0.4)node[anchor=east,font=\scriptsize]{$-a+\frac{l}{2}$};
    \draw(0,-1)node[anchor=east,font=\scriptsize]{$-a-\frac{l}{2}$};
    \draw[dashed] (1.0,-1.5) -- (1.0,1.5);
    \draw(1.5,0)node[anchor=north]{$\ \ z_0$};
    \draw[->] (-0.5,0) -- (2.0, 0)node[anchor=west]{$z$};
    \draw[->] (0,-1.5)node[anchor=north]{CFT$_{(1)}$} -- (0,1.5)node[anchor=south]{$x$};
  \end{scope}  
  \begin{scope}[yshift=-6cm]
    \begin{scope}[xshift=0cm]]
    \draw(-1.5,2.0)node[anchor=south]{(iii)};
    \draw(90:1.6) arc (90:-90:0.6cm);
    \draw(90:-0.4) arc (90:-90:0.6cm);
    \draw[red,semithick](0,-1.6) -- (0,-0.4);
    \draw(0,1.6)node[anchor=east,font=\scriptsize]{$a+\frac{l}{2}$};
    \draw(0,0.4)node[anchor=east,font=\scriptsize]{$a-\frac{l}{2}$};
    \draw[red,semithick](0,0.4) -- (0,1.6);
    \draw[dashed] (1,-2) -- (1,2);
    \draw(1,-2.5)node[anchor=north]{$S_1$};
    \draw(1.1,0)node[anchor=north]{$\ \ z_0$};
    \draw(0,-0.4)node[anchor=east,font=\scriptsize]{$-a+\frac{l}{2}$};
    \draw(0,-1.4)node[anchor=east,font=\scriptsize]{$-a-\frac{l}{2}$};
    \draw[->] (-0.5,0) -- (2.0, 0)node[anchor=west]{$z$};
    \draw[->] (0,-2)node[anchor=north]{CFT$_{(1)}$} -- (0,2)node[anchor=south]{$x$};
  \end{scope}
  \begin{scope}[xshift=4cm]
    \draw(90:1.3) arc (90:0:1cm);
    \draw(0,-0.7) -- (0,0.7);
    \draw(90:-1.3) arc (-90:0:1cm);
    \draw(90:0.1) arc (90:-90:0.1cm);
    \draw[red,semithick](0,-1.3) -- (0,-0.1);
    \draw(0,1.3)node[anchor=east,font=\scriptsize]{$a+\frac{l}{2}$};
    \draw(0,0.4)node[anchor=east,font=\scriptsize]{$a-\frac{l}{2}$};
    \draw[red,semithick](0,0.1) -- (0,1.3);
    \draw[dashed] (1,-2) -- (1,2);
    \draw(1,-0.4) -- (1,0.3);
    \draw(1,-2.5)node[anchor=north]{$S_3$};
    \draw(1.1,0)node[anchor=north]{$\ \ z_0$};
    \draw(0,-0.3)node[anchor=east,font=\scriptsize]{$-a+\frac{l}{2}$};
    \draw(0,-1.3)node[anchor=east,font=\scriptsize]{$-a-\frac{l}{2}$};
    \draw[->] (-0.5,0) -- (2.0, 0)node[anchor=west]{$z$};
    \draw[->] (0,-2)node[anchor=north]{CFT$_{(1)}$} -- (0,2)node[anchor=south]{$x$};
  \end{scope}
  \begin{scope}[xshift=11cm]
  \draw(-1.5,2.0)node[anchor=south]{(iv)};
    \draw(90:1.8) arc (90:0:0.6cm);
    \draw(0.6,1) -- (0.6,1.2);
    \draw(90:0.4) arc (-90:0:0.6cm);
    \draw(-90:1.8) arc (-90:0:0.6cm);
    \draw(0.6,-1) -- (0.6,-1.2);
    \draw(-90:0.4) arc (90:0:0.6cm);
    \draw[red,semithick](0,-1.8) -- (0,-0.4);
    \draw(0,1.8)node[anchor=east,font=\scriptsize]{$a+\frac{l}{2}$};
    \draw(0,0.4)node[anchor=east,font=\scriptsize]{$a-\frac{l}{2}$};
    \draw[red,semithick](0,0.4) -- (0,1.8);
    \draw[dashed] (0.6,-2.0) -- (0.6,2.0);
    \draw(1,-2.5)node[anchor=north]{$S_1'$};
    \draw(0.8,0)node[anchor=north]{$\ z_0$};
    \draw(0,-0.4)node[anchor=east,font=\scriptsize]{$-a+\frac{l}{2}$};
    \draw(0,-1.8)node[anchor=east,font=\scriptsize]{$-a-\frac{l}{2}$};
    \draw[->] (-0.5,0) -- (1.7, 0)node[anchor=west]{$z$};
    \draw[->] (0,-2.0)node[anchor=north]{CFT$_{(1)}$} -- (0,2.0)node[anchor=south]{$x$};
  \end{scope}
  \begin{scope}[xshift=15cm]
    \draw(90:1.5) arc (90:0:0.6cm);
    \draw(0.6,-0.9) -- (0.6,0.9);
    \draw(90:0.1) arc (90:-90:0.1cm);
    \draw(-90:1.5) arc (-90:0:0.6cm);
    \draw[red,semithick](0,-1.5) -- (0,-0.1);
    \draw(0,1.3)node[anchor=east,font=\scriptsize]{$a+\frac{l}{2}$};
    \draw(0,0.3)node[anchor=east,font=\scriptsize]{$a-\frac{l}{2}$};
    \draw[red,semithick](0,0.1) -- (0,1.5);
    \draw[dashed] (0.6,-2) -- (0.6,2);
    \draw(1,-2.5)node[anchor=north]{$S_3$};
    \draw(0.8,0)node[anchor=north]{$\ z_0$};
    \draw(0,-0.3)node[anchor=east,font=\scriptsize]{$-a+\frac{l}{2}$};
    \draw(0,-1.3)node[anchor=east,font=\scriptsize]{$-a-\frac{l}{2}$};
    \draw[->] (-0.5,0) -- (1.7, 0)node[anchor=west]{$z$};
    \draw[->] (0,-2.0)node[anchor=north]{CFT$_{(1)}$} -- (0,2.0)node[anchor=south]{$x$};
  \end{scope}
  \end{scope}
  \end{scope}
\end{tikzpicture}
\caption{EE with two regions in same side}
\label{Fig:S1S2S3}
\end{figure}
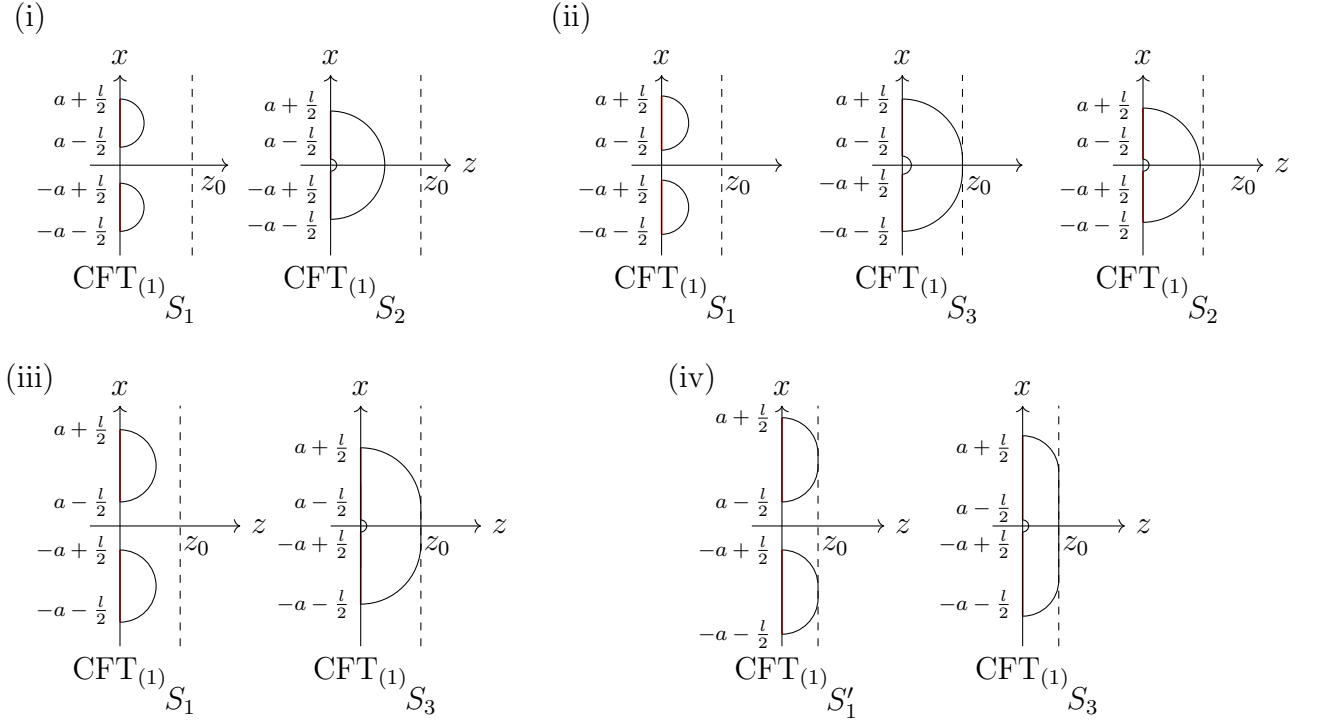

Now we turn to the case where the two intervals are on the same side $A=\left[a-\frac{l}{2},a+\frac{l}{2}\right]$ and 
$B=\left[-a-\frac{l}{2},-a+\frac{l}{2}\right]$. In this case, we always have $a>\frac{l}{2}$. The resulting entropy is found as follows (refer to Fig.\ref{Fig:S1S2S3} for a sketch of each phase $S_1,S_1',S_2$ and $S_3$).When (i) $l<\frac{2z_0}{1+ \sqrt{2}}$, the entropy is the same in the case of pure AdS,
\ba
S_{AB}= 
\left\{
\begin{array}{ll}
S_1=\frac{2c}{3}\log\frac{l}{\ep} &\ \ \   (a>\frac{l}{\sqrt{2}})\\
S_2=\frac{2c}{3}\qty[\log\frac{4(a-\frac{l}{2})(a+\frac{l}{2})}{\ep^2}]&\ \ \  (a<\frac{l}{\sqrt{2}})
\end{array}
\right.
\ea
When (ii) $\frac{2z_0}{1+ \sqrt{2}}<l<z_0$, we find
\ba
S_{AB}= 
\left\{
\begin{array}{ll}
S_1=\frac{2c}{3}\log\frac{l}{\ep} &\ \ \   (a>a_1)\\
S_3=\frac{c}{3}[\log\frac{4(a-\frac{l}{2})z_0}{\ep^2} + \frac{a+\frac{l}{2}}{z_0}-1]&\ \ \  (z_0-\frac{l}{2}<a<a_1)\\
S_2=\frac{c}{3}\qty[\log\frac{4(a-\frac{l}{2})(a+\frac{l}{2})}{\ep^2}]&\ \ \  (a<z_0-\frac{l}{2})
\end{array}
\right.
\ea
and when (iii) $z_0<l<2z_0$, we obtain
\ba
S_{AB}= 
\left\{
\begin{array}{ll}
S_1=\frac{2c}{3}\log\frac{l}{\ep} &\ \ \   (a>a_1)\\
S_3=\frac{c}{3}[\log\frac{4(a-\frac{l}{2})z_0}{\ep^2} + \frac{a+\frac{l}{2}}{z_0}-1]&\ \ \  (a<a_1)
\end{array}
\right.
\ea
Finally, when (iv) $l>2z_0$ we have,
\ba
S_{AB}= \left\{
\begin{array}{ll}
S_1'=\frac{2c}{3}\log\frac{2z_0}{\ep}+ \frac{c}{3} \frac{l-2z_0}{z_0} &\ \ \   (a>a_2)\\
S_3=\frac{c}{3}[\log\frac{4(a-\frac{l}{2})z_0}{\ep^2} + \frac{a+\frac{l}{2}}{z_0}-1]&\ \ \  (a<a_2),
\end{array}
\right.
\ea
In the above we defined $a_2\equiv \frac{l}{2}+ \alpha_* z_0$, where $\alpha_*$ satisfies $\log\alpha_* + \alpha_* + 1 = 0$ ($\alpha_* \simeq 0.278\cdots$). Another constant $a_1$ is defined as follows. When we consider a single AdS or $z_0=\infty$, we find the phase transition between the connected and disconnected configuration happens at $l = \sqrt{2}a$.
In our AdS wormhole, the phase transition point occurs at $a=a_1$, where $a_1$ is the solution to $\log\frac{l^2}{4(a-l/2)z_0} = \frac{a+l/2}{z_0} - 1$. We can show $a_1<\frac{l}{\sqrt{2}}$. The phase separation between (i) and (ii) is given by $l=\frac{2z_0}{1+ \sqrt{2}}$ and can be obtain by setting $a+\frac{l}{2}=z_0$ at $l=\s{2}a$, where the well-known phase transition \cite{Headrick:2007km} between the connected and disconnected phase occurs.


\subsubsection{Lorentzian wormhole}
Finally we would like to consider the holographic entropy in the Lorentzian AdS wormhole by the Wick rotation $x^0\to it$. For simplicity, we choose two semi infinite subsystems in the different CFTs: $A=[0,\infty]$ at $t=0$ in CFT$_{(1)}$ and 
$B=[b,\infty]$ at $t=t_b$ in CFT$_{(2)}$. We can calculate $S_{AB}$ from (\ref{geodesicB}) as follows:
\ba
S_{AB}=\frac{c}{3}\log \frac{\frac{b^2-t^2}{4}+z_0^2}{\epsilon z_0}.
\ea
This is real and positive valued when $t^2<b^2+4z_0^2$, where the geodesic $\Gamma_{AB}$ is space-like. At $t^2=b^2+4z_0^2$, the geodesic becomes null and the entropy gets vanishing, which is possible because two boundaries in a traversable wormhole are causally connected.  
Moreover, when $t^2>b^2+4z_0^2$, it becomes time-like and $S_{AB}$ becomes complex valued:
\ba
S_{AB}=\frac{c}{3}\log \frac{\frac{t^2-b^2}{4}-z_0^2}{\epsilon z_0}+\frac{c}{3}\pi i. \label{WHTEE}
\ea
This behavior looks very similar to the time-like entanglement entropy \cite{Doi:2022iyj,Doi:2023zaf,Liu:2022ugc} in AdS/CFT. Indeed, as we will argue in the next section, we can view (\ref{WHTEE}) as a general case of time-like entanglement entropy in one of two possible CFT duals. The complex values of entanglement entropy also appears in dS$_3/$CFT$_{(2)}$ \cite{Hikida:2021ese,Hikida:2022ltr,Doi:2022iyj,Narayan:2022afv}, which can be regarded as a Wick rotation of the time-like entanglement entropy.

\subsection{Construction of wormholes by gluing BTZ black holes}\label{sec:BTZ}
As the second example of the AdS wormhole, we consider gluing two BTZ black holes. The global extension of the BTZ black hole 
\ba
ds^2=a^2 \qty[-\frac{1-\frac{z^2}{a^2}}{z^2}dt^2+\frac{dz^2}{z^2\left(1-\frac{z^2}{a^2}\right)}+\frac{dx^2}{z^2}],  \label{BTZm}
\ea
can be described by the Kruskal coordinate
\ba
ds^2=\frac{-4a^2 dudv+ (1-uv)^2dx^2}{(1+uv)^2}. \label{BTZnull}
\ea
They are related via
\ba
u=-\s{\frac{a-z}{a+z}}e^{-t/a},\ \ \ \ v=\s{\frac{a-z}{a+z}}e^{t/a}.
\ea

\subsubsection{Gluing procedure}
Consider a two dimensional surface $v=f(u)$ in the BTZ black hole and consider the right half region $v>f(u)$. We impose the Neumann boundary condition at the internal boundary $v=f(u)$, which is called the EOW brane (end of the world-brane) \cite{Takayanagi:2011zk,Fujita:2011fp}:
\ba
K_{ab}-h_{ab}K=-\mathcal{T}h_{ab}, \label{EOW}
\ea
where $K_{ab}$ is the extrinsic curvature, $h_{ab}$ is the induced metric on the EOW brane, and $\mathcal{T}$ is the tension, which takes values in the range $|\mathcal{T}|\leq 1$.
The normal vector $N^a$ on the EOW brane can be chosen as $(N^u,N^v,N^x)=\left(\frac{1+uv}{2\s{v'}},-\frac{1+uv}{2}\s{v'},0\right)$.

We can find the following solution by solving (\ref{EOW}):
\ba
v=\frac{u-\frac{T}{\s{1-\mathcal{T}^2}}}{\frac{\mathcal{T}}{\s{1-\mathcal{T}^2}}u+1}. 
\ea

\begin{figure}[ht]
		\centering
		\includegraphics[width=8cm]{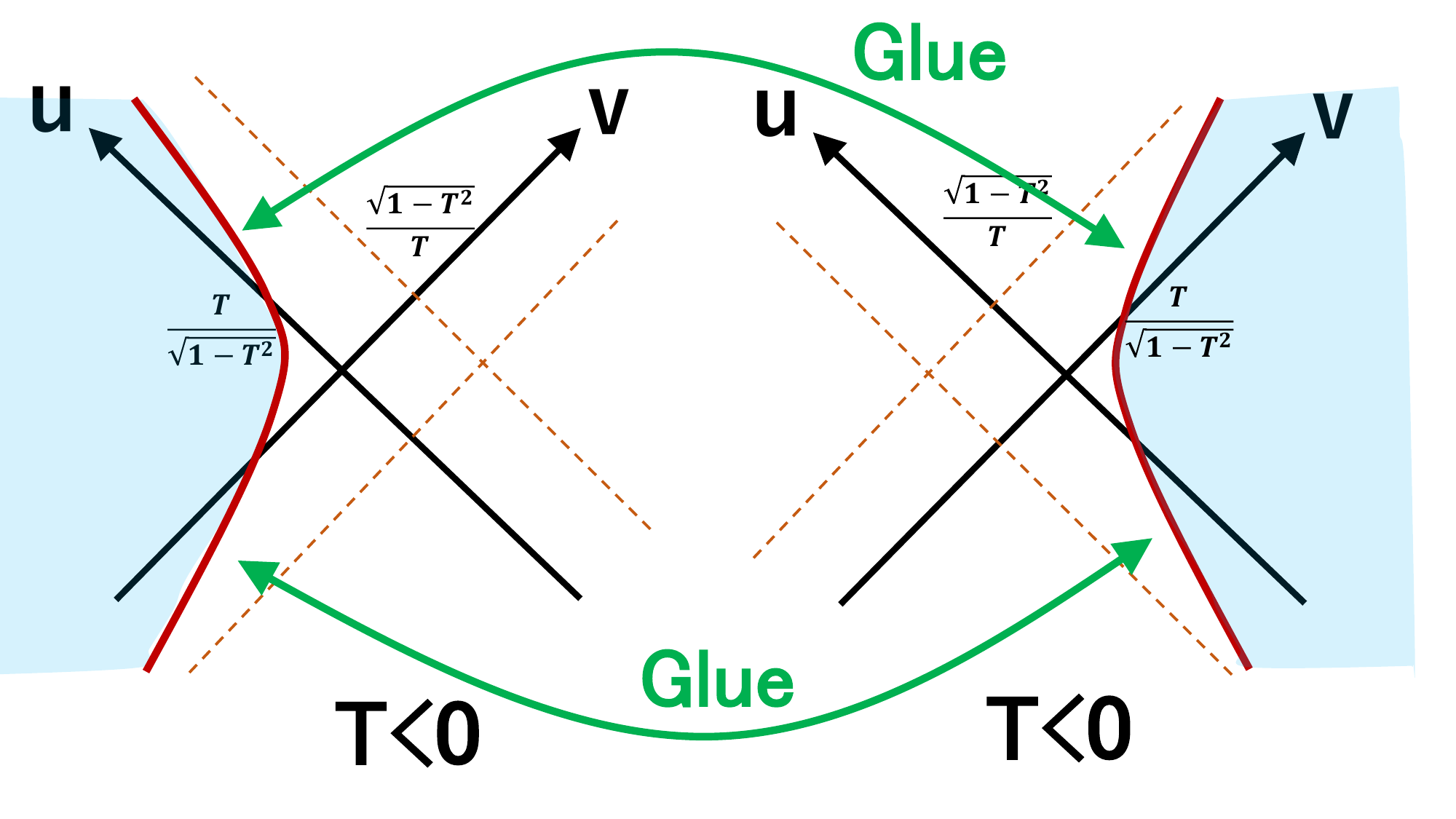}
		\caption{Gluing two geometries. Comment: Tension $\mathcal{T}$ is its absolute value.} 
		\label{fig:BTZglue}
\end{figure}

Now we pick up two copies of the geometry with the negative tension $\mathcal{T}<0$ and glue along the EOW brane as depicted in Fig.\ref{fig:BTZglue}, where the EOW brane boundary in the left and right half is located, respectively: 
\ba
v=\frac{u-\lambda}{\lambda u+1},\ \ \mbox{and},\ \ u=\frac{v-\lambda}{\lambda v+1},\ \ 
\left(\lambda=-\frac{\mathcal{T}}{\s{1-\mathcal{T}^2}}>0\right).  \label{EOWa}
\ea

\subsubsection{Geodesic length}

We consider the geodesic for which one endpoint of the geodesic is in the left boundary at $(t,x)=(t_1,0)$ and the other one is in the right boundary at $(t,0)=(t_2,0)$. We leave the detailed calculation in the appendix \ref{ap:BTZ} and below we only show the final results. 

When $t_1+t_2=0$, the geodesic length $D_{12}$ is computed as follows
\begin{equation}
  D_{12} = 2 a \log \qty[\frac{2a}{\epsilon} \cosh(\frac{t_1}{a})]+2\log\qty[-\lambda + \sqrt{\lambda^2  + 1}].\label{ttt1}
\end{equation}
This shows the familiar linear $t$ growth at late time as in the original BTZ black hole \cite{Hartman:2013qma}. The second constant term can be regarded as the g-function or boundary entropy as in the AdS/BCFT \cite{Takayanagi:2011zk,Fujita:2011fp}.

On the other hand, when $t_1=t_2$, we find
\begin{equation}
  D_{12} = 2 a \log \qty[\frac{2a}{\epsilon} \qty(- \lambda \cosh \qty(\frac{t_1}{a}) + \sqrt{1 + \lambda^2})]. \label{ttt2}
\end{equation}
This shows the characteristic feature of the traversable wormhole in that $D_{12}$ vanishes when 
$- \lambda \cosh \qty(\frac{t_1}{a}) + \sqrt{1 + \lambda^2}=0$. Notice that in the original BTZ black hole (i.e. at $\lambda=0$), $D_{12}$ does not depend on $t_1$ because it is invariant under the Schwarzwald time translation. For a positive tension wormhole deformation $\lambda>0$, $D_{12}$ decreases under the time evolution and the geodesic becomes null at a time.

\subsubsection{Two point functions in glued BTZ} \label{2ptBTZ}

Here we calculate holographic two point functions in a glued BTZ geometry. 
We first solve the equation of motion of the bulk scalar field, then calculate the bulk-to-bulk propagator and bulk-to-boundary propagator. Finally we get the two point functions.

We rewrite the standard BTZ metric
\begin{align}
  ds^2 =& - \qty(r^2 - a^2) d t^2 + \frac{1}{(r^2 / a^2) - 1} dr^2 + r^2 dx^2, \label{BTZ_sta}
\end{align}
into the AdS$_2$ sliced metric (see Appendix.\ref{corcha} for coordinate transformations):
\begin{align}
ds^2 =& d \rho^2 + a^2 \cosh^2 \qty(\frac{\rho}{a}) \qty(\frac{- d\tau^2 + dy^2}{y^2}). \label{BTZ_glo_Poi}
\end{align}
The EOW brane is located at
\begin{equation}
  \sinh^2 \qty(\frac{\rho}{a}) = \lambda^2 \label{EOWloc}
  ~~ \Leftrightarrow ~~ \rho = a \log (\lambda + \sqrt{\lambda^2 + 1}) =: \rho_0.
\end{equation}

The Klein-Gordon equation in the metric eq.\eqref{BTZ_glo_Poi} reads
\begin{equation}
  \qty[\pdv[2]{\rho} + \frac{2}{a} \tanh(\frac{\rho}{a}) \pdv{\rho} + \frac{y^2}{a^2 \cosh(\rho / a)} \qty(- \pdv[2]{\tau} + \pdv[2]{y}) - m^2] \phi(\rho, \tau, y) = 0 .\label{EOM_BTZ}
\end{equation}
From now on, we set $\displaystyle m^2 = -\frac{3}{4 a^2}$ for simplicity. Separating valuables as
\begin{equation}
  \phi(\rho, \tau, y) = \tilde{\phi}(\rho) \varphi (\tau, y),
\end{equation}
and solving equation of motion, we get following independent solutions labeled by a constant $\nu$; 
\begin{align}
  \tilde{\phi}_\nu (\rho) &= 
    \left\{
      \begin{aligned}
        \frac{\sin(2 \nu \cot^{-1}(e^{\frac{\rho}{a}}))}{\sqrt{2 \cosh(\rho/a)}} , \\
        \frac{ \cos(2 \nu \cot^{-1}(e^{\frac{\rho}{a}}))}{\sqrt{2 \cosh(\rho/a)}}.
      \end{aligned}
    \right. \label{sincos},\;
  \varphi_\nu (\tau, y) = 
  \left\{
    \begin{aligned}
        \int \frac{\dd{\omega}}{2 \pi} e^{-i \omega \tau} \sqrt{y} J_{\nu} (\omega y), \\
        \int \frac{\dd{\omega}}{2 \pi} e^{-i \omega \tau} \sqrt{y} Y_{\nu} (\omega y) .
    \end{aligned}
  \right.
\end{align}
The detailed procedure of solving equation of motion is summarized in Appendix.\ref{soluti}

On the right boundary $(\rho \rightarrow \infty)$, sine corresponds to normalizable mode and cosine corresponds to non-normalizable mode, although both of them converge at the boundary due to negative mass square. 
To get bulk-to-bulk propagator, We demand that sine should also be normalizable on the left boundary.
This requirement is satisfied when $\nu$ is non-negative integer, if there is no EOW brane cut. 
With EOW brane case, we should quantize $\nu$ as
\begin{equation}
  2 \nu \cot^{-1}(e^{\rho_0 / a}) = \frac{n}{2} \pi 
  ~~\Leftrightarrow ~~ \nu = \frac{\pi}{4 \cot^{-1}(e^{\rho_0 / a})} n =: N n,\; n\in \mathbb{Z}_{>0}
\end{equation}
$\rho_0$ is the location of EOW brane \eqref{EOWloc}. 
Note that $N$ is a constant larger than $1$ for the gluing along the negative tension EOW brane.

Bulk-to-bulk propagator can be constructed by the following combinations of the solutions \cite{Chiodaroli:2016jod};
\begin{equation}
  \begin{aligned}
    G_{\Delta}(\rho, \tau, y; \rho', \tau', y') =  \int_{-\infty}^{\infty} \frac{\dd{\omega}}{2\pi} &e^{-i \omega (\tau-\tau')}
    \sum_{n=0}^{\infty} \tilde{G}_{\nu(n)}(y,y')
    \frac{\sin(2 \nu \cot^{-1}(e^{\frac{\rho}{a}}))}{\sqrt{2 \cosh(\rho/a)}} \frac{\sin(2 \nu \cot^{-1}(e^{\frac{\rho'}{a}}))}{\sqrt{2 \cosh(\rho'/a)}} .
  \end{aligned}
\end{equation}
\begin{equation}
  \begin{aligned}
    \tilde{G}_{\nu(n)}(y,y') = \frac{\pi}{-2i} &\left[\theta(y-y')  \sqrt{y} H^{(1)}_{\nu(n)} (|\omega| y) \sqrt{y'} J_{\nu(n)} (|\omega| y')+\text{($y\leftrightarrow y'$)}  \right].
  \end{aligned}
\end{equation}
This is the propagator for the case when two points are located in right side ($\rho > 0$) of the EOW brane. 
When they are located in left side ($\rho < 0$), we should replace $\sin(2 \nu \cot^{-1}(e^{\frac{\rho}{a}}))$ to  $(-1)^{n-1} \sin(2 \nu \tan^{-1}(e^{\frac{\rho}{a}}))$. 

Bulk-to-boundary propagator can be obtained by the following limit
\begin{equation}
  K_{\Delta} (\rho, \tau, y; \tau', y'_{\bot }) = \lim_{z' \rightarrow 0} \frac{2 \Delta - 2}{z'^\Delta} G_{\Delta} (\rho, \tau, y; z'(\rho',y'), \tau', y')
\end{equation}
where $z$ and $y_{\bot}$ are Poincar\'e coordinate and boundary CFT coordinate respectively, which are related to current coordinates by
\begin{align}
  z =& \frac{y}{\cosh(\rho / a)} \simeq \frac{2y}{e^{+\rho / a}}, \quad y_{\bot} = + y \qquad \text{(right region)} \\
  z =& \frac{y}{\cosh(\rho / a)} \simeq \frac{2y}{e^{- \rho / a}}, \quad y_{\bot} = - y \qquad \text{(left region)}
\end{align}
Take the limit and we get
\begin{equation}
  \begin{aligned}
    K_{\Delta} (\rho, \tau, y; \tau', y'_{\bot}) = \int_{-\infty}^{\infty} \frac{d\omega}{2\pi} e^{-i \omega (\tau-\tau')}
    \sum_{n=0}^{\infty} \tilde{G}_{\nu(n)}(y,y'_{\bot}) \frac{2 \nu(n)}{|2y'_{\bot}|^{3/2}}  \frac{\sin(2 \nu(n) \cot^{-1}(e^{\frac{\rho}{a}}))}{\sqrt{2 \cosh(\rho/a)}}
  \end{aligned}
\end{equation}
The solution of wave equation with usual AdS/CFT boundary conditions is given by
\begin{equation}
  \phi(\rho, \tau, y) = \int \dd{\tau'} \dd{y'_{\bot}} K_{\Delta} (\rho, \tau, y; \tau', y'_{\bot}) \phi_0 (\tau', y'_{\bot})
\end{equation}
and this has the following asymptotic form
\begin{equation}
  \phi(\rho, \tau, y) \xrightarrow[z \rightarrow 0]{} z^{\frac{1}{2}}\phi_0(\tau, y) + z^{\frac{3}{2}} \ev{\mathcal{O}(\tau,y)}_{\phi_0} \label{asympt}
\end{equation}
$\phi_0$ is a source in right side boundary. Note that we are considering $m^2 = -\frac{3}{4} \Leftrightarrow \Delta = \frac{3}{2}$ case. 
The two point function are given by functional derivative
\begin{equation}
  \ev{\mathcal{O}_1 (\tau, y_{\bot}) \mathcal{O}_1 (\tau', y'_{\bot})} = \left. \frac{\delta}{\delta \phi_0(\tau,y_{\bot})} \ev{\mathcal{O}(\tau',y'_{\bot})}_{\phi_0} \right|_{\phi_0 = 0} \label{2ptfcn}
\end{equation}
Combining eq.\eqref{asympt}, \eqref{2ptfcn}, two point function of the same side points are gained from the bulk-to-boundary propagator;
\begin{align}
  \ev{\mathcal{O}_1 (\tau, y_{\bot}) \mathcal{O}_1 (\tau', y'_{\bot})} &= \lim_{z \rightarrow 0} z^{-\frac{3}{2}} K_{\Delta} (\rho, \tau, y_{\bot}; \tau', y'_{\bot}) \\
  &= \int_{-\infty}^{\infty} \frac{\dd\omega}{2\pi} e^{-i \omega (\tau-\tau')}
  \sum_{n=0}^{\infty} \tilde{G}_{\nu(n)}(y_{\bot}, y'_{\bot}) \frac{4 \nu(n)^2}{|4 y_{\bot} y'_{\bot}|^{3/2}} \label{2ptf}
\end{align}
In the case of $\ev{\mathcal{O}_1 \mathcal{O}_2}$, i.e. two points in each other sides, we replace $\sin(2 \nu \cot^{-1}(e^{\frac{\rho}{a}}))$  to  $(-1)^{n-1} \sin(2 \nu \tan^{-1}(e^{\frac{\rho}{a}}))$, and take $\rho \rightarrow - \infty$ limit; 
\begin{align}
  \ev{\mathcal{O}_1 (\tau, y_{\bot}) \mathcal{O}_2 (\tau', y'_{\bot})} = \int_{-\infty}^{\infty} \frac{d\omega}{2\pi} e^{-i \omega (\tau-\tau')}
  \sum_{n=0}^{\infty} (-1)^{n-1} \tilde{G}_{\nu(n)}(y_{\bot}, y'_{\bot}) \frac{4 \nu(n)^2}{|4 y_{\bot} y'_{\bot}|^{3/2}} \label{2ptf2}
\end{align}
Performing the Fourier transformation \cite{Chiodaroli:2016jod},
\begin{equation}
    \int_{-\infty}^{\infty} \dd{\omega} e^{-i \omega (\tau-\tau')} \tilde{G}_{\nu(n)}(y, y') = Q_{\nu(n) - \frac{1}{2}}  \qty(\frac{y^2 + y'^2 - (\tau - \tau')^2}{2 y y'})
\end{equation}
where $Q_\nu$ is the Legendre function of the second kind.
Inserting this into eq.\eqref{2ptf}, We get
\begin{align}
  \ev{\mathcal{O}_1 (\tau, y_{\bot}) \mathcal{O}_1 (\tau', y'_{\bot})} &= \frac{1}{4 \pi |y_{\bot} y'_{\bot}|^{\frac{3}{2}}} \sum_{n=0}^{\infty} \qty(\nu(n))^2 Q_{\nu(n) - \frac{1}{2}} \qty(\xi) \\
  \ev{\mathcal{O}_2 (\tau, y_{\bot}) \mathcal{O}_1 (\tau', y'_{\bot})} &= \frac{1}{4 \pi |y_{\bot} y'_{\bot}|^{\frac{3}{2}}} \sum_{n=0}^{\infty} (-1)^{n-1} \qty(\nu(n))^2 Q_{\nu(n) - \frac{1}{2}} \qty(|\xi|) \\
  \xi &:= \frac{y^2_{\bot} + y'^2_{\bot} - (\tau - \tau')^2}{2 |y_{\bot} y'_{\bot}|}
\end{align}

\begin{figure}[ht]
  \centering
  \includegraphics[scale = 0.6]{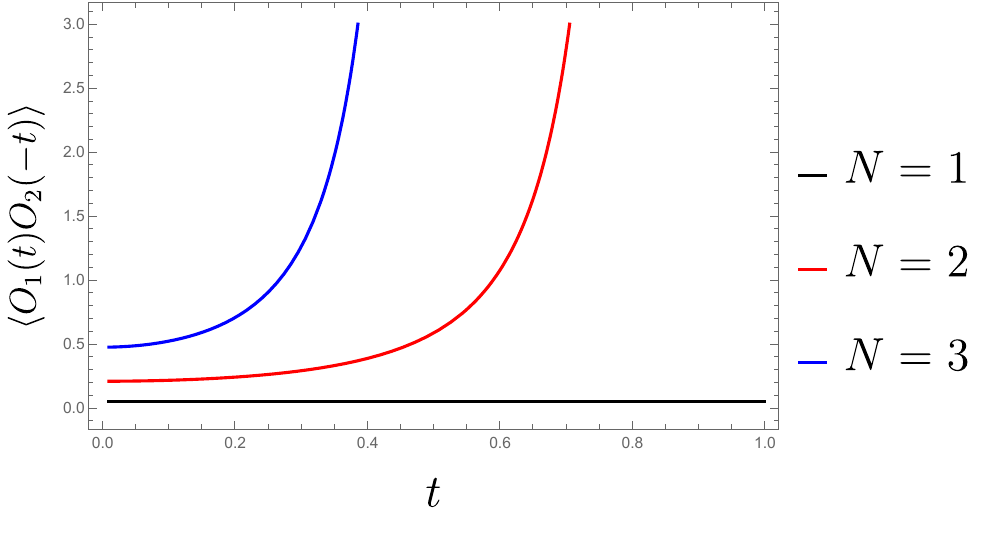}
  \caption{Plot of the two points function $\ev{\mathcal{O}_2(t,0) \mathcal{O}_1(-t,0)}$ in $N = 1, 2, 3$ cases. The functions have divergent behavior when the geometry has wormhole structure ($N > 1$). We are ignoring the overall factor here. }
  \label{o1o2_1}
\end{figure}

Finally we have to rewrite two point functions in the original coordinates \eqref{BTZ_sta}. 
They are related via
\begin{equation}
  \left\{
    \begin{aligned}
      \tau = \pm e^{x} \sinh t \\
      y_{\bot} = \pm e^{x} \cosh t     
    \end{aligned}
  \right. \qquad\text{($+ / -$ for right/left region)}
\end{equation}
The Jacobian is
\begin{equation}
  J := \mathrm{det} \begin{pmatrix}
     \partial_\tau t & \partial_{y_{\bot}} t \\
     \partial_\tau x & \partial_{y_{\bot}} t
  \end{pmatrix} = \frac{a^2}{e^{2 x}}
\end{equation}
Substituting this coordinate and multiplying the conformal factor $|J|^{-3/4}$, we get 
\begin{align}
  \ev{\mathcal{O}_1 (t, x) \mathcal{O}_1 (t', x')} &= \frac{1}{4 \pi a^3 (\cosh t \cosh t')^{\frac{3}{2}}} \sum_{n=0}^{\infty} \qty(\nu(n))^2 Q_{\nu(n) - \frac{1}{2}} \qty(\xi) \\
  \ev{\mathcal{O}_2 (t, x) \mathcal{O}_1 (t', x')} &= \frac{1}{4 \pi a^3 (\cosh t \cosh t')^{\frac{3}{2}}} \sum_{n=0}^{\infty} (-1)^{n-1} \qty(\nu(n))^2 Q_{\nu(n) - \frac{1}{2}} \qty(\xi) \\
  \xi &= \frac{\cosh(x-x') \pm \sinh t \sinh t'}{\cosh t \cosh t'}
\end{align}
We show the time dependence of $\ev{\mathcal{O}_2(t,0) \mathcal{O}_1(t,0)}$ in Fig.\ref{o1o2_1}. For a traversable wormhole ($N > 1$), it develops a divergence as the geodesic between two boundaries can become null. In $N = 2$ case, this emerges at $t = \log(1 + \sqrt{2})$. In $N=3$, this is at $ \displaystyle t = \frac{1}{2} \log 3$. They are exactly the time that the geodesic between two points becomes null:
\begin{equation}
  \cosh t = \coth (\rho_0 / a).
\end{equation}

\section{Two candidates of CFT duals and their toy models}
\label{sec:twosetups}

\begin{figure}[ttt]
		\centering
		\includegraphics[width=14cm]{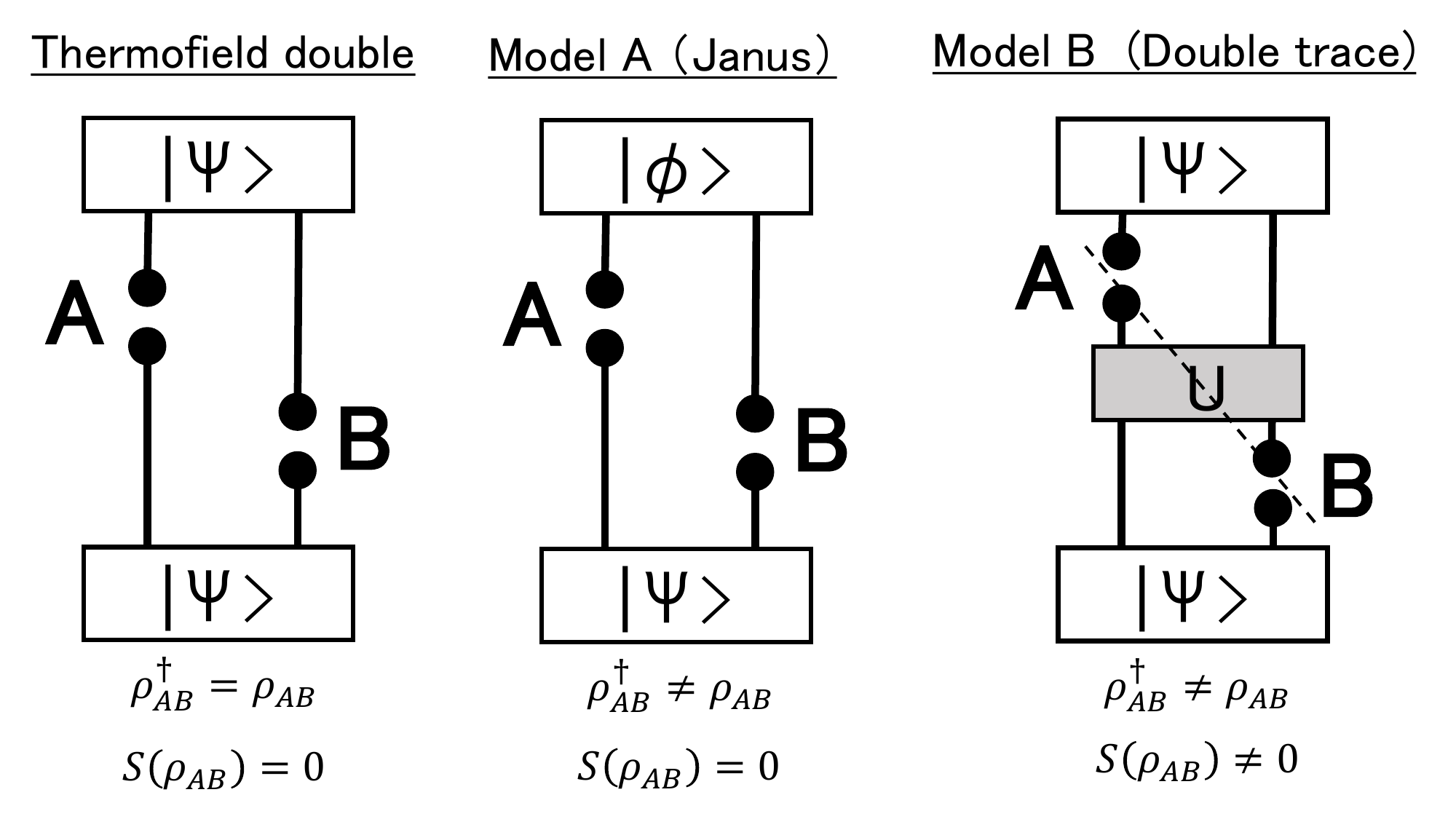}
		\caption{CFT duals of AdS wormholes. The left panel shows the standard thermofield double state dual to a non-traversable wormhole. The middle and right panel show the model A (Janus deformation) and model B (double trace deformation), both dual to traversable AdS wormholes. The dotted diagonal line describes a time slice where we can purify the state. This requires the extra Hilbert space due to the presence of unitary transformation $U$, which represents the double trace interactions between the two CFTs. This shows $S(\rho_{AB})\neq 0$.} 
		\label{fig:Model}
\end{figure}

Now we would like to give constructions of CFT duals of the traversable AdS wormhole backgrounds, which were presented in section \ref{sec:AdSWH}. In this paper, we propose two different models of CFT duals, both of which are given by certain deformations of a pair of CFTs, namely CFT$_{(1)}$ and CFT$_{(2)}$. One is the model A, described by a complex valued Janus deformation of the CFTs, where the initial state differs from the final state. The other is the model B, obtained by a double trace deformation of CFTs, which leads to interactions between two CFTs. They are sketched in Fig.\ref{fig:Model}, where the time slices of  CFT$_{(1)}$ and CFT$_{(2)}$ are denoted by $A$ and $B$, respectively.

A holography between traversable wormholes and double trace deformations \cite{Aharony:2001pa,Witten:2001ua} has been already much explored \cite{Gao:2016bin,Maldacena:2017axo,Maldacena:2018gjk,Maldacena:2018lmt}.
Though the model B is closely related to the previous works, a creation of eternal wormholes from two originally disconnected AdS geometries, has not been well studied before. Moreover, we will point out an important quantum information theoretical property which has not been noticed before. On the other hand, the model A is new and has not been discussed in earlier works. 

An important feature common to both the model A and B, is that the ``quantum state" $\rho_{AB}$ on  a time slice of CFT$_{(1)}$ (i.e.$A$) and that of CFT$_{(2)}$ (i.e.$B$) becomes non-hermitian $\rho^\dagger_{AB}\neq \rho_{AB}$. Due to the interaction between the two CFTs, the pseudo entropy $S_{AB}=-\mbox{Tr}[\rho_{AB}\log\rho_{AB}]$ gets non-vanishing in the model B, while it is vanishing in the model A.

Below we will explain the model A and model B, heuristically by providing toy examples. We will present more details how the model A and B are dual to the traversable wormholes in section \ref{sec:Janus} and section \ref{sec:DT}, \ref{sec:TTbar}, respectively.

\subsection{Model A: Janus deformation}

First we explain the construction of model A. Consider a thermofield double (TFD) state dual to an eternal AdS black hole \cite{Maldacena:2001kr}. The TFD state $|\mbox{TFD}(\beta)\lb$ is written as 
\ba
|\mbox{TFD}(\beta)\lb={\cal N} e^{-\frac{\beta}{2}(H_1+H_2)}\sum_{n}|n\lb_1|n\lb_2,
\ea
where ${\cal N}$ is the normalization factor. Here we wrote the Hamiltonian of CFT$_{(i)}$ as $H_i$ and the energy eigenstate of $H_{i}$ as $|n\lb_{i}$ for $i=1,2$. A TFD state is prepared by a path-integral in the Euclidean time interval: $-\frac{\beta}{4}\leq \tau \leq \frac{\beta}{4}$.

We generalize the TFD states to a family of states which depend on one more parameter $\gamma$, expressed as $|\mbox{TFD}(\beta,\gamma)\lb$. We choose $\gamma$ to be a deformation of TFD state by inserting a conformal interface at $\tau=0$ in the path-integral. This is dual to the traversable AdS wormhole solution in the presence of bulk scalar field \cite{Bak:2007jm,Bak:2007qw}. 

The quantum state of the time slice $A$ at $t=t_1$  in CFT$_{(1)}$ and the other one $B$ at $t=t_2$ in  CFT$_{(2)}$ is described by the density matrix
\ba
\rho_{AB}=e^{-iH_1t_1}e^{-iH_2t_2}|\mbox{TFD}(\beta,\gamma)\lb\la\mbox{TFD}(\beta,\gamma)|e^{iH_1t_1}e^{iH_2t_2}. \label{Janustdfsa}
\ea
When $\gamma$ is real valued, it is obvious that this is hermitian $\rho^\dagger_{AB}= \rho_{AB}$. However, if we continue this state to imaginary $\gamma$, then $\rho_{AB}$ is no longer hermitian because 
$|\mbox{TFD}(\beta,\gamma)\lb^\dagger=\la\mbox{TFD}(\beta,\gamma^*)|$.
Thus $\rho_{AB}$ is a transition matrix (instead of a density matrix) when $\gamma$ is imaginary in the sense of \cite{Nakata:2021ubr}. Namely, the initial state and final state are different. It is useful to note that its pseudo entropy vanishes $S(\rho_{AB})=0$.

To see how traversable wormhole emerges, consider the following toy example:
\ba
\rho_{AB}=|0\lb\ \la\mbox{TFD}(\beta)| e^{-iH_1t_1}e^{-iH_2t_2},
\ea
which is highly non-hermitian. Consider a two dimensional CFT and calculate the two point function 
$\la \mO_1(-t_1,x_1)\mO_2(-t_2,x_2)\lb$ by inserting a primary operator $\mO$ (dimension $\Delta$) on $A$ and another one on $B$. This two point function can be expressed as the two point function on a complex plane ${\mathbb{C}}$ and is evaluated as follows:
\ba
\la \mO_1(-t_1,x_1)\mO_2(-t_2,x_2)\lb&=&\mbox{Tr}[\rho_{AB}\mO_1\mO_2]
=\left\la \mO\left(\frac{\beta}{4}+it_2,x_2\right)\mO\left(-\frac{\beta}{4}-it_1,x_1\right)\right\lb_{{\mathbb{C}}}\no
&=&\left[(x_1-x_2)^2+\left(\frac{\beta}{2}+i(t_1+t_2)\right)^2\right]^{-\Delta}.  \label{tpfwjerhr}
\ea
In the high temperature limit $\beta\to 0$, the two point function gets divergent at $|x_1-x_2|=|t_1+t_2|$. This is because two points are light-like separated if we take into account the reflection at 
$t=0$. Indeed, the signal from $(-t_1,x_1)$ propagates at the speed of light till $t=0$ and then is converted from CFT$_{(1)}$ to CFT$_{(2)}$ via the TFD state. Eventually it travels till $(-t_2,x_2)$ in the CFT$_{(2)}$. 
In this example, initial state $|0\lb$ does not have any entanglement between the two CFTs. Then it experiences the time evolution by $H_1+H_2$, which does not have any interactions between the CFTs. Thus we expect that the two point function $\la \mO_1\mO_2\lb$ vanishes. However, we perform the final state projection at $t=0$ on to the TFD state. This violates the naive causality and causes the two point function to be enhanced. As this toy example shows, when the initial and final state are different, when $\rho_{AB}$ is non-hermitian, we can have causality violating propagation between two CFTs even if there is no interactions between them. 

The gravity dual of this setup can be found as follows. Consider a two dimensional torus dual to a BTZ or thermal AdS$_3$. We take the periodicity of thermal circle to be $\frac{\beta}{2}+L$ and take the limit $L\to\infty$. We analytically continue the Euclidean times $\tau_{1,2}$ to the Lorentzian ones by $\tau_1=\frac{\beta}{4}+it_1$ and $\tau_2=-\frac{\beta}{4}-it_2$ in each of the two operators. This reproduces the two point function (\ref{tpfwjerhr}). However, the metric in terms of the Lorentzian time, becomes complex valued. Indeed, (\ref{tpfwjerhr}) gets also complex valued. 

We argue a similar effect can be found in the Janus deformed state (\ref{Janustdfsa}) when $\gamma$ is imaginary and this is dual to a traversable wormhole with a real valued metric. We regard $\gamma$ as a deformation of TFD state by a scalar operator. Such a deformation is dual to a bulk scalar field in the AdS/CFT. When $\gamma$ is real, we do not expect any traversable wormhole behavior because the final state and initial state are the same. A class of gravity duals of such deformed TFD states can be found by considering black holes \cite{Bak:2007jm,Bak:2007qw} in the Janus solutions \cite{Bak:2003jk}.
A Janus solution is dual to an interface in a holographic CFT and is constructed by tuning on a kink configuration of a bulk scalar field in AdS. We will show that in the AdS$_3$ setup, the solution becomes an traversable wormhole when $\gamma$ is imaginary and provide a relevant CFT analysis in section \ref{sec:Janus}.

\subsection{Model B: double trace deformation}

The other class of CFT models, dual to traversable AdS wormholes, is the double trace deformation. Consider two identical and independent $d$ dimensional CFTs: CFT$_{(1)}$ and CFT$_{(2)}$ and introduce the double trace interaction. The total action looks like
\ba
S_{\mathrm{tot}}=S_{\mathrm{CFT}(1)}+S_{\mathrm{CFT}(2)}+\frac{1}{2}\sum^{M}_{s=1}\int d^dxd^dy\ \lambda^{[s]}(x,y)\mO^{[s]}_1(x)\mO^{[s]}_2(y), \label{DTla}
\ea
where $\mO^{[s]}_{i}(x)$ are operators in CFT$_{(i)}$ for $i=1,2$ and $s$ labels different primary operators. The coefficient $\lambda^{[s]}(x,y)$ measures the strength of the non-local double trace deformation. We impose the translational invariance such that $\lambda^{[s]}(x,y)$ only depends on $x-y$, which allows us to perform the Fourier transformation
\ba
\lambda^{[s]}(x,y)=\int dk\; \lambda^{[s]}(k)e^{ik(x-y)}.  \label{DTlb}
\ea

We are interested in $\lambda^{[s]}(k)$ which decays fast at high momenta 
$k\gg \Lambda$. This is because we would like to have two asymptotically AdS boundaries. This requires the presence of the UV region where the double trace deformation should be negligible. The double trace interactions cause the wormhole connection in the interior of the AdS i.e. the IR region.
When we compare this argument with the toy gravity model discussed in section \ref{sec:AdSWH}, we expect that this UV cut off scale $\Lambda$ correspond to $\sim 1/z_0$.

As discovered in \cite{Gao:2016bin} (refer also to e.g.
\cite{Maldacena:2017axo,Maldacena:2018gjk,Maldacena:2018lmt,Harvey:2023oom}), the mixed boundary condition in the AdS gravity induced by the double trace interaction, leads to the negative Casimir energy, whose back-reaction makes the wormhole traversable. To have a macroscopic size of wormhole, $M$ needs to be very large $O(N^2)$ or $\lambda^{[k]}$ are taken to be very large. Though we will not analyze the back-reaction in this paper, we will examine the behaviors of two point functions to probe wormhole geometries.  Another way to find a $O(N^2)$ effect is to consider the double trace deformation of energy stress tensors, which will be discussed later.

\subsubsection{Properties of quantum states}\label{sec:PQS}
\begin{figure}[ht]
		\centering
		\includegraphics[width=5cm]{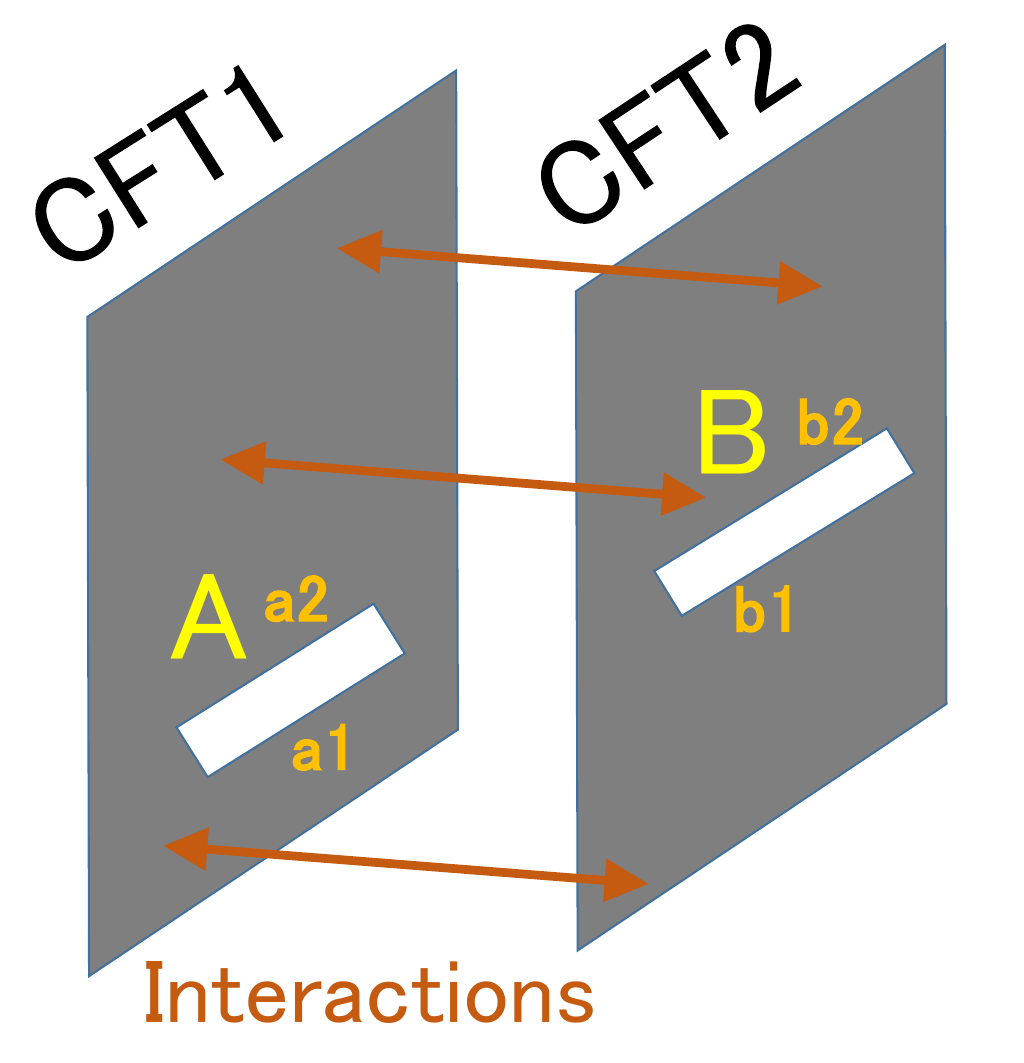}
        \hspace{5mm}
           \includegraphics[width=6cm]{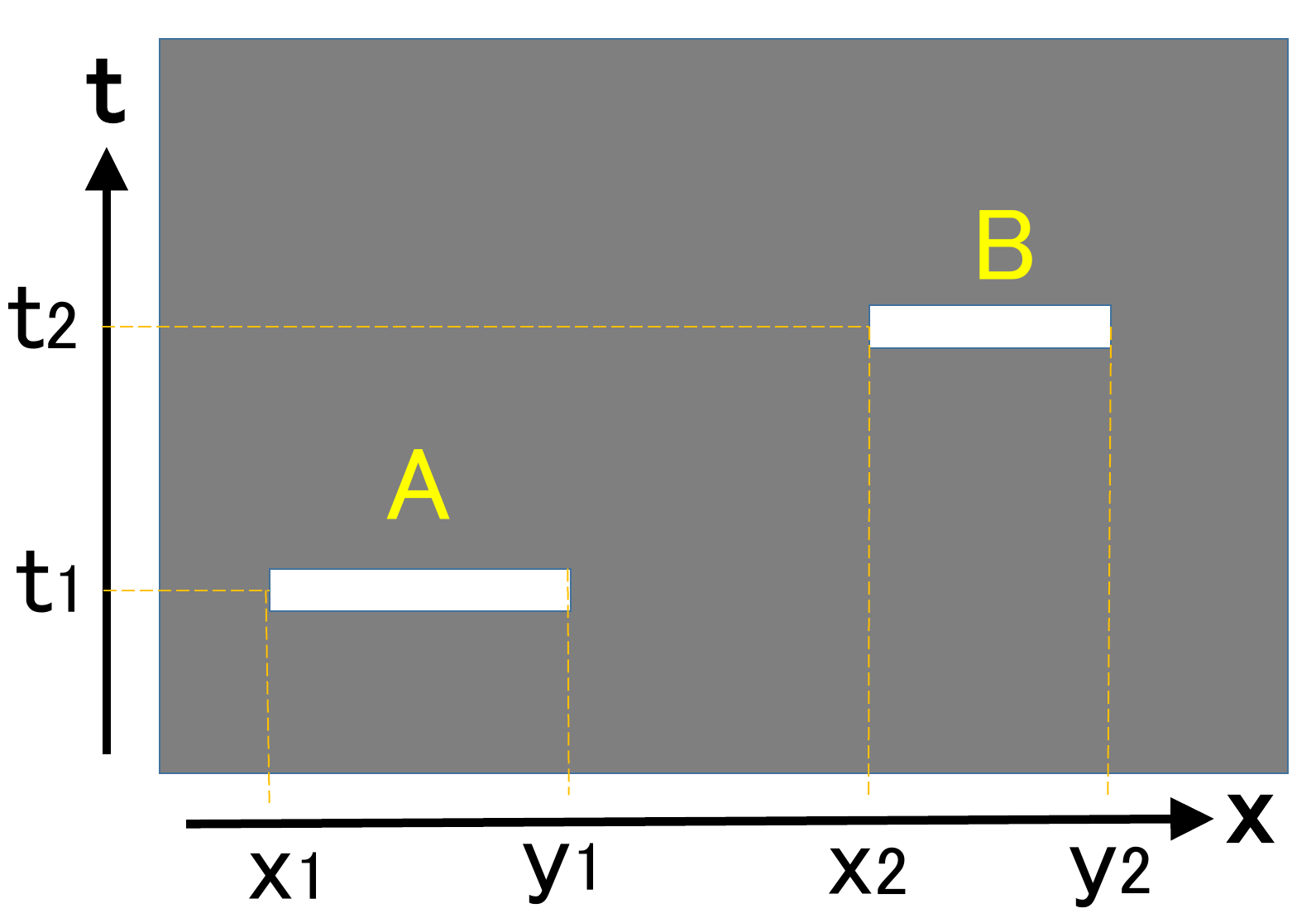}
		\caption{Calculation of pseudo entropy in the model B (left) and in a single CFT with double intervals (right).} 
		\label{fig:modelB}
\end{figure}

Let us take a subsystem $A$ at $t=t_1$ in CFT$_{(1)}$ and the other subsystem $B$ at $t=t_2$ in CFT$_{(2)}$ and consider the "entanglement entropy $S_{AB}$" by tracing out their complement, as depicted in the left panel of Fig.\ref{fig:modelB}. Since the two CFTs are interacting, we should think that $A$ and $B$ are time-like separated. Therefore the Hilbert space on $A$ and that on $B$ are not independent in the standard sense of relativistic quantum field theories. The situation is similar to the time-like entanglement entropy \cite{Doi:2022iyj,Doi:2023zaf} or in other words is a special examples of pseudo entropy \cite{Nakata:2021ubr}.  The density matrix $\rho_{AB}$ looks like
\ba
[\rho_{AB}]^{a_2,b_2}_{a_1,b_1}=\la \Psi_0|_{12}\cdot \left(|b_2\lb\la b_1|\right)_{2}\cdot \mathcal{P}e^{-i\int^{t_2}_{t_1}dt H_{12}(t)}\cdot 
(|a_2\lb\la a_1|)_{1}\cdot |\Psi_0\lb_{12},
\ea
where $H_{12}$ is the total Hamiltonian which includes the interactions between the two CFTs; $|\Psi_0\lb$ is the initial quantum state, which coincides with the final state. For example, we can choose it to be the ground state of $H_{12}$. Note that $\rho_{AB}$ is again not hermitian due to the interacting real time evolution. 

This situation is analogous to the calculation of $S_{AB}$ in a single CFT for the double intervals $A$ and $B$, sketched in Fig.\ref{fig:modelB}. When $A$ and $B$ are space-like separated i.e.
$(x_2-y_1)^2>(t_2-t_1)^2$, $S_{AB}$ is well-defined as the ordinary entanglement entropy. However, when $(x_2-y_1)^2<(t_2-t_1)^2$, $\rho_{AB}$ is not hermitian and is not a regular density matrix but a transition matrix \cite{Kusuki:2017jxh}. As a simple example, we can explicitly calculate $S_{AB}$ for the Dirac fermion CFT in two dimension \cite{Casini:2005rm} in the setup of right panel of Fig.\ref{fig:modelB}. Introducing the light cone coordinates as $(u_i,\bar{u}_i)=(x_i-t_i,x_i+t_i)$
and $(v_i,\bar{v}_i)=(y_i-t_i,y_i+t_i)$, the pseudo entropy reads
\ba
S_{AB}&=&\frac{c}{6}\log\frac{|v_1-u_1|^2}{\ep^2}+\frac{c}{6}\log\frac{|v_2-u_2|^2}{\ep^2}+\frac{c}{6}\log\frac{|v_1-u_2|^2}{\ep^2}+\frac{c}{6}\log\frac{|v_2-u_1|^2}{\ep^2}\no
&&-\frac{c}{6}\log\frac{|u_1-u_2|^2}{\ep^2}-
\frac{c}{6}\log\frac{|v_1-v_2|^2}{\ep^2}.
\ea
From this, we find that $S_{AB}$ becomes complex valued when $|v_1-u_2|^2=(x_2-y_1)^2-(t_2-t_1)^2<0$, i.e. the right end point of $A$ and the left one of $B$ becomes time-like, when both $A$ and $B$ cannot be on any space-like surface.

\subsubsection{Toy model: two coupled harmonic oscillators}
\label{tchpb}
As a toy model of double trace deformation of paired CFTs, let us consider a simple model of two coupled harmonic oscillators $A$ and $B$, whose detailed calculations are presented in appendix \ref{HarmoB}. It is defined by the Hamiltonian 
\ba
H=\frac{1}{\s{1-\lambda^2}}\left[a^\dagger a+b^\dagger b+\lambda(a^\dagger b^\dagger +ab)+1-\s{1-\lambda^2}\right].
\label{Hamosw}
\ea
Via the Bogoliubov transformation (we set $\lambda=\tanh2\theta$),
\ba
&& \ti{a}=\cosh\theta\ a+\sinh\theta\ b^\dagger,\ \ \ \ \ti{b}=\sinh\theta\ a^\dagger +\cosh\theta\ b,  \label{bgla}
\ea
the Hamiltonian is diagonalized as
\ba
H=\ti{a}^\dagger \ti{a}+\ti{b}^\dagger \ti{b}. \label{eeedai}
\ea
Thus the ground state $|\Psi\lb$ is found to be
\ba
|\Psi\lb_{AB}\!=\!|\ti{0}\lb_{AB}\!=\!\frac{1}{\cosh\theta}e^{-\tanh\theta a^\dagger b^\dagger}|0\lb_A|0\lb_B
\!=\!\frac{1}{\cosh\theta}\sum_{k=0}^\infty (-\tanh\theta)^k |k\lb_A |k\lb_B,
\label{gsthdf}
\ea
where we introduce the number states $|n\lb_A=\frac{(a^{\dagger})^n}{\s{n!}}|0\lb_A$ and $|m\lb_B=\frac{(b^\dagger)^m}{\s{m!}}|0\lb_B$ as usual. It it helpful to note that if we write 
\ba
a=\frac{x_1+ip_1}{\s{2}},\ \ \ a^\dagger=\frac{x_1-ip_1}{\s{2}},\ \ \ b=\frac{x_2+ip_2}{\s{2}},\ \ \ b^\dagger=\frac{x_2-ip_2}{\s{2}},
\ea
then the Hamiltonian (\ref{Hamosw}) looks like
\ba
H=\frac{1}{2}(p_1^2+x_1^2)+\frac{1}{2}(p_2^2+x_2^2)+\lambda(x_1x_2-p_1p_2)+(\mbox{const.}).  \label{QQQP}
\ea

\begin{figure}[t]
		\centering
		\includegraphics[width=7cm]{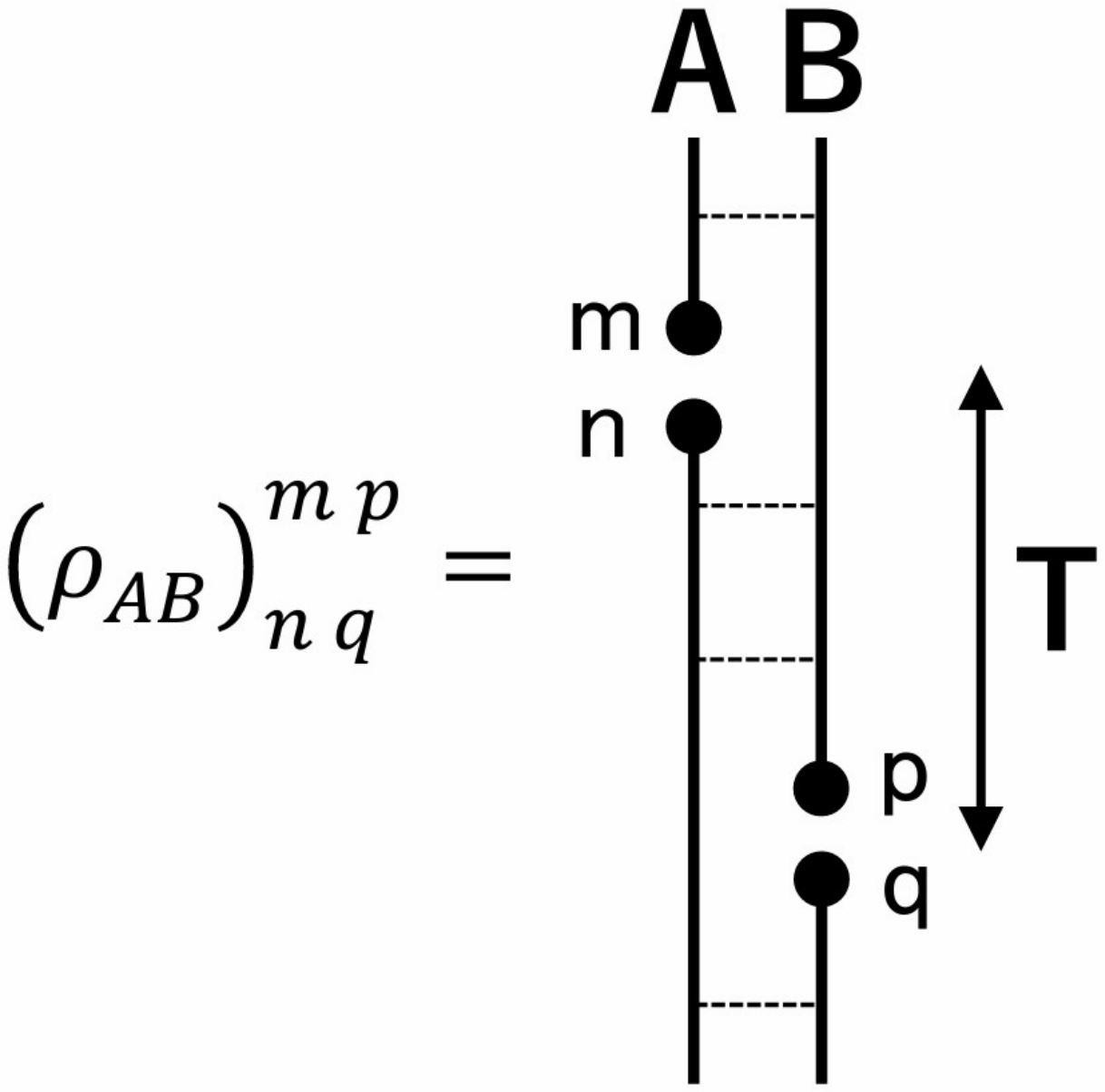}
		\caption{A sketch of the setups of two coupled harmonic oscillators.} 
		\label{fig:TCH}
\end{figure}

Consider the transition matrix $\rho_{AB}$ by focusing on at $t=T$ for the harmonic oscillator $A$ and at $t=0$ for $B$ as described in Fig.\ref{fig:TCH}. This is described as 
\ba
 \left[\rho_{AB}\right]^{mp}_{np} &=&\la \Psi||m\lb_A\la n|e^{-iHT}|p\lb_B\la q||\Psi\lb\no
&=& \frac{1}{\cosh^2\theta}(-\tanh\theta)^{m+q}\la n|_A\la m|_B
e^{-iHT}|q\lb_A|p\lb_B. \label{rabb}
\ea
This is non-vanishing only when $n+p=m+q$. It is straightforward to see that $\rho_{AB}$ is non hermitian  when 
$T\notin \pi {\mathbb{Z}}$.
We can also easily confirm 
$\mbox{Tr}\rho_{AB}=\la \Psi|e^{-iHT}|\Psi\lb=1.$

The characteristic feature of the model B (as opposed to the model A) is that the pseudo entropy for the total system is non-vanishing. 
Indeed, we can evaluate the second Renyi pseudo entropy
as follows (see appendix \ref{HarmoB} for the derivation):
\ba
S^{(2)}_{AB}=\log\left[\frac{1+e^{-2iT}+(1-e^{-2iT})\cosh 4\theta}{2}\right]. \label{PEHOQ}
\ea
We find this is vanishing either when $\theta=0$ (no coupling) or $T\in\pi{\mathbb{Z}}$.
In general, this entropy takes complex values.

The reduced transition matrix $\rho_A$, by tracing out $B$ reads
\ba
[\rho_A]^m_n&=&\delta_{n,m}\sum_{p=0}^\infty\frac{1}{\cosh^2\theta}
(-\tanh\theta)^{m+p}\la m|_A\la m|_B e^{-iHT}|p\lb_A|p\lb_B\no
&=&\delta_{n,m}\frac{1}{\cosh^2\theta}(\tanh\theta)^{2m}, \label{rhoa}
\ea
which is hermitian and time-independent. Thus the pseudo entropy for $\rho_A$ is the same as the standard entanglement entropy $S_A$ between the two harmonic oscillators:
\ba
S_A=\cosh^2\theta\log\cosh^2\theta-\sinh^2\theta\log\sinh^2\theta.  \label{EEsc}
\ea

Finally we can compute the two point function $\la x_A(T) x_B(0)\lb$ as follows:
\ba
\la x_A(T)x_B(0)\lb &=& \la \Psi|\frac{a+a^\dagger}{\s{2}}
e^{-iHT}\frac{b+b^\dagger}{\s{2}}|\Psi\lb \no
&=&\sum_{m,n,p,q=0}^{\infty}\la m|\frac{a+a^\dagger}{\s{2}}|n\lb_A
\la q|\frac{b+b^\dagger}{\s{2}}|p\lb_B \cdot [\rho_{AB}]^{mp}_{nq}.
\ea
After some algebras (refer to appendix \ref{HarmoB} for details), we obtain
\ba
\la x_A(T)x_B(0)\lb =-e^{-iT}\sinh\theta\cosh\theta.  \label{TPFH}
\ea

These results in the toy model imply not only that the correlation between two CFTs is generated by the double trace interaction, but also that the total state $\rho_{AB}$ of the two CFTs is not pure. The latter feature is characteristic to the model B and is missing in the model A. It is clear that we cannot explain $S_{AB}\neq 0$ from the dual classical geometry of wormhole because usually we regard a complete time slice in a given geometry as a pure state as follows from the calculation of holographic pseudo/entanglement entropy. Instead, the entropy comes from extra microscopic wormholes due to quantum correction in the gravity dual as we will argue in section \ref{sec:CoupledCFT}.

\section{Janus deformations and traversable wormholes}
\label{sec:Janus}
As one of realizations of a traversable wormhole in AdS, we consider a Janus solution with the Janus deformation parameter taken to be imaginary valued (i.e. model A). This model has an advantage that we can clearly understand its CFT dual in terms of the transition matrix of two different thermofield double states.

To show this, let us recall the AdS$_3$ Janus solution \cite{Bak:2007jm,Bak:2007qw}.
For this we start with the three dimensional action for the 3d metric and the dilaton field $\varphi$:
\ba
I_{\mathrm{grav}}=\frac{1}{16\pi \GN}\int d^3x\sqrt{g}\left[R[g]-g^{ab}\de_a\vp\de_b\vp+2\right].\label{actionJ}
\ea
The solution ansatz looks like
\ba
ds^2=f(\mu)(d\mu^2+ds^2_{\mathrm{AdS}_2}),\ \ \ \ \vp=\vp(\mu).\label{janusol}
\ea
The equations of motion for the scalar field $\vp$ and the metric $f(\mu)$ leads to 
\be
 \frac{d\vp(\mu)}{d\mu}=\frac{\gamma}{\s{f(\mu)}},\ \ \  \ \ 
 \frac{df(\mu)}{d\mu}=\s{f(4f^2-4f+2\gamma^2)},
\ee
where $\gamma$ is a constant which parametrizes the Janus deformation.

We choose the integration constant $\mu_0$ such that $-\mu_0\leq \mu\leq \mu_0$, given by
\ba
\mu_0=\s{\frac{2}{1+\chi}}\int^{1}_0 \frac{ds}{\s{\qty(1-s^2)\qty(1-\frac{1-\chi}{1+\chi} s^2)}}= \s{\frac{2}{1+\chi}}\EllipticK{\frac{1-\chi}{1+\chi}},
\ea
where we introduced $\chi=\sqrt{1-2\gamma^2}$, and $\EllipticK{z}$ is the complete elliptic integral of the first kind.
At $\mu=\pm\mu_0$, $f(\mu)$ gets divergent and these two correspond to two different AdS boundaries. When $\gamma=0$ i.e. without the Janus deformation, we have $f(\mu)=\frac{1}{\cos^2\mu}$ and $\mu_0=\frac{\pi}{2}$.
 At $\mu=\pm\mu_0$, the scalar field approaches 
\ba
\vp\to \pm \frac{1}{2\s{2}}\log \left[\frac{1+\s{2}\gamma}{1-\s{2}\gamma}\right].
\ea

The coordinate transformation $y=\int^{\mu}_0 ds\s{f(s)}$ rewrites the solution as follows
\ba\label{eq:Janus_coordinates}
&& ds^2=dy^2+f(y)ds^2_{\mathrm{AdS}_2},\no
&& f(y)=\frac{1}{2}(1+\s{1-2\gamma^2}\cosh 2y),\no
&& \vp(y)=\vp_0+\frac{1}{\s{2}}\log \left[\frac{1+\s{1-2\gamma^2}+\s{2}\gamma \tanh y}{1+\s{1-2\gamma^2}-\s{2}\gamma \tanh y}\right].
\ea

\subsection{Traversable wormholes from Janus solutions}

Now we consider the case where the AdS$_2$ given by an AdS$_2$ black hole \cite{Bak:2007jm,Bak:2007qw} (see also \cite{Bak:2011ga,Bak:2013uaa,Bak:2018txn}
\ba
ds^2_{\mathrm{AdS}_2}=-d\tau^2+r_0^2\cos^2\tau d\theta^2,\ \ \ (-\frac{\pi}{2}\leq \tau\leq \frac{\pi}{2}).  \label{JBH}
\ea

\begin{figure}[ht]
    \centering
    \begin{minipage}[h]{0.333\linewidth}
        \centering
      \resizebox{1\textwidth}{!}{%
\begin{circuitikz}
\tikzstyle{every node}=[font=\LARGE]
\draw(6.5,9.5)node[above]{\huge{$\gamma=0\;(\chi=1)$}};
\draw [short] (8.75,9) -- (8.75,4.25);
\draw [short] (4.25,9) -- (4.25,4.25);
\draw[domain=4.25:8.75,samples=100,smooth] plot (\x,{0.1*sin(10*\x r -10.75 r ) +9});
\draw [dashed] (4.25,4.25) -- (8.75,9);
\draw [dashed] (4.25,9) -- (8.75,4.25);
\draw[domain=4.25:8.75,samples=100,smooth] plot (\x,{0.1*sin(10*\x r -10.75 r ) +4.25});
\draw(3.5,8.6)node[left]{\huge{$\tau$}};
\draw [short,->] (3.55,4.25) -- (3.55,9);
\draw(9,3)node[below]{\huge{$\mu,y$}};
\draw [short,->] (3.8,3.3) -- (9,3.3);
\draw(9,3.4)node[above]{\large{$\mu_0$}};
\draw(4,3.4)node[above]{\large{$-\mu_0$}};
\draw[->] (8,5.75) to [out=135,in=-135] (8,7.75);
\draw(8,8)node[right]{\huge{$t$}};
\draw[->] (8,6.75)to [out=0,in=180](8.5,6.75);
\draw(8.5,6.5)node[below]{\huge{$\rho$}};
\draw[ultra thick,red,->](4.25,6)--(7,9);
\end{circuitikz}
}%
    \end{minipage}
    \begin{minipage}[h]{0.33\linewidth}
        \centering
      \resizebox{1\textwidth}{!}{%
\begin{circuitikz}
\tikzstyle{every node}=[font=\LARGE]
\draw(6.5,9.5)node[above]{\huge{$\gamma^2>0\;(\chi<1)$}};
\draw [short] (10,9) -- (10,4.25);
\draw [short] (4.25,9) -- (4.25,4.25);
\draw[domain=4.25:10,samples=100,smooth] plot (\x,{0.1*sin(10*\x r -10.75 r ) +9});
\draw [dashed] (5.5,4.25) -- (10,9);
\draw [dashed] (5.5,9) -- (10,4.25);
\draw [dashed] (4.25,9) -- (8.75,4.25);
\draw [dashed] (4.25,4.25) -- (8.75,9);
\draw[domain=4.25:10,samples=100,smooth] plot (\x,{0.1*sin(10*\x r -10.75 r ) +4.25});
\draw(3.5,8.6)node[left]{\huge{$\tau$}};
\draw [short,->] (3.55,4.25) -- (3.55,9);
\draw(10,3)node[below]{\huge{$\mu,y$}};
\draw [short,->] (3.8,3.3) -- (10,3.3);
\draw(10,3.4)node[above]{\large{$\mu_0$}};
\draw(4,3.4)node[above]{\large{$-\mu_0$}};
\draw[ultra thick,red,->](4.25,6)--(7,9);
\end{circuitikz}
}%
    \end{minipage}
     \begin{minipage}[h]{0.333\linewidth}
        \centering
      \resizebox{1\textwidth}{!}{%
\begin{circuitikz}
\tikzstyle{every node}=[font=\LARGE]
\draw(6.5,9.5)node[above]{\huge{$\gamma^2<0\;(\chi>1)$}};
\draw [short] (8.75,9) -- (8.75,2.25);
\draw [short] (4.25,9) -- (4.25,2.25);
\draw[domain=4.25:8.75,samples=100,smooth] plot (\x,{0.1*sin(10*\x r -10.75 r ) +9});
\draw [dashed] (4.25,4.25) -- (8.75,9);
\draw [dashed] (4.25,9) -- (8.75,4.25);
\draw [dashed] (4.25,2.25) -- (8.75,7);
\draw [dashed] (4.25,7) -- (8.75,2.25);
\draw[domain=4.25:8.75,samples=100,smooth] plot (\x,{0.1*sin(10*\x r -10.75 r ) +2.25});
\draw(2.7,8.6)node[left]{\huge{$\tau$}};
\draw [short,->] (2.7,2.25) -- (2.7,9);
\draw(9,1)node[below]{\huge{$\mu,y$}};
\draw [short,->] (3.8,1.3) -- (9,1.3);
\draw(9,1.4)node[above]{\large{$\mu_0$}};
\draw(4,1.4)node[above]{\large{$-\mu_0$}};
\draw[ultra thick,red,->](4.25,3.25)--(8.75,7.75);
\draw[blue,ultra thick](4.25,4)to [out=10,in=190]  (8.75,7);
\draw(8.75,7)node[right]{\huge{$\tau_b^{(2)}$}};
\draw(4.25,4)node[left]{\huge{$\tau_b^{(1)}$}};
\end{circuitikz}
}%
    \end{minipage}
     \caption{The Penrose diagram of the BTZ black hole (above,left, $\gamma^2=0$), time dependent Janus (above right $\gamma^2>0$) and the traversable wormhole (below $\gamma^2<0$). The red line denote the null geodesics. For $\gamma^2<0$ case, the null geodesics from left reaches to the right boundary. The blue line denotes the geodesics connecting between two points locates in left and right boundaries. }
    \label{fig:Janus}
\end{figure}
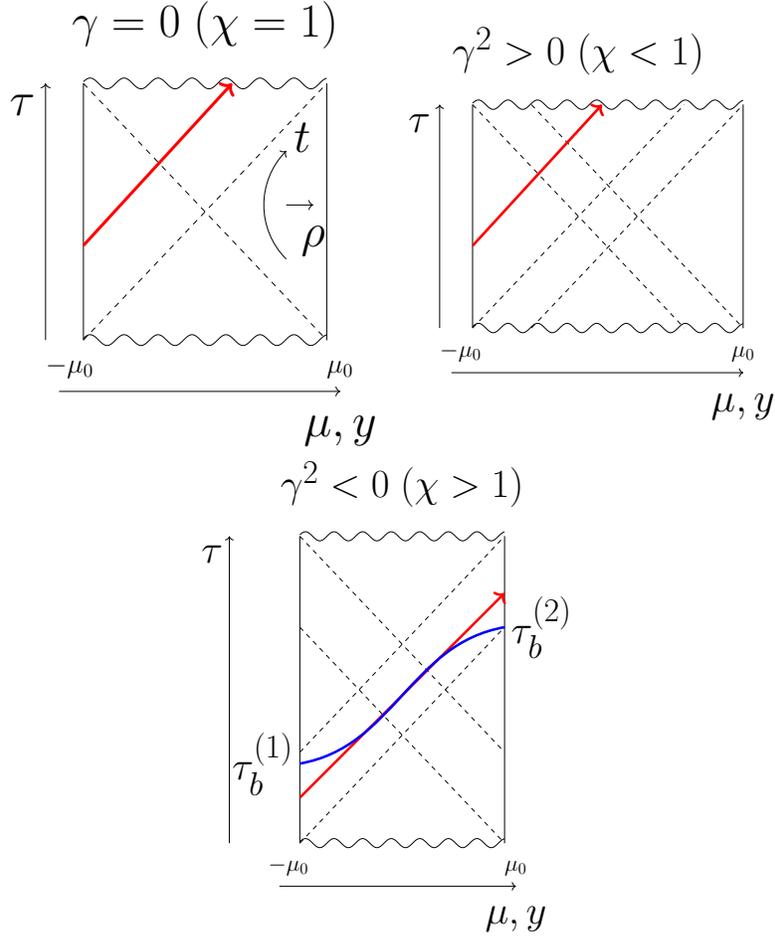

In order to keep the metric and scalar real valued, we need to require
$0\leq \gamma^2\leq \frac{1}{2}$. The trivial case without the Janus deformation  corresponds to $\gamma=0$, which coincides with the BTZ solution. In this case, we have $\mu_0>\frac{\pi}{2}$ and the left and right boundary are not causally connected, as depicted in the upper right picture in Fig.\ref{fig:Janus}. 

Now we allow the scalar field to take imaginary values, though we require the metric remains to be real valued. This condition is satisfied when 
$\gamma$ is imaginary i.e. $\gamma^2<0$. In this extended case, we can easily show 
$\mu_0\leq \frac{\pi}{2}$. Thus the left and right boundary is now causally connected and it becomes a traversable wormhole as depicted in the lower right panel of Fig.\ref{fig:Janus}. In this case the scalar field takes imaginary values \footnote{The complex dilaton (axion) solution appears also in the Giddings-Strominger Euclidean wormhole solution \cite{GIDDINGS1988890}. In the dilaton or axion gravity, we have dual field representation, that is the theory with $p$-form gauge field $B_p$ which field strength $H_{p+1}$ is related to axion such that $H_{p+1}=*d\vp$. In the Giddings-Strominger case, the solution of the dual field is real \cite{Hebecker:2018ofv}. On the other hand, in our case, the dual field strength $H_2$ is given by $H_2 \propto \sqrt{2}\gamma \cos{\tau}d\tau \wedge d\theta$ and also pure imaginary. Notice that in Euclidean signature, the dual field is real similar to the Giddings-Strominger case.}.
This leads to the negative energy background which allows the traversable wormhole.\footnote{In \cite{Bak:2018txn}, the Janus black hole solution for a real value of $\gamma$ can be made traversable if we further deform it by a double trace deformation. In this paper we do not consider any double trace deformation when we talk about Janus solutions.}

\subsection{Computing geodesic length}
To explicitly confirm that the solution describes a traversable wormhole when $\gamma$ takes imaginary value, let us compute the geodesic length between two points, each at one of the two different boundaries. Before we discuss the geodesics length we need to rewrite our coordinates \eqref{eq:Janus_coordinates} because the time $\tau$ is not the proper boundary time. To see this, let us going back the BTZ black hole case, \ie $\gamma=0$. In this case, the convenient coordinate is the following familiar BTZ metric
\begin{equation}
     \begin{split}
         ds^2&=-r_0^2\sinh^2{\rho}dt^2+d\rho^2+r_0^2\cosh^2{\rho}\;d\theta^2.
     \end{split} 
 \end{equation}
This coordinate and the Janus coordinates are related by the transformations
 \begin{equation}
     \begin{split}
         \sinh{y}&=\sinh{\rho}\cosh{r_0 t},\\
         \cosh{y}\sin{\tau}&=\sinh{\rho}\sinh{r_0t},\\
         \cosh{y}\cos{\tau}&=\cosh{\rho}.
     \end{split}
 \end{equation}

\begin{figure}[ht]
    \centering
    \includegraphics[width=0.75\linewidth]{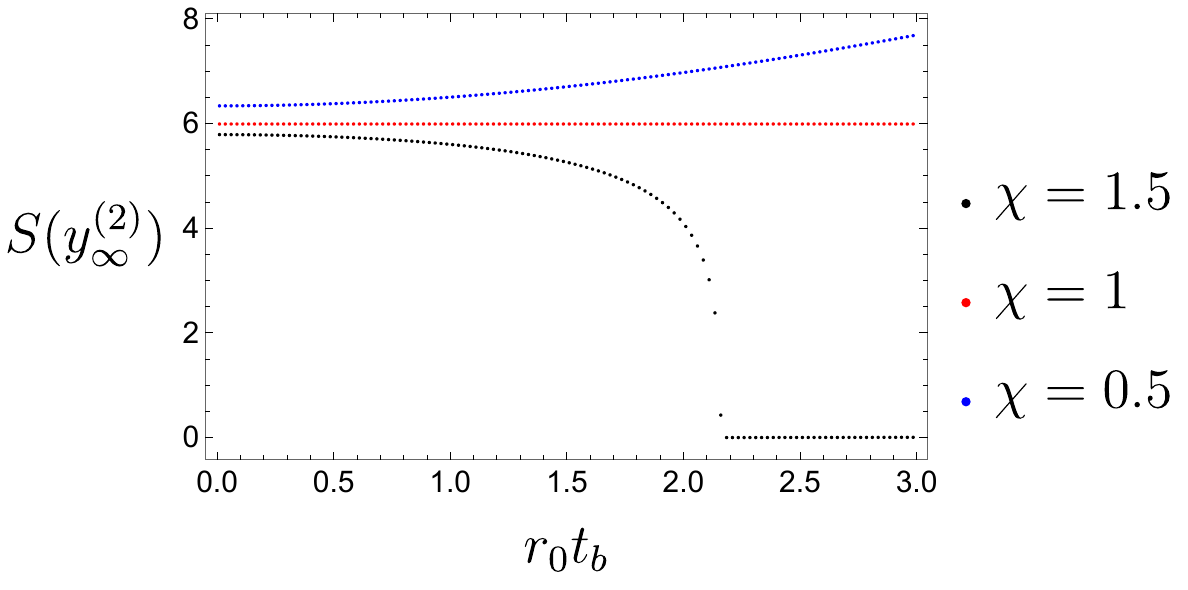}
    \caption{Plots of the geodesics length with  $\chi=1.5$ (black), $\chi=1$ (red), $\chi=0.5$ (blue). We set $\ep=0.01$. For $\chi=1.5>1$, we observed that the geodesic length getting zero when $r_0 t^{\text{crit}}_b\approx 2.16$. }
    \label{fig:Janus_EE}
\end{figure}

 Then we consider the geodesics connecting two point $(\tau^{(2)}_b,y_\infty)$ and $(\tau^{(1)}_b,-y_\infty)$. See Fig.\ref{fig:Janus} for intuition. The asymptotic values $\tau_\infty$ and $y_\infty$ are related to the boundary values $\frac{e^{\rho_\infty}}{2}\approx\frac{2}{\ep},\; \tau_\infty\approx t_b$ by 
 \begin{equation}
     \begin{split}
         f(y_\infty) &\approx \frac{1}{4}e^{2\rho_\infty}\cosh^2{r_0 t_b}\\
         \tan{\tau_\infty}&\approx \sinh{r_0 t_b}
     \end{split}
     \end{equation}
     where $t_b$ is the time at which the subsystem is defined. 
     Then, we consider the geodesics which connects two asymptotic boundaries. The geodesics is along $\theta=\const$. Thus, we consider the curve $\tau=\tau(y)$. The geodesics is determined by varying the functional
     \begin{equation}
         S_{\mathrm{geodesics}}[y]=\int_{-y_{\infty}}^{y_{\infty}} dy \sqrt{1-f(y)\qty(\frac{d\tau}{dy})^2}.
     \end{equation}
     The equation of motion is simply obtained by $\delta S_{\mathrm{geodesics}}[y]=0$ and 
     \begin{equation}
     \begin{split}
         \frac{d}{dy}\qty(\frac{f(y)\tau'(y)}{\sqrt{1-f(y)(\tau'(y))^2}})=0, \Leftrightarrow \frac{f(y)\tau'}{\sqrt{1-f(y)\tau'^2}}=\sqrt{C}
     \end{split}
     \end{equation}
     where $C$ is a positive integration constant. By solving the equation, we obtain,
     \begin{equation}
         \begin{split}
             \tau(y)&=B+A(y;C),\\
               A(y;C)&= \int^y dy\sqrt{\frac{C}{f(y)(f(y)+C)}}\\
               &= \frac{2\sqrt{C}}{\sqrt{(1+\chi)(1+\chi+2C)}}\int_0^{\tanh{y}}\frac{ds}{\sqrt{\qty(1-\frac{1-\chi}{1+\chi}s^2)\qty(1-\frac{1-\chi+2C}{1+\chi+2C}s^2)}}.
         \end{split}
     \end{equation}
     Importantly, $A(-y,C)=-A(y,C)$ and we find that
\begin{equation}
    \begin{split}
        \tau(y\to\infty) &=B + A(\infty;C)=\tau^{(2)}_b\\
        \tau(y\to-\infty) &=B -A(\infty;C)=\tau^{(1)}_b\\
        \Leftrightarrow B= \frac{\tau^{(2)}_b+\tau^{(1)}_b}{2}&,\; A(\infty;C) = \frac{\tau^{(2)}_b-\tau^{(1)}_b}{2}.
    \end{split}
\end{equation}
The conditions with function $A(\infty;C)$ determines the value of $C$ in terms of the boundary time. 
     The simple case will be $\tau^{(1)}_b=\tau^{(2)}_b=\tau_b$ and in this case $B=\tau_b,C=0$ and thus $\tau(y)=\tau_0$ even for non zero $\gamma$.\par 
     Next we discuss the on-shell action. It is analytically obtained as
     \begin{equation}
         \begin{split}
             \underline{S}_{\mathrm{geodesics}} &= \int_{-y^{(2)}_\infty}^{y^{(2)}_\infty} dy \sqrt{\frac{f(y)}{f(y)+C}}\\
             &= 2S(y^{(2)}_\infty;C)
         \end{split}
     \end{equation}
    where we introduce 
     \begin{equation}
        \begin{split}
           S(y;C)
            &= \int^y dy\sqrt{\frac{f(y)}{f(y)+C}}\\
            &=\sqrt{\frac{1+\chi}{1+\chi+2C}}\int_0^{\tanh{y_\infty}}\frac{ds}{1-s^2}\sqrt{\frac{1-\frac{1-\chi}{1+\chi}s^2}{1-\frac{1-\chi+2C}{1+\chi+2C}s^2}}.
        \end{split}
     \end{equation}
     One simple case is  $t^{(1)}_b=t^{(2)}_b=t_b$ and we have $C=0$ and $\tau(y)= \tan^{-1}{\sinh{r_0 t_b}}$. In this case,
     \begin{equation}
         \underline{S}_{\mathrm{geodesics}}= 2y^{(2)}_\infty = 2\log\qty({\frac{1}{\chi}\frac{2}{\ep}\cosh{r_0 t_b}}).
     \end{equation}
     This results is similar to the linear growth in \cite{Hartman:2013qma}.
     The interesting case is when we set $t^{(1)}_b=-t^{(2)}_b=t_b$. In this case, we find that at some critical time the geodesics length approaching to zero (See Fig. \ref{fig:Janus_EE}) for $\chi>1$. This is because that we have a traversable wormhole in the bulk and we can connect the two points causally. When $\chi=1$, $S(y_\infty;C)=\rho_\infty=\const$ because of the isometry of the BTZ blackhole or thermofield double state is invariant under the time translation with opposite sign. On the other hand $\chi<1$ case (e.g. real valued Janus deformation), we find the geodesic length monotonically increasing because the ER bridge getting long. 
     
     For $\chi>1$, the critical time can be analytically derived as follows. Since the critical time is the time that the geodesics getting null, we demand $\frac{f(y)}{f(y)+C}=0$ for any $y\in [-y^{(2)}_\infty,y^{(2)}_{\infty}]$. Since $f(y)>0$ for $\chi>0$, this is a case when $C=+\infty$. In this case, the expression of $\tau_b=A(\infty,C)$ simplifies to 
     \begin{equation}
         \tau_b^{\text{crit}} = A(\infty;C\to\infty)= \sqrt{\frac{2}{1+\chi}}K\qty(\frac{1-\chi}{1+\chi})= \mu_0.
     \end{equation}
Thus the critical time is given by 
\begin{equation}
    r_0 t^{\text{crit.}}_b = \arcsinh\left(\tan\mu_0\right).
\end{equation}
We see that $t^{\text{crit.}}_b$ is positive finite only when $\chi>1$. 

\subsection{Janus thermofield double state in free scalar CFT}
\label{sec:scalarjanus}

In the well-known example of the AdS$_3/$CFT$_{(2)}$ \cite{Maldacena:1997re}, the gravity on AdS$_3$ is dual to the so called D1-D5 CFT, which is defined by the supersymmetric two dimensional CFT with the central charge $c=6Q_1Q_5$ for $Q_1$ D1 -branes and $Q_5$ D5-branes. Its bosonic part is described by a symmetric product CFT Sym$(T^4)^{Q_1Q_5}$ with $c=4Q_1Q_5$, where $T^4$ represents a CFT which consists of four compactified massless scalar fields. This duality arises because the near horizon limit of the D1-D5 system is type IIB string on AdS$_3\times$S$^3\times T^4$.

The CFT dual to the classical gravity on the AdS$_3$ is obtained in the strongly interacting limit. Similarly the BTZ black hole is dual to the thermofield double (TFD) state of the D1-D5 CFT, where the radii of $T^4$ are identical in the CFT$_{(1)}$ and CFT$_{(2)}$.

As a deformation of this AdS$_3/$CFT$_{(2)}$, the AdS$_3$ Janus solution (\ref{janusol}) is dual to an asymmetric TFD state of the D1-D5 CFT \cite{Bak:2007jm,Bak:2007qw,Nakaguchi:2014eiu}, where the radius of $T^4$ (denoted by $R_1$) in the CFT$_{(1)}$ and that ($R_2$) in the CFT$_{(2)}$ differs. As a tractable toy example, here we would like to examine a $c=1$ free scalar compactified on a circle. We consider a TFD state in the $c=2$ CFT which consists of a scalar $\phi^{(1)}$ (radius $R_1$) and another one $\phi^{(2)}$ (radius $R_2$). The Janus deformation is measured by the parameter $\theta$ by defined by
\ba
\tan\theta=\frac{R_2}{R_1}.
\ea
For $\theta=\frac{\pi}{4}$, we have $R_1=R_2$, which corresponds to the TFD state before the Janus deformation. The difference $\theta-\frac{\pi}{4}$ quantifies the Janus deformation. In the D1-D5 CFT, such an asymmetric radius shift corresponds to turning on the real valued bulk scalar field $\vp$ in (\ref{actionJ}), which is proportional to the parameter $\gamma$. On the other hand, we are interested in the Janus deformation which leads to a non-hermitian transition matrix. This corresponds to  an imaginary value of $\gamma\left(\propto \theta-\frac{\pi}{4}\right)$.

\begin{figure}[ttt]
		\centering
		\includegraphics[width=9cm]{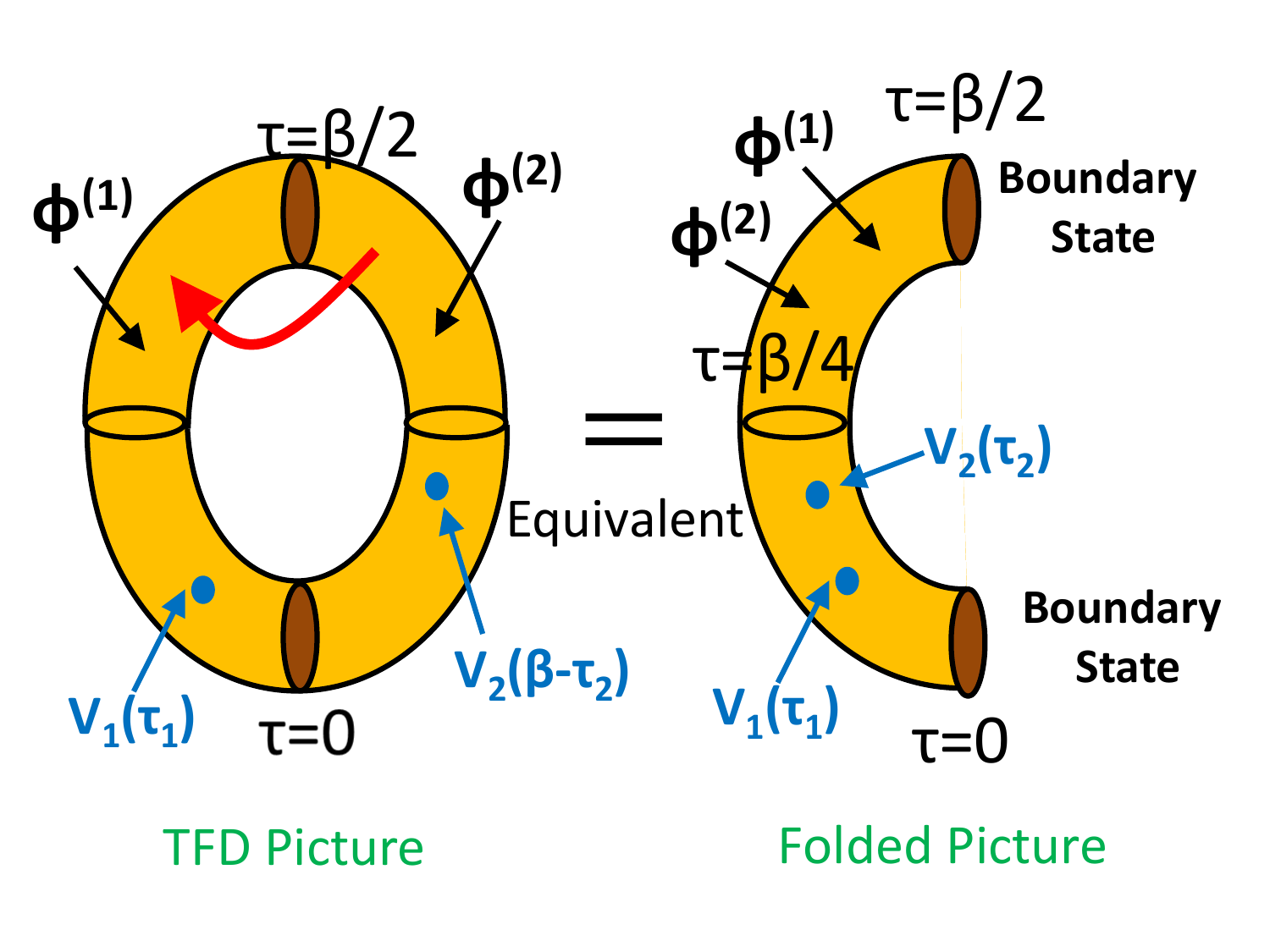}
		\caption{A sketch of the Euclidean description of the asymmetric TFD state (left) and its equivalent description using the doubling trick (right). In the left the two point function is computed for a $c=1$ CFT on a torus, while in the right it is computed for a $c=2$ CFT on a cylinder with two boundaries.} 
		\label{fig:Fold}
\end{figure}

Consider the CFT lives on a cylinder whose Euclidean time and space coordinate are written by $\tau$ and $\sigma$. The latter obeys $2\pi$ periodicity. Such a TFD state can be constructed by path-integration from $\tau=-\frac{\beta}{2}$ to $\tau=\frac{\beta}{2}$ where we inserted a conformal interface at $\tau=0$ such that the scalar field on $-\f{\beta}{2}\leq \tau< 0$ is $\phi^{(1)}$, while that on  $0<\tau\leq \f{\beta}{2}$ is the other scalar field $\phi^{(2)}$ as depicted in Fig.\ref{fig:Fold}. This setup is equivalent to 
the setup of two scalar fields on a cylinder, via the doubling method \cite{Bachas:2001vj,Sakai:2008tt} as in the right panel of Fig.\ref{fig:Fold}. For the detailed calculations, refer to the appendix \ref{ap:januscft}.

As a probe of the TFD state, we analyze the two point function 
$\la V_1(\tau_1)V_2(\tau_2)\lb$, by choosing the vertex operators $V_1$ and $V_2$ to be \ba
&& V_1=e^{i\lambda_+\phi^{(1)}_L(\tau_1)+i\lambda_-\phi^{(1)}_R(\tau_1)},\ \ \ \  V_2=e^{i\mu_+\phi^{(2)}_L(\tau_2)+i\mu_-\phi^{(2)}_R(\tau_2)}, \label{vextoa}
\ea where $V(\tau)=e^{\tau H}V(0)e^{-\tau H}$. We express the left and right moving part of the massless scalar field $\phi^{(i)}$ as $\phi^{(i)}_L$ and $\phi^{(i)}_R$ for $i=1,2$. For simplicity we insert them at the same spacial coordinate 
$\sigma_1=\sigma_2=0$. By imposing zero mode condition (\ref{qqq}), in terms of the winding number $w$ and momentum number $n$ we have  \ba
&& \lambda_{\pm}=\frac{n}{R_1}\pm \frac{wR_1}{2},
\ \ \ \ \mu_{\pm}=\frac{n}{R_2}\mp \frac{wR_2}{2}.\label{vextob}
\ea

As shown in the appendix \ref{ap:januscft}, the two point function $\la V_1(\tau_1)V_2(\tau_2)\lb$ is found as:
\ba
&& \la V_1(\tau_1)V_2(\tau_2)\lb \no
&&=\frac{\sum_{\ti{w},\ti{n}\in Z} e^{-\frac{\beta}{2}\left[\frac{\ti{w}^2(R_1^2+R_2^2)}{4}+\ti{n}^2\left(\frac{1}{R_1^2}+\frac{1}{R_2^2}\right)\right]+\left(\frac{2n\ti{n}}{R_1^2}+\frac{w\ti{w}R_1^2}{2}\right)\tau_1+\left(\frac{2n\ti{n}}{R_2^2}+\frac{w\ti{w}R_2^2}{2}\right)\tau_2}}{ \sum_{\ti{w},\ti{n}\in Z} e^{-\frac{\beta}{2}\left[\frac{\ti{w}^2(R_1^2+R_2^2)}{4}+\ti{n}^2\left(\frac{1}{R_1^2}+\frac{1}{R_2^2}\right)\right]}}\no
&& \cdot \left[\frac{\theta_1\left(\frac{i\tau_1}{\pi}\Bigr|\frac{i\beta}{2\pi}\right)}{\eta\left(\frac{i\beta}{2\pi}\right)^3}
\right]^{\left[\left(\frac{n}{R_1}\right)^2-\left(\frac{wR_1}{2}\right)^2\right]\cos2\theta}\cdot 
 \left[\frac{\theta_1\left(\frac{i\tau_2}{\pi}\Bigr|\frac{i\beta}{2\pi}\right)}{\eta\left(\frac{i\beta}{2\pi}\right)^3}
\right]^{\left[-\left(\frac{n}{R_2}\right)^2+\left(\frac{wR_2}{2}\right)^2\right]\cos2\theta}\no
&&  \cdot\left[\frac{\theta_1\left(\frac{i(\tau_1+\tau_2)}{2\pi}\Bigr|\frac{i\beta}{2\pi}\right)}{\eta\left(\frac{i\beta}{2\pi}\right)^3}
\right]^{-2\left[\frac{n^2}{R_1R_2}+\frac{w^2 R_1R_2}{4}\right]\sin2\theta}. \label{VV}
\ea

If we take the limit $\tau_1\to 0$ and $\tau_2\to 0$, we find
\ba
\la V_1(\tau_1)V_2(\tau_2)\lb\propto \frac{1}{|\tau_1|^{\Delta_1}|\tau_2|^{\Delta_2}}
\cdot \left[\frac{(\tau_1+\tau_2)^2}{4\tau_1\tau_2}
\right]^{-\left(\frac{n^2}{R_1R_2}+\frac{w^2R_1R_2}{4}\right)\sin2\theta}.
\ea
This fits nicely with the general form of two point functions in interface CFTs \cite{Chiodaroli:2016jod}. Moreover, if we set $\theta=\frac{\pi}{4}$ or equally $R_1=R_2$, then we can confirm that (\ref{VV}) is reduced to the known two point function on a torus of a single scalar CFT. 

We are interested in the high temperature limit $\beta\to 0$ in order to compare our results with those of the BTZ black hole. We obtain the following expression of two point function in the high temperature limit (refer to the appendix \ref{ap:januscft} for details):
\ba
&& \la V_1(\tau_1)V_2(\tau_2)\lb \no
&&\simeq \left[\frac{\beta}{\pi}\cdot \sin\left(\frac{2\pi \tau_1}{\beta}\right) \right]^{\left[\left(\frac{n}{R_1}\right)^2-\left(\frac{wR_1}{2}\right)^2\right]\cos2\theta}
\cdot \left[\frac{\beta}{\pi}\cdot \sin\left(\frac{2\pi \tau_2}{\beta}\right) \right]^{\left[-\left(\frac{n}{R_2}\right)^2+\left(\frac{wR_2}{2}\right)^2\right]\cos2\theta}\no
&&
\ \ \ \  \cdot\left[\frac{\beta}{\pi}\cdot \sin\left(\frac{\pi(\tau_1+\tau_2)}{\beta}\right) \right]^{-2\left[\frac{n^2}{R_1R_2}+\frac{w^2 R_1R_2}{4}\right]\sin2\theta}. \label{hightempqx}
\ea
\vspace{5mm}

When $\tau_1=\tau_2(\equiv\tau)$, we find 
\ba
\la V_1(\tau_1)V_2(\tau_2)\lb\propto 
 \left[\frac{\beta}{\pi}\cdot \sin\left(\frac{2\pi \tau}{\beta}\right)\right]^{-\Delta_1-\Delta_2}.
\ea
We can obtain the Lorentzian time evolution by setting
\ba
\tau_1=\frac{\beta}{4}+it,\ \ \ \ \tau_2=\frac{\beta}{4}+it.
\ea
Then we find 
\ba
\la V_1(\tau_1)V_2(\tau_2)\lb\propto 
 \left[\frac{\beta}{\pi}\cdot \cosh\frac{2\pi }{\beta}t  \right]^{-\Delta_1-\Delta_2}.
\ea
This behavior does not depend on the Janus parameter $\theta$. This is consistent with the time evolution of the geodesic length in the BTZ (\ref{t1t2}) via the standard identification $\la V_1V_2\lb\sim e^{-(\Delta_1+\Delta_2)D_{12}}$, though the BTZ result is shifted by the boundary entropy term. 

On the other hand, when $\tau_1=\tau$ and $\tau_2=\frac{\beta}{2}-\tau$, we find 
\ba
\la V_1(\tau_1)V_2(\tau_2)\lb\propto 
 \left[\frac{\beta}{\pi}\cdot \sin\left(\frac{2\pi \tau}{\beta}\right)\right]^{\eta},
\ea
where
\ba
\eta
&=&-\frac{(R^2_1-R^2_2)^2}{(R_1^2+R_2^2)R_1R_2}\cdot \left(\frac{n^2}{R_1R_2}+\frac{w^2R_1R_2}{4}\right).
\ea
Notice that $\eta<0$ for non-zero Janus deformation $R_1\neq R_2$.
The Lorentzian continuation can be computed by setting
\ba
\tau_1=\frac{\beta}{4}+it,\ \ \ \ \tau_2=\frac{\beta}{4}-it.
\ea
In this case the two point function is suppressed in the high temperature region $\beta\ll 2\pi$:
\ba
\la V_1(t_1)V_2(t_2)\lb\propto \left[\frac{\beta}{\pi}\cdot\cosh\frac{2\pi }{\beta}t\right]^{\eta}. \label{WWs}
\ea
This looks qualitatively agreeing with the Janus gravity dual for real values of $\gamma$, where the internal region expands to increase the geodesic length and to reduce the two point function, by remembering the global geometry in the final panel of Fig.\ref{fig:Janus}.

Let us extend the Janus deformation to the case where $\theta-\frac{\pi}{4}$ is imaginary. In the holographic context, this is expected to correspond to the Janus solution with an imaginary value of the bulk scalar field, i.e. imaginary $\gamma$. We set $\theta=\frac{\pi}{4}-i\delta/2$, where $\delta$ is the imaginary Janus deformation parameter. Then we get
\ba
\cos2\theta=i\sinh\delta,\ \ \ \ \sin2\theta=\cosh\delta. \label{wickq}
\ea
In terms of radius we find
\ba
R_1=\frac{R}{\s{\tan\theta}}, \ \ \ \ R_2=\s{\tan\theta}R,
\ea
where $R$ is the radius before the Janus deformation. Note that under the imaginary Janus deformation, the radii $R_1$ and $R_2$ take complex values. In this case, we find
\ba
\eta=\frac{2\sinh^2\delta}{\cosh\delta}\cdot \left(\frac{n^2}{R^2}+\frac{w^2R^2}{4}\right).
\ea
Since $\eta>0$ for the imaginary Janus deformation, we can conclude that the two point function (\ref{WWs}) increases under the time evolution and this qualitatively agrees with the gravity dual expectation of traversable wormhole as computed in (\ref{t1nt2}).

\subsection{Entanglement entropy versus pseudo entropy}
As a final analysis of the Janus wormhole, let us look at the pseudo entropy in the dual CFTs. First note that the pseudo entropy for the total system of two CFTs i.e. $S_{AB}$ does vanish because the total state is pure. The pseudo entropy $S_A$ is non-trivial and should be related to some sort of entanglement between CFT$_{(1)}(=A)$ and
CFT$_{(2)}(=B)$. Note that when the state is described by a hermitian density matrix (i.e. $\gamma^2\geq 0$),  $S_{A}$ becomes the usual entanglement entropy.

In the gravity solution (\ref{eq:Janus_coordinates}) and (\ref{JBH}), we can compute $S_A$ from the area of the minimal surface which divides the time slice $\tau=0$ into the left and right side, as follows
\ba
S_A=S_B=\frac{2\pi r_0f(0)}{4G_N}=\frac{\pi}{6}cr_0\left(1+\s{1-\gamma^2}\right).
\label{JBHS}
\ea
This is monotonically decreasing under the usual Janus deformation $\gamma^2>0$, while it is increasing under the imaginary Janus deformation $\gamma^2<0$. 

The Janus deformation of the thermofield double state in the free scalar interface CFT introduced in \ref{sec:scalarjanus} is described by the quantum state (i.e. the boundary state), given by (\ref{BST}). This is schematically written as 
\ba
|\mathrm{TFD}(\beta,\gamma)\lb={\cal \ti{N}}\exp\left[\sum_{i=1}^{\infty}e^{-\frac{\beta}{2}E_i}\left(\sin2\theta~ a^\dagger_ib^\dagger_i
+\cos2\theta\left((a^\dagger_i)^2-(b^\dagger_i)^2\right)\right)\right]|0\lb,
\label{tfdi}
\ea
where ${\cal \ti{N}}$ is a normalization constant. $a^\dagger_i\in A$ and  $b^\dagger_i\in B$ are the infinitely many creation operators corresponding to each mode in the free CFT and $E_i$ are their energies. It is obvious that when $\theta=\frac{\pi}{4}$ (i.e. no deformation $\gamma=0$), the state gets maximally entangled and $S_A$ reaches its maximum as a function of $\theta$. It is obvious that as $\theta$ deviates from $\frac{\pi}{4}$, the amount of quantum entanglement should decrease as $S_A=S_{max}-s(\theta-\frac{\pi}{4})^2+\ddd$, where $s$ is a positive constant. By identifying $\theta-\frac{\pi}{4}\propto \gamma$, we realize $S_A$ is a monotonically decreasing function of $\gamma^2$, agreeing with (\ref{JBHS}).

Then we turn to the imaginary Janus deformation $\gamma^2<0$. The gravity result (\ref{JBHS}) argues that $S_A$ becomes larger than the $\gamma=0$ case. One may get confused because $\gamma=0$ is the maximally entangled state and it does not seem to be able to increase the entropy. However, it is known that pseudo entropy can exceed the maximum value \cite{Nakata:2021ubr,Ishiyama:2022odv}. The Janus geometry is dual to $\rho_{AB}=|\mathrm{TFD}(\beta,\gamma)\lb\la \mathrm{TFD}(\beta,\gamma)|$ and importantly, this is not hermitian when $\gamma^2<0$, where the initial state $|\mathrm{TFD}(\beta,\gamma)\la$ is given by (\ref{tfdi}) and the final state
$\la \mathrm{TFD}(\beta,\gamma)|$ is defined by
\ba
\la \mathrm{TFD}(\beta,\gamma)|={\cal \ti{N}}\la 0|\exp\left[\sum_{i=1}^{\infty}e^{-\frac{\beta}{2}E_i}\left(\sin2\theta~ a_i b_i+\cos2\theta\left((a_i)^2-(b_i)^2\right)\right)\right]. \label{tfdf}
\ea
When $\theta$ is complex valued such that $\theta=\frac{\pi}{4}-i\frac{\delta}{2}$ as in (\ref{wickq}), it is clear that the hermitian conjugate of (\ref{tfdi}) does not coincide with  (\ref{tfdf}) and thus $\rho_{AB}$ is not hermitian. Since $S_A$ can be obtained by the analytical continuation of the result for real $\theta$ and $S_A$ is an even function of $\theta-\frac{\pi}{4}$, we expect that $S_A$ is a monotonically increasing function of $\delta^2\propto -\gamma^2$. This explains the gravity dual behavior (\ref{JBHS}).

\section{Double trace deformation and traversable AdS wormholes}
\label{sec:DT}

Now we explore another way to get CFT duals of traversable AdS geometries, i.e. double trace deformations. For scalar operators, the double trace deformation of two CFTs is described by the action (\ref{DTla}), which we will focus on below. The main purpose of here is to show that two point functions in a double trace deformed CFT can reproduce holographic correlation functions in an traversable AdS wormhole. We will also present a free CFT model which shares basic properties of the double trace deformation.

Essentially the same argument can be done for the double trace deformation by energy stress tensor. This allows us to make a more universal argument such that the gravity sector lives in a traversable wormhole, which will discussed in section \ref{sec:TTbar}.

\subsection{Double trace deformation of scalar operator and two point functions}\label{sec:deformb}
Consider the double trace deformation of a single operator $\mO$ with the dimension $\Delta$:
\ba
S=S_{\mathrm{CFT}(1)}+S_{\mathrm{CFT}(2)}+\frac{1}{2}\int d^dxd^dy\ \lambda(x,y)\mO_1(x)\mO_2(y),  \label{doubleab}
\ea
where $\mO_1$ is a primary operators in the CFT$_{(1)}$ and $\mO_2$ is its copy in the CFT$_{(2)}$. We assume the translational invariance and perform the Fourier transformation:
\ba
\lambda (x,y)=\int d^dk e^{ik(x-y)} \lambda(k).
\ea
As mentioned before we are interested in $\lambda(k)$ which decays fast in the UV in order to have a wormhole localized in the IR region $z>z_0$ in the AdS.

Consider a pair of asymptotically AdS$_{d+1}$ geometries which are originally disconnected such that they are dual to CFT$_{(1)}$ and CFT$_{(2)}$. The scalar field in the first and second AdS are denoted by $\Phi^{(1)}$ and $\Phi^{(2)}$. 
Before the deformation, the sources $\ap^{(i)}$ and expectation values $\beta^{(i)}(=\la \mO_i\lb)$ in the two CFTs $i=1,2$, are found from the near boundary behavior of the scalar fields as obeys from the general formulation \cite{Klebanov:1999tb}: 
\ba
\Phi^{(i)}\simeq \ap^{(i)}z_i^{d-\Delta}+\beta^{(i)}z_i^\Delta\ \ \ (z_1,z_2\to 0),
\ea
where $z_1$ and $z_2$ are the radial coordinate of the two AdS geometries. By requiring the regular behavior in the IR limit $z_{1,2}\to \infty$, we find the relation between $\ap^{(i)}$ and $\beta^{(i)}$
\ba
\frac{\beta^{(i)}}{\ap^{(i)}}=-G(k), \label{relgbe}
\ea
Note that $G(k)$ is the Fourier transformation of the standard two point function in the undeformed theory. When we consider the Poincar\'e AdS$_{d+1}$: $ds^2=z^{-2}(dz^2+dx_\mu dx^\mu)$, this is explicitly given by 
\ba
 G_p(k)\equiv \frac{\Gamma(1-\nu)}{\Gamma(1+\nu)}\left(\frac{k}{2}\right)^{2\nu}.
 \label{PoinG}
\ea

In the presence of the double trace deformation \cite{Aharony:2001pa,Witten:2001kn} (see also 
\cite{Mueck:2002gm,Gubser:2002vv,Gubser:2002zh,Fujita:2008rs} for further calculations useful for arguments below), the relation between the sources and the expectation values are deformed. If we write the actual sources after the deformation as $J^{(i)}$, the relation reads 
\ba
J^{(1)}=\ap^{(1)}-\lambda \beta^{(2)},\ \ \ J^{(2)}=\ap^{(2)}-\lambda \beta^{(1)}. \label{bcdefz}
\ea
By using (\ref{relgbe}) and (\ref{bcdefz}), we obtain 
\ba
&& \beta^{(1)}=-\frac{G}{1-\lambda^2 G^2}J^{(1)}+\frac{\lambda G^2}{1-\lambda^2 G^2}J^{(2)}, \no
&&\beta^{(2)}=-\frac{G}{1-\lambda^2 G^2}J^{(2)}+\frac{\lambda G^2}{1-\lambda^2 G^2}J^{(1)}.  \label{DTbet}
\ea
Thus the two point functions are found to be
\ba
&& \la \mO_1 (k)\mO_1 (-k)\lb=\la \mO_2(k)\mO_2(-k)\lb=\frac{G}{1-\lambda^2 G^2}, \no
&& \la \mO_1 (k)\mO_2(-k)\lb=\frac{\lambda G^2}{1-\lambda^2 G^2}.  \label{tpfdtq}
\ea

Let us compare these two point functions (\ref{tpfdtq}) with those computed from the traversable wormhole (\ref{Qtpwh}). Indeed, we can choose $G$ and $\lambda$ such that both agree with each other. We obtain
\ba
&& G(k)= \frac{P(k)^2-Q(k)^2}{P(k)}=\left\{
\begin{array}{ll}
 & \frac{\Gamma(1-\nu)}{\Gamma(1+\nu)}\left(\frac{k}{2}\right)^{2\nu}\ \ \  (kz_0\gg 1)\\
 &\frac{d^2-4\nu^2}{(d+2\nu)d}\cdot \frac{1}{z_0^{2\nu}}\ \ \   (kz_0\ll 1).
\end{array}
\right.\label{GK}\no
&& \lambda(k)= \frac{Q(k)}{P(k)^2-Q(k)^2}=\left\{
\begin{array}{ll}
 &\frac{2\sin\pi\nu\Gamma(1+\nu)}{\Gamma(1-\nu)}\left(\frac{k}{2}\right)^{-2\nu}e^{-2kz_0}\ \ \  ((kz_0\gg 1)\\
&\frac{2(d+2\nu)\nu}{d^2-4\nu^2}\cdot z_0^{2\nu}\ \ \   (kz_0\ll 1).
\end{array}
\right.\label{LK}
\ea
The behavior of $G(k)$ shows that the IR region of the original AdS should be deformed so that two point functions in the traversable wormhole model are reproduced.
This means that we start with a non-conformal deformation of the CFT.
The UV behavior of $G(k)$, which is identical to  (\ref{PoinG}), fits nicely with the expectation that it is asymptotically AdS.
The result (\ref{LK}) shows that the double trace interaction is suppressed for $kz_0\gg 1$.

If we continue our setup to the Lorentzian theory, the result (\ref{LK}) for $kz_0\gg 1$ leads to the following behavior of the double trace interaction:
\be
\lambda(t_1,x_2,t_2,x_2)\propto \frac{1}{\left(-(t_1-t_2)^2+(x_1-x_2)^2+4z_0^2\right)^{d-2\nu-\frac{1}{2}}}.
\label{interx}
\ee
Note that the coefficient of the interaction gets divergent at $-(t_1-t_2)^2+(x_1-x_2)^2+4z_0^2=0$ where the two point function $\la \mO_1 \mO_2\lb$ gets divergent as in (\ref{shortq}) and where the two points are connected by a null geodesic in the dual wormhole spacetime. This divergence is only visible in the Lorentzian signature. 

It is useful to revisit the calculations (\ref{DTbet}) and rewrite it in the form:
\ba
&& J^{(1)}+\left(\frac{1}{G}-\lambda\right)\beta^{(1)}=J^{(2)}+\left(\frac{1}{G}-\lambda\right)\beta^{(2)},\no
&& J^{(1)}+\left(\frac{1}{G}+\lambda\right)\beta^{(1)}+J^{(2)}+\left(\frac{1}{G}+\lambda\right)\beta^{(2)}=0.  \label{deformb}
\ea
We can view this as the gluing conditions (\ref{gluingcnfda}) and (\ref{gluingcnfdb}) in the traversable AdS wormhole.
We write the scalar field solutions in the left and right half of the wormhole as $\Phi(z)=J^{(1)}\eta_{-}(z)+\beta^{(1)}\eta_{+}(z)$ and $\Psi(w)=J^{(2)}\eta_{-}(w)+\beta^{(2)}\eta_{+}(w)$, where $\eta_{\pm}$ are the solutions on the equation of motion of the scalar field in the wormhole geometry such that $\eta_{\pm}(z)\simeq z^{\frac{d}{2}\pm\nu}$ in the boundary limit $z\to 0$. Then we can identify
\ba
&& \frac{1}{G}=-\frac{1}{2}\left[\frac{\eta_{+}(z)}{\eta_{-}(z)}+\frac{\de_z\eta_{+}(z)}{\de_z\eta_{-}(z)}\right]_{z=z_0},
\ \ \ \  \lambda=\frac{1}{2}\left[\frac{\eta_{+}(z)}{\eta_{-}(z)}-\frac{\de_z\eta_{+}(z)}{\de_z\eta_{-}(z)}\right]_{z=z_0},\label{bcwhq}
\ea
where $z=z_0$ is the gluing point. For the pure Poincar\'e AdS, we have 
\ba
\eta^P_\pm(z)=\mp \Gamma(1\pm \nu)\left(\frac{k}{2}\right)^{\pm\nu}z^{\frac{d}{2}}I_{\mp \nu}(kz),
\ea
and this leads to the two point functions (\ref{LK}) via (\ref{bcwhq}). 

Since we have  $(I_+(kz)-I_-(kz))/I_+(kz)\simeq O(e^{-2kz})$ in the limit $kz\to \infty$, the double trace interaction $\lambda$ is indeed exponentially suppressed $\lambda(k)=O(e^{-2kz_0})$ in the high momentum limit as in (\ref{LK}). Moreover, we can apply (\ref{bcwhq}) to more general traversable wormholes which are asymptotically AdS. If we assume that when $z<z_1$ (we have $z_1<z_0$ in general),  we can approximate the geometry by the pure AdS, then it is straightforward to see that the double trace interaction $\lambda$ is suppressed as $\lambda(k)=O(e^{-2kz_1})$ at high momentum. Thus we can rewrite the boundary condition in the double trace deformation into the gluing condition of traversable wormhole.

Before we go on, we would like to mention the limitation of our argument here, which is qualitative and heuristic. This is because we cannot apply the above results to the two point functions for other operators behave under the deformation (\ref{doubleab}). To analyze this we need to work out back-reactions due to the quantum effects. In the case of an eternal AdS black hole, the deformation of boundary condition like (\ref{bcdefz}) leads to the negative Casimir energy and this modifies the bulk metric as shown in \cite{Gao:2016bin}. Though we expect a similar effect in our model, for our purpose, we need to start with two separated gravitational theories on the AdS and cannot apply the argument of perturbation of non-traversable wormhole into traversable one, done in e.g.\cite{Gao:2016bin}.

Nevertheless, at least for the operator $\mO$ in (\ref{doubleab}), the geometry looks like a traversable wormhole. In principle, we can deform the CFTs by all operators as in (\ref{DTla}) so that the wormhole gets traversable for any operators. It is still possible that the metrics probed by each operator are different. This implies that the gravity theory around  $z\sim z_0$ gets highly quantum or non-local, which may not be described by classical gravity. However, near AdS boundary regions can be described by classical gravity as the effect of double trace interactions is reduced as we have seen in the above. It is exciting that this provide a novel model of quantum gravity, which deserves future studies. 

One more helpful setup to study the traversable wormhole is the double trance deformation by the energy stress tensors $\sim \int d^dx T^{(1)}(x)_{ab}T^{(2)ab}(x)$. This might be regarded as an extension  of $\TTbar$ deformation for the doubled CFT \cite{Zamolodchikov:2004ce,McGough:2016lol}, where the $\TTbar$ interaction bridges the two CFTs. The advantage of this approach is that the interaction leads to the modification of the gravity background at the classical level. Since the metric coupled to matter fields universally, we expect that this $\TTbar$ double trace interaction can give a more universal argument. Indeed, the analysis of the deformation of boundary condition (\ref{bcdefz}), two point functions (\ref{tpfdtq}) and their agreement with the calculations in AdS wormholes can be done equally for the double trace deformation for energy stress tensors. In section \ref{sec:TTbar}, we will present more detailed analysis of this deformation from different viewpoints.

\subsection{Toy CFT model: two coupled free CFTs}\label{sec:CoupledCFT}

To see basic properties in the double trace deformed CFTs, here we analyze a pair of free scalar CFTs as a toy example. Consider two free massless scalar fields $\phi_1$ and $\phi_2$ in two dimensions, coupled via an exactly marginal interactions (we choose  $-1<\lambda<1$ ):
\ba
S=-\int dtd\sigma\frac{1}{2\s{1-\lambda^2}}\left[\de_a\phi_1\de^a\phi_1+\de_a\phi_2\de^a\phi_2
+2\lambda\de_a\phi_1\de^a\phi_2\right].\label{intCFTs}
\ea
We compactify the spacial coordinate $\sigma$ as $\sigma\sim \sigma+2\pi$.
The canonical momenta read
\ba
&&\Pi_1=\frac{\delta S}{\delta \dot{\phi_1}}=\frac{1}{\s{1-\lambda^2}}(\dot{\phi_1}+\lambda\dot{\phi_2}),\no
&& \Pi_2=\frac{\delta S}{\delta \dot{\phi_2}}=\frac{1}{\s{1-\lambda^2}}(\dot{\phi_2}+\lambda\dot{\phi_1}).
\ea

The Hamiltonian is computed as
\ba
H=\frac{1}{2\s{1-\lambda^2}}\int^{2\pi}_0 d\sigma\left[\Pi_1^2+\Pi_2^2+(\de_x\phi_1)^2+(\de_x\phi_2)^2+2\lambda(\de_x\phi_1\de_x\phi_2-\Pi_1\Pi_2)\right].\ \ \ \label{hamcftt}
\ea
As we will show soon, this looks identical to (\ref{QQQP}) after mode expansions in $x$ direction.

\subsubsection{Canonical quantization}
The canonical commutation relation at the same time is
\begin{align}
  [\phi^{(i)}(t=0,\sigma), \Pi^{(j)}(t=0,\sigma')] = i \delta^{(2)}(\sigma-\sigma')\delta_{ij}
\end{align}
If we expand
\begin{align}
  \phi^{(i)}(t=0,\sigma) = \sum_{n=-\infty}^{\infty} \phi^{(i)}_n e^{in\sigma}, \quad
  \Pi^{(i)}(t=0,\sigma) = \sum_{n=-\infty}^{\infty} \Pi^{(j)}_n e^{in\sigma},
\end{align}
the commutation relation becomes
\begin{align}
  [\phi_n^{(i)}, \Pi_m^{(j)}] = \frac{i}{2\pi}\delta_{n+m,0}\delta_{i,j}
\end{align}
Then the Hamiltonian is expressed as  
\begin{align}
H=  \frac{\pi}{\sqrt{1-\lambda^2}}\sum_{n=-\infty}^{\infty}\qty[\Pi^{(1)}_n \Pi^{(1)}_{-n} + \Pi^{(2)}_{n} \Pi^{(2)}_{-n} 
  + n^2 \phi^{(1)}_{n} \phi^{(1)}_{-n} + n^2 \phi^{(2)}_{n} \phi^{(2)}_{-n} 
  + 2\lambda(n^2 \phi^{(1)}_{n} \phi^{(2)}_{-n} - \Pi^{(1)}_{n} \Pi^{(2)}_{-n})].\nonumber
\end{align}
We perform the following basis change $(n>0,\ i,j=1,2)$:
\be
  \phi^{(i)}_{n} = \frac{x^{(i)}_n + i y^{(i)}_n}{\sqrt{4\pi}}, \quad
  \Pi^{(i)}_{n} = \frac{p^{(i)}_n + i q^{(i)}_n}{\sqrt{4\pi}}, \quad
  \phi^{(i)}_{-n} (= {\phi^{(i)}_{n}}^\dagger) = \frac{x^{(i)}_n - i y^{(i)}_n}{\sqrt{4\pi}}, \quad
  \Pi^{(i)}_{-n} (= {\Pi^{(i)}_{n}}^\dagger) = \frac{p^{(i)}_n - i q^{(i)}_n}{\sqrt{4\pi}},\nonumber
\ee
which leads to the familiar commutation relations
\begin{align}
  [x_n^{(i)}, p_m^{(j)}] = i \delta_{i,j}\delta_{n,m}, \ \ \ [y_n^{(i)}, q_m^{(j)}] = i\delta_{i,j}\delta_{n,m}, \cdots
\end{align}
Now the Hamiltonian is rewritten in the form of coupled harmonic oscillators:
\begin{align}
  H = \frac{1}{\sqrt{1-\lambda^2}}\sum_{n=1}^{\infty} & \left[\frac{1}{2}({p^{(1)}_n}^2 + {q^{(1)}_n}^2 + {p^{(2)}_n}^2 + {q^{(2)}_n}^2) 
  + \frac{n^2}{2} ({x^{(1)}_n}^2 + {y^{(1)}_n}^2 + {x^{(2)}_n}^2 + {y^{(2)}_n}^2) \right. \nonumber\\
  &\left.+ \lambda n^2 (x^{(1)}_{n} x^{(2)}_{n} + y^{(1)}_{n} y^{(2)}_{n}) - \lambda (p^{(1)}_{n} p^{(2)}_{n} + q^{(1)}_{n} q^{(2)}_{n})\right] +H_0+(\text{const.}),\nonumber
\end{align}
where the zero mode part $H_0$ is given by $H_0=\frac{\pi}{\s{1-\lambda^2}}\left[(\Pi^{(1)}_0)^2+(\Pi^{(2)}_0)^2-2\lambda
\Pi^{(1)}_0\Pi^{(2)}_0\right]$.

We introduce the creation and annihilation operators:
\begin{align}
  a^{(i)}_n = \frac{n x^{(i)}_n + i p^{(i)}_n}{\sqrt{2n}}, \quad
  {a^{(i)}_n}^\dagger =  \frac{n x^{(i)}_n - i p^{(i)}_n}{\sqrt{2n}}, \quad
  b^{(i)}_n = \frac{n y^{(i)}_n + i q^{(i)}_n}{\sqrt{2n}}, \quad
  {b^{(i)}_n}^\dagger =  \frac{n y^{(i)}_n - i q^{(i)}_n}{\sqrt{2n}},\nonumber
\end{align}
which satisfy the commutation relations
\begin{align}
  [a^{(i)}_n, {a^{(j)}_m}^\dagger] = \delta_{n,m}\delta_{i,j},\ \ \   [b^{(i)}_n, {b^{(j)}_m}^\dagger] = \delta_{n,m}\delta_{i,j}.
\end{align}
Finally the Hamiltonian become
\begin{align}
  H &= H_0+\frac{1}{\sqrt{1-\lambda^2}}\sum_{n=1}^{\infty}  
  n\left[{a^{(1)}_n}^\dagger a^{(1)}_n + {b^{(1)}_n}^\dagger b^{(1)}_n + {a^{(2)}_n}^\dagger a^{(2)}_n + {b^{(2)}_n}^\dagger b^{(2)}_n  \right] \nonumber\\
  &\ \ + \lambda\sum_{n=1}^{M}n\left[ \qty(a^{(1)}_{n} a^{(2)}_{n} + {a^{(1)}_{n}}^\dagger {a^{(2)}_{n}}^\dagger 
  + b^{(1)}_{n} b^{(2)}_{n} +  {b^{(1)}_{n}}^\dagger {b^{(2)}_{n}}^\dagger)\right] +H_0+(\text{const.}).
  \label{hamfing}
\end{align}
Indeed, this is a sum of infinitely many copies of the two coupled harmonic oscillators defined by (\ref{Hamosw}). To regulate the interaction between CFT$_{(1)}$ and CFT$_{(2)}$, we imposed the UV cut off in (\ref{hamfing}) such that the summation over $n$ is terminated at $n=M(\sim \frac{1}{z_0})$.

\subsubsection{Pseudo entropy and two point function}
We consider the ground state of this system, denoted by $|\Omega\lb$. This can be found by performing the Bogoliubov transformation for each oscillators $a^{(i)}_n$ and $b^{(i)}_n$, which takes exactly the same form as (\ref{bgla}), by setting $\lambda=\tanh2\theta$. For the zero mode sector, the ground state is simply the zero momentum state $\Pi^{(1)}_0|\Omega\lb=\Pi^{(2)}_0|\Omega\lb=0$.

Now let us calculate the pseudo entropy for $\rho_{AB}$ where $A$ is the total system of CFT$_{(1)}$ at $t=T$, while $B$ is that of CFT$_{(2)}$ at $t=0$. As in the quantum mechanical example of section \ref{tchpb}, $\rho_{AB}$ is in general not hermitian due to the interactions between the two CFTs. The calculation of second Renyi pseudo entropy $S^{(2)}_{AB}$ can be done as in section \ref{tchpb}. Indeed, by summing (\ref{PEHOQ}) over each mode $n$, we obtain:
\begin{align}
  S_{AB}^{(2)} = 2\sum_{n=1}^{M} \log\qty[\frac{1+e^{-2inT}+(1-e^{-2inT})\cosh 4\theta}{2}]. \label{PEWHq}
\end{align}
It is easy to see that starting from $S^{(2)}_{AB}=0$ at $T=0$, it grows rapidly and reaches $S^{(2)}_{AB}=O(M)$ at the time $T=O(1/M)$. This implies that for a more general CFT with the central charge $c$ on the same circle, the pseudo entropy reaches $S_{AB}=O\left(\frac{c}{z_0}\right)$.

In the AdS wormhole geometry, the standard calculation of holographic entanglement entropy implies that the cross section of the wormhole, which is clearly $O\left(\frac{c}{z_0}\right)$, thinking of an AdS$_3$ example of (\ref{WHR}), should be identified with the entanglement entropy between the two CFTs, namely $S_A$. On the other hand, we normally consider $\rho_{AB}$ as a pure state because $A$ in CFT$_{(1)}$ and $B$ in CFT$_{(2)}$ are situated on a common global time slice in the wormhole spacetime. This should lead to $S_{AB}=0$. 

\begin{figure}[ttt]
		\centering
		\includegraphics[width=10cm]{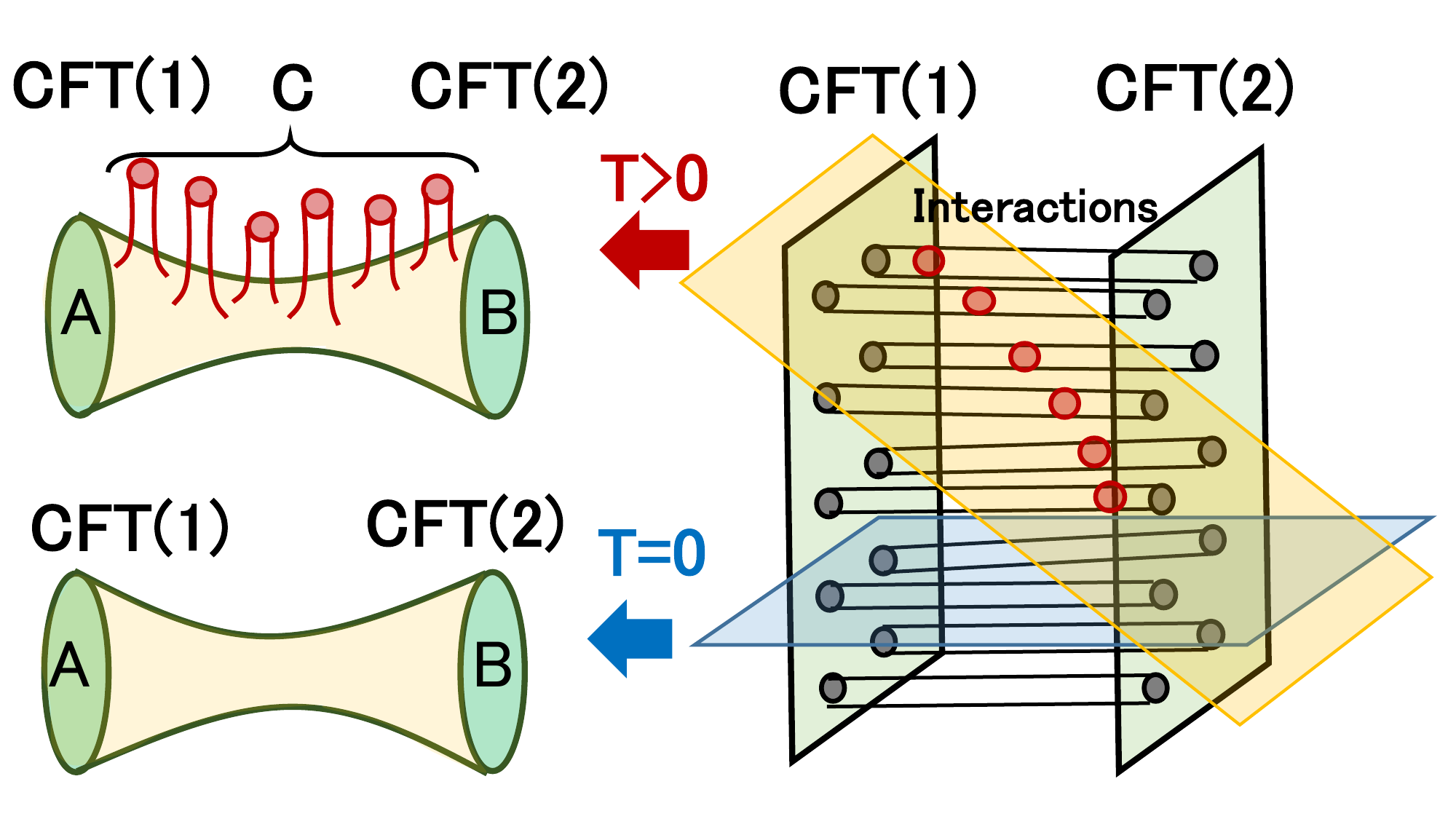}
		\caption{A sketch of gravity dual of a double trace deformation of doubled CFTs. We consider the time slice $A$
       at $t=T$ in CFT$_{(1)}$ and $B$ at $t=0$ in CFT$_{(2)}$ as shown in the right panel. When $T=0$, in the dual gravity (blue plane) we expect a macroscopic wormhole, sketched in the lower left panel. When $T>0$, the bulk time slice (orange plane) which connects $A$ and $B$ intersects with microscopic wormholes created by double trace interactions. Thus its gravity dual includes many microscopic wormholes connected to extra region $C$, depicted in the upper left panel. This causes non-vanishing $S_{AB}$. This extra subsystem $C$ is exactly the one is need to add to purify $\rho_{AB}$ for $T>0$.} 
		\label{fig:MWH}
\end{figure}

In our free model,  $S_A$ is simply given by $M$ times
the harmonic oscillator entropy (\ref{EEsc}) and thus 
$S_A=O(M)=O\left(\frac{c}{z_0}\right)$, agreeing with the holographic expectation. However, the pseudo entropy as we computed in (\ref{PEWHq}) turns out to be non-vanishing and is the same order $S_{AB}=O\left(\frac{c}{z_0}\right)$. This strongly suggests that the gravity dual of the double trace deformation is highly quantum and cannot simply be described by a classical wormhole geometry. It is still possible that for coarse-grained observables such as correlation functions or $S_A$ we can trust the classical gravity picture. However, we need a full quantum gravity treatment to understand global quantities such as $S_{AB}$. This issue arises in any examples of wormhole induced by double trace deformations, including \cite{Gao:2016bin,Maldacena:2019cbz}. Our CFT result of $S_{AB}$ implies that the double trace interactions produce a lot of microscopic wormholes as depicted in the upper left panel of Fig.\ref{fig:MWH}. Indeed if we try to purify $\rho_{AB}$, then we encounter extra Hilbert space $C$, induced by cutting links created by double trace interactions as in the right panel of Fig.\ref{fig:MWH} and of Fig.\ref{fig:Model}. The sum of cross sections of such wormholes should explain the whole $S_{AB}$. The ground state of a double trace deformed Hamiltonian should have quantum entanglement between the two CFTs as in our free model it is described by the TFD state 
(\ref{gsthdf}). This is expected to generate a macroscopic wormhole. However, in addition, due to the time-dependent dynamics due to the interactions between two CFTs, we need to take into account microscopic wormholes, where we consider a state at different times between two CFTs. This is special to  the model B (double trace deformation) and does not happen in the model A (Janus deformation) as we have emphasized before.

Finally we evaluate the two point function. Via the same calculation (\ref{TPFH}) for the coupled harmonic oscillators, we obtain 
\begin{align}
  \ev{x^{(1)}_n(T) x^{(2)}_n(0)} &= \bra{\Omega}\frac{a^{(1)}_n+{a^{(1)}_n}^\dagger}{\sqrt{2n}}\cdot
  e^{-iHnT}\cdot \frac{a^{(2)}_n+{a^{(2)}_n}^\dagger}{\sqrt{2n}}\ket{\Omega} \nonumber\\
  &= -\frac{e^{-inT}}{n}\sinh\theta\cosh\theta. \label{sseqkge}
\end{align}
Similarly we can calculate $\ev{y^{(1)}_n(T) y^{(2)}_n(0)}$ as in (\ref{sseqkge}) and obtain:
\ba
\ev{\phi^{(1)}(T,\sigma)\phi^{(2)}(0,0)}
&=&-\frac{\sinh 2\theta}{2\pi}\sum_{n=1}^M \frac{1}{n}\cos\left(n\sigma\right)e^{-inT}.
\ea
In the limit $M\to\infty$, which means that the interaction between the two CFTs is not suppressed in the UV, it becomes
\ba
\ev{\phi^{(1)}(T,\sigma)\phi^{(2)}(0,0)}=
\frac{\sinh 2\theta}{4\pi}\cdot \log(1-e^{-i(T-\sigma)})
(1-e^{-i(T+\sigma)}).
\ea
This develops the divergence at $T\pm \sigma=2\pi m$, where $m$ is an arbitrary integer. This is the light cone singularity in a standard two dimensional CFT with the compactification $\sigma\sim \sigma+2\pi$. In this $M\to\infty$ limit, the two CFTs are interacting at any scale and can be regarded as a single CFT. A signal can propagate from one of the CFTs to the other immediately. At finite $M$, this interaction is limited to the energy scale up tp $O(M)$ and the sharp propagation is missing. We cannot find the singularity like (\ref{shortq}), which corresponds to the choice (\ref{LK}) of the double trace deformation. This is not surprising partly because we did not introduce the cut off for the energy $k^0$ and also because we analyzed the free CFT for which we do not expect a bulk locality in its gravity dual.

\section{Non-local $\TTbar$ deformation and AdS wormhole}\label{sec:TTbar}

In the previous section, we considered the double trace deformations by a primary operator. As discussed in \cite{Gao:2016bin}, they give a traversable wormhole in the bulk but at the 1-loop order. Though in principle we can consider the double trace deformation by $O(N^2)=O(G_N^{-1})$ operators including stringy operators to have a macroscopic wormhole, this analysis gets too complicated.
Instead, here we try to create a wormhole at classical level $O(G_N^{-1})$ by considering a certain class of double trace deformation by the boundary stress tensors in two dimensional CFTs. One well-known examples of a double trace deformation by stress tensors is the $\TTbar$ deformation:
\begin{equation}\label{eq:TTbar}
  \frac{\partial}{\partial \mu} S_{[\mu]} =  \int d^2 x \sqrt{\gamma}\qty[\qty(\Tmu)_{ab}\qty(\Tmu)^{ab}-\qty(\qty(\Tmu)_a^a)^2],
\end{equation}
where $\Tmu_{ab}$ is a stress tensor of the theory $S_{[\mu]}$,
 \begin{equation}
     \delta S_{[\mu]} = -\frac{1}{2}\int d^2 \sqrt{\gamma}\delta\gamma_{ab}\Tmu^{ab}.
 \end{equation}

The $\TTbar$ deformation is well studied in the two dimensional CFTs \cite{Zamolodchikov:2004ce,Smirnov:2016lqw,Cavaglia:2016oda}, including its relations to two dimensional gravity \cite{Conti:2018tca,Gorbenko:2018oov} and its holography \cite{McGough:2016lol,Guica:2019nzm,Kawamoto:2023wzj} (See for reviews and lectures \cite{Jiang:2019epa,Monica_Guica_TTbar,Luis_Apolo_TTbar}). Since the stress tensor is a quasi primary operator and shows $\LandauO{c}=\LandauO{1/\GN}$ effect in their correlation functions, we expect the deformations gives a huge back reaction. However, the usual $\TTbar$ deformation is a deformation by the irrelevant operator and thus their gravitational effects occurs around the conformal boundary, \ie UV region of the boundary theories. Indeed, it is discussed that the holographic dual of $\TTbar$ deformed CFT is considered to be a finite cut-off theory \cite{McGough:2016lol,Guica:2019nzm}, which will be quickly reviewed in section \ref{reviewTTb}. 

In this section, we consider a family of generalizations of $\TTbar$ deformation to the doubled CFTs. One may first come up with the following $T_1T_2$ deformation:
\begin{equation}
    \frac{\partial}{\partial \mu} S_{[\mu]} = \frac{1}{2}\int d^2 x \qty({T_{[\mu]}^{(1)}}_{ab}{T_{[\mu]}^{(2)}}^{ab}-{T_{[\mu]}}^{(1)}T_{[\mu]}^{(2)}),
\end{equation}
where $T^{(1)}_{ab}, T^{(2)}_{ab}$ denote their stress tensors in two holographic CFTs, respectively. The deformations of the energy spectrum are discussed in \cite{Ferko:2022dpg}. This deformations will glue the two AdS boundaries \footnote{In \cite{Bzowski:2020umc}, similar deformation and its gravity dual is discussed. They put the mixed boundary conditions for the metric as an assumption and find the one solutions of the Einstein equation. However, these solutions are different from our expectation and we discuss other gravity dual in which two theories are related by the asymptotic boundaries. }. \par 

 Instead, we would like to consider a non-local generalization of $T_1T_2$ deformations to make that our deformations affect only the bulk interior. We consider following $\TTbar$ deformation
\begin{equation}
  \frac{\partial}{\partial \mu} S_{[\mu]}  =\frac{1}{2}\int d^2 x\sqrt{\gamma} \ep_{ac}\ep_{bd}(\Tmu^{(1)})^{ab}F(-\Box_\gamma)(\Tmu^{(2)})^{cd},
\end{equation}
where $\ep_{ab}$ is a Levi-Civita tensor for two dimensional metric $\gamma_{ab}$ and $F(x)$ is an analytic function such that $\lim_{x\to \infty}F(x)=0$. Before discuss the property of this deformation and its gravity dual we give a quick review of usual $\TTbar$ deformation (\ref{eq:TTbar}).

\subsection{Short review of usual $\TTbar$ deformation}\label{reviewTTb}
Here we review usual $\TTbar$ deformation based on \cite{Conti:2018tca,Guica:2019nzm,Cardy:2018sdv,Hirano:2020nwq}. We consider the one parameter theories $S_{[\mu]}$ with flow \eqref{eq:TTbar}.
In terms of the path-integral, this can be written as 
 \begin{equation}
     \begin{split}
         \int [dX]e^{-S_{[\mu+\Delta \mu]}[X;\gamma_{[\mu+\Delta\mu]}]}= \int [dX] e^{-S_{[\mu]}[X;\gamma_{[\mu]}]}e^{-\frac{\Delta\mu}{2}S_{T\overline{T}}}.
     \end{split}
 \end{equation}
 Then, by doing the Hubbard-Stratonovich transformation \cite{Cardy:2018sdv}, 
 \begin{equation}
     \begin{split}
         e^{\frac{\Delta\mu}{2}\int d^2 x \sqrt{\gam} \ep_{ac}\ep_{bd}\Tmu^{ab}\Tmu^{cd}} &\propto \int [dh] e^{+\int d^2x \sqrt{\gam} \frac{1}{2} \Tmu^{ab}h_{ab}-\frac{1}{8\Delta\mu}\int d^2 x \sqrt{\gam}\ep_{[\mu]}^{ac}\ep_{[\mu]}^{bd}h_{ab}h_{cd}}.
     \end{split}
 \end{equation}
 Since we are thinking $\Delta \mu$ is small, we are using the saddle point approximation for $h$-integral. The saddle point $h^*_{ab}$ is given by
 \begin{equation}\label{eq:SAP_sol}
     \begin{split}
         \Tmu^{ab}&=-\frac{1}{2\Delta \mu}\ep_{[\mu]}^{ac}\ep_{[\mu]}^{bd}h^*_{cd},\\
         h^*_{ab}&=-2\Delta \mu (\Hat{T}_{[\mu]})_{ab}\equiv-2\Delta \mu \qty(\Tmu_{ab}-(\gam)_{ab}\Tmu).
     \end{split}
 \end{equation}
 By substituting the saddle point in the exponent we see the original $\TTbar$ action. Also we are thinking that $h_{ab}=\LandauO{\Delta\mu}$ and 
 \begin{equation}
     \begin{split}
         e^{-W_{[\mu+\Delta\mu]}[\gamma_{[\mu+\Delta\mu]}]}
        &\approx e^{-W_{[\mu]}[\gamma_{[\mu]}+h^*]-\frac{1}{8\Delta\mu}\int d^2 x \sqrt{\gam}\ep_{[\mu]}^{ac}\ep_{[\mu]}^{bd}(h^*)_{ab}(h^*)_{cd}+\LandauO{(\Delta\mu)^2}}.
     \end{split}
 \end{equation}
We may read off the flow of the metric as 
 \begin{equation}\label{eq:flow_eq_random}
 \begin{split}
     (\gamma_{[\mu+\Delta\mu]})_{ab}&= (\gam)_{ab}+(h^*)_{ab}\; 
     \Leftrightarrow\partial_{\mu}(\gam)_{ab}=-2\qty(\Hat{T}_{[\mu]})_{ab}.
 \end{split}
 \end{equation}
 This matches the results in \cite{Guica:2019nzm}. To determine the flow equation for $\Tmu$ we need to compare the effective action.
 \begin{equation}\label{eq:flow_random_eff_action1}
 \begin{split}
 W_{[\mu+\Delta\mu]}[\gamma_{[\mu+\Delta\mu]}] &= W_{[\mu]}[\gamma_{[\mu]}]-\frac{1}{2}\int d^2 x \sqrt{\gam}(T_{[\mu]})^{ab}h^*_{ab}+\frac{1}{8\Delta\mu}\int d^2 x \sqrt{\gam}\ep_{[\mu]}^{ac}\ep_{[\mu]}^{bd}(h^*)_{ab}(h^*)_{cd}\\&=
  W_{[\mu]}[\gam]+\frac{\Delta\mu}{2}\int d^2 x \sqrt{\gam}\OTT.
 \end{split}
 \end{equation}
 Then, we consider the deformation of $\gam$, $\gam \to \gam +\delta\gamma$ which does not change the flow equation \eqref{eq:flow_eq_random}, \ie
 \begin{equation}
     \partial_\mu (\delta\gamma_{ab}) = -2 \delta (\Hat{T}_{[\mu]})_{ab}=0.
 \end{equation}
 Then, we take the variation of \eqref{eq:flow_random_eff_action1} in this direction, 
 \begin{equation}
     \begin{split}
         \delta\qty(W_{[\mu+\Delta\mu]}-W_{[\mu]})&= -\frac{\Delta\mu}{2}\delta\qty(\int d^2 x \sqrt{\gam}\ep_{[\mu]}^{ac}\ep_{[\mu]}^{bd}\qty(\Hat{T}_{[\mu]})_{ab}\qty(\Hat{T}_{[\mu]})_{cd}).
     \end{split}
 \end{equation}
 After small algebra we find the flow equations for the stress tensors 
\begin{equation}
    \begin{split}
        \partial_\mu\qty(\sqrt{\gam}(\Tmu)_{ab})
        &=\sqrt{\gam}\qty(\frac{1}{2}\gam_{ab}\qty(\Tmu^2-\Tmu^{ab}\Tmu_{ab})+2\qty(\Tmu_{ab}\Tmu-\Tmu_{ac}\Tmu^c_b)).\nonumber
    \end{split}
\end{equation}
  From these, we can read off the following equations,
\begin{equation}\label{eq:flow_equation}
    \begin{split}
        \partial_{\mu}\gam_{ab}&=-2 (\Hat{T}_{[\mu]})_{ab},\;
        \partial_{\mu} (\Hat{T}_{[\mu]})_{ab} = -(\Hat{T}_{[\mu]})_{ac}(\Hat{T}_{[\mu]})^{c}_b,\;
        \partial_{\mu}\qty((\Hat{T}_{[\mu]})_{ac}(\Hat{T}_{[\mu]})^{c}_b)=0.
    \end{split}
\end{equation}
By combining the equations, we find 
\begin{equation}\label{eq:TTbar_three_deriv}
    \partial_\mu^3 \gam_{ab} =0.
\end{equation}
This means the solutions of the flow equations are given by the quadratic order in $\mu$,
\begin{equation}
    \gam_{ab}= \gaz_{ab}- 2\mu ({\Hat{T}_{[0]}})_{ab}+ \mu^2 ({\Hat{T}_{[0]}})_{ac}({\Hat{T}_{[0]}})^{c}_b.
\end{equation}
Now we consider the gravity duals. To this end let us consider the Fefferman-Graham coordinates of the bulk metric,
\begin{equation}
\begin{split}
    ds^2 &= dr^2 + e^{2r}\gamma_{ab}(r,x^a)dx^a dx^b,\\
    \gamma_{ab}(r,x^a) &= \gz_{ab}(x^a)+e^{-2r}\gt_{ab}(x^a)+e^{-4r}\gf_{ab}(x^a)+\cdots. 
\end{split}
\end{equation}
If the bulk gravity is the three dimensional Einstein theory, by solving the Einstein equation order by order for $e^{-2r}$, we find 
\begin{equation}
      \gf = \frac{1}{4} \gt (\gz)^{-1} \gt,
\end{equation}
and the expansion ends at the order of $\LandauO{e^{-4r}}$ \cite{Skenderis:1999nb}.
As in the standard prescription in the multi trace deformation in AdS/CFT \cite{Witten:2001ua}, by using the relation for the un-deformed theory, we obtain the holographic dictionary for the $\TTbar$ deformed theory and its gravity dual. To this end, notice that
\begin{equation}
  \gz_{ab}= \gaz_{ab},\;  \gt_{ab}= 8\pi \GN 
  \qty(\Hat{T}_{[0]})_{ab}.
\end{equation}
Then, we find that 
\begin{equation}\label{eq:finite_cutoff}
    \gam= \gz+ e^{-2r_c}\cdot\gt +e^{-4r_c}\cdot\gf. 
\end{equation}
where
\begin{equation}\label{eq:cut-off_coupling}
    e^{-2r_c} = - \frac{\mu}{4\pi \GN}.
\end{equation}
This boundary conditions for a given source $\gam$ is same as the one we assign the Dirichlet boundary conditions on the finite cut-off surface $\rho=\rho_c$. Thus, we see that the finite cut-off proposal: the $\TTbar$ deformed holographic CFTs with negative deformed parameter $\mu$ are dual to the Einstein gravity with a finite cut-off boundary at $r=r_c$. \par 

Also we can see the $\TTbar$ deformation appears as a finite cut-off AdS/CFT at least perturbatively. There is a systematic method for holographic renormalization with only using the induced field on the cut-off surface \cite{Papadimitriou:2004ap,Papadimitriou:2004rz,Papadimitriou:2016yit}. In this method, we take the ADM decomposition with radial time $r$,
\begin{eqnarray}
    ds^2= dr^2 + h_{ab}(r,x^a)dx^a dx^b.
\end{eqnarray}
Here we set the lapse to be one and shift vectors to be zero. We denote the constant $r$ surface by $\Sigma_r$. The induced metric on $\Sigma_r$,  $h_{ab}$, is related to the Fefferman-Graham coordinate with $h_{ab}=e^{2r}\gamma_{ab}$.
. For the asymptotic AdS geometry it is useful to consider the following dilation operator,
  \begin{eqnarray}
      \delta_D := \int_{\Sigma_r} d^2 x \; 2 h_{ab}\frac{\delta}{\delta h_{ab}}.
      \end{eqnarray}
    This operator is basically equivalent to the $r$ derivative for the large $r$. In this note, we consider the pure gravity case. The task we do is solving the Hamilton-Jacobi equation. To this end let us coincide the Hamilton's principal action $\mathcal{S}$ in the form,
    \begin{eqnarray}
        \mathcal{S}[\gamma] = \int_{\Sigma_r} d^2 x \sqrt{h} \mathcal{L}.
    \end{eqnarray}
    We consider the spectrum decomposition for the dilation operator,
    \begin{eqnarray}
        \mathcal{S} = \sum_{k=0} \mathcal{S}_{2-2k},\; \delta_D \mathcal{S}_{2k} = 2k\; \mathcal{S}_{2k} . 
    \end{eqnarray}
    Also we have the spectrum decomposition for the Lagrangian density
    \begin{eqnarray}
        \mathcal{L} = \sum_{k=0} \mathcal{L}_{2k},
    \end{eqnarray}
    where each $\mathcal{L}_{(\alpha_k)}$ is defined up to the total derivative. From the general discussion in the Hamilton-Jacobi formalism, we have the identity,
    \begin{eqnarray}\label{eq:recursive_identity1}
        \Pi^{ab}\delta_D h_{ab}= 2 \Pi=\frac{1}{\sqrt{h}}\delta_D (\sqrt{h}\mathcal{L})+\qty(\text{total derivative}).
    \end{eqnarray}
    The total derivative term can be absorbed in to the Lagrangian density and we do not care about it. Also we have for each eigenvalue sector,
    \begin{eqnarray}
        \Pi_{(2k)}^{ab}\delta_D h_{ab}= \frac{1}{\sqrt{h}}\delta_D (\sqrt{h}\mathcal{L}_{(2k)}).
    \end{eqnarray}
    In the pure gravity case, it is known that we have the following decomposition
    \begin{equation}
        \begin{split}
            \mathcal{L} &= \mL_{(0)}+ \mtL{2} \log{e^{-2r}}+ \mL_{(2)}+\mL_{(4)}+\cdots,\\
            \delta_D \mL_{(n)}&= -n \mL_{(n)},\; n\neq 2, \; \delta_D \mtL{2} = -2 \mtL{2}\\
            \delta_D \mL_{(2)}&= - 2 \mL_{(2)}-2 \mtL{2}.
        \end{split}
    \end{equation}
    This is because we have the formula $\delta_D \gamma_{ab}= 2 \gamma_{ab}$. The transformation law of $\mL_{(2)}$ shows $\mL_{(2)}$ cannot be the polynomial of the induced fields and thus non-local functional. As mentioned the dilatation operator $\delta_D$ is roughly the $r$ derivative, thus we can read off the following scaling $\mL_{(n)} \sim e^{-nr},\; \sqrt{h}\sim e^{2r}$. Thus we define the counter term, 
    \begin{eqnarray}
        S_{\mathrm{ct}} = -\int_{\Sigma_r}d^2 x \sqrt{h}\qty( \mL_{(0)}+ \mtL{2} \log{e^{-2r}}).
    \end{eqnarray}
    Also we have the renormalized on-shell action,
    \begin{eqnarray}
        S = S_{\mathrm{reg}}+  S_{\mathrm{ct}}  = \int_{\Sigma_r}d^2 x \sqrt{h} \mL_{(2)}. 
    \end{eqnarray}
    \par 
    \noindent
            \textbf{Recursive Relation for $\mL_{(n)},\mtL{2}$ }\par
            
            Let us introduce the method to compute the $\mL_{(n)}\; (n<2),\mtL{2}$. To this end we consider the expansion for the canonical momenta,
            \begin{equation}
                \Pi^{ab}:= \frac{1}{\sqrt{h}}\frac{\delta}{\delta h_{ab}} \int_{\Sigma_r} d^2 x \sqrt{h} \mathcal{L} =\Pie{0}^{ab} + \Pite{2}^{ab} \log{e^{-2r}}+ \Pie{2}^{ab}+\Pie{4}^{ab}+\cdots
            \end{equation}
            Note that $\delta_D \Pie{n}^a_b = -n \Pie{n}^a_b$. By using this relation and the identity \eqref{eq:recursive_identity1} and the Hamilton constraint we can have the recursive algorithm for the $\mL_{(n)},\mtL{2}$. 
            \begin{enumerate}
                \item  We know that from the Einstein action the canonical momentum is defined as\footnote{There is a difference of volume $\sqrt{h}$ from usual definition. } 
                \begin{eqnarray}
                    \Pi^{ab} = -\frac{1}{16\pi \GN}\qty(K h^{ab}-K^{ab}).
                \end{eqnarray}
                    In the Fefferman-Graham coordinates
                     \begin{eqnarray}
                         K_{ab} = \frac{1}{2}\frac{\partial}{\partial r}\gamma_{ab}. 
                     \end{eqnarray}
                     Also we know that the $\gamma_{ab}$ behaves as 
                     \begin{eqnarray}
                         h_{ab}\sim e^{2r} \gz_{ab}.
                     \end{eqnarray}
                     This shows the leading term
                     \begin{eqnarray}
                         \Pie{0}^{ab}= -\frac{1}{16\pi \GN}h^{ab}.
                     \end{eqnarray}
                     \item By using the identity \eqref{eq:recursive_identity1}, we have
                     \begin{equation}
                     \begin{split}
                          &2\qty(\Pie{0}+\Pie{2}+ \Pite{2} \log{e^{-2r}}+ \Pie{2}+\cdots)\\
                          &= 2\mL_{(0)}+0\cdot \mtL{2}\log{e^{-2r}} -2 \mtL{2}+ 0\cdot \mL_{(2)}-2\mL_{(4)}+\cdots 
                     \end{split}
                     \end{equation}
                     By picking up equal weight term we have equations,
                     \begin{equation}\label{eq:conformal_anomaly1}
                     \begin{split}
                         \mL_{(0)} &= \Pie{0}= -\frac{1}{8\pi\GN},
                         \Pite{2}=0,\\
                         \mtL{2}&= -\Pie{2},
                         \mL_{(4)}= -\Pie{4},
                         \cdots
                     \end{split}
                     \end{equation}
                     Note that there is no equation for determine $\mL_{(2)}$.
                     \item To determine $\Pie{n}$ from $\Pie{0}^{ab}$ we should have the equation between $\Pie{n}$ s. This is obtained from the Hamilton constraint,
                     \begin{eqnarray}
                         0=\mathcal{H}=16\pi \GN \qty(\Pi_a^b\Pi_b^a -  \Pi^2) + \frac{1}{16\pi \GN} R[h]+\frac{1}{8\pi \GN}.
                     \end{eqnarray}
                     By expanding this equation in the dilatiton operator, we have 
                     \begin{align}
                         16\pi \GN \qty(\Pie{0}_a^b\Pie{0}_b^a -\Pie{0}^2) + \frac{1}{8\pi \GN}=&0,\\
                         \frac{1}{16\pi \GN}R[h]+32\pi\GN\qty(\Pie{0}^{ab}\Pie{2}_{ab}-\Pie{0}\Pie{2})&=0,\\
            2\Pie{4}_a^b\Pie{0}_b^a+\Pie{2}_a^b\Pie{2}_b^a-2\Pie{4}\Pie{0}-\Pie{2}^2&=0.  
                     \end{align}
                     \item From $\Pie{n}$ we can read off the Lagrangian density $\mathcal{L}_{(n)}$. And from these $\mathcal{L}_{(n)}$, we can determine $\Pie{n}^{ab}$. 
            \end{enumerate}
         
            From the second equation of the Hamiltonian constraint we obtain, 
            \begin{eqnarray}
                \Pie{2} = -\frac{1}{32\pi\GN}R[h],\; \mtL{2} = \frac{1}{32\pi\GN}R[h].
            \end{eqnarray}
            Thus we have a counter term
            \begin{eqnarray}
                S_{\mathrm{ct}}= -\frac{1}{8\pi\GN}\int_{\Sigma_r}d^2 x \sqrt{h}\qty(-1 +\frac{1}{4}R[h]\log{e^{-2r}}).
            \end{eqnarray}
            We will also have 
            \begin{equation}
                \Pite{2}^{ab}  = \frac{1}{64\pi\GN} \qty(R[h]^{ab}-\frac{1}{2}R[h]h^{ab})=0.
            \end{equation}
            This is consistent with the condition. For later use let us also consider the finite correction. In the same manner we find,
            \begin{align}
                \Pie{4} & = -8\pi\GN \qty(\Pie{2}^2-\Pie{2}_a^b\Pie{2}_b^a),\\
                \mL_{(4)} &=  -\Pie{4} = -8\pi\GN \qty(\Pie{2}_a^b\Pie{2}_b^a-\Pie{2}^2).
            \end{align}
 Finally we can write down the effective action for the boundary theory at $r=r_c$. 
               \begin{equation}
                   \begin{split}
                       S_{\mathrm{CFT}}&:= \lim_{r_c\to\infty}\int_{\Sigma_{r_c}} d^2x \sqrt{h} \mL_{(2)},\\
                       S_{\text{finite}} &:= \int_{\Sigma_{r_c}} d^2x \sqrt{h} \qty(\mL_{(2)}+\mL_{(4)}).
                   \end{split}
               \end{equation}
               Then we have two stress tensors,
\begin{equation}
    \begin{split}
        \ev{T^{ab}}_{\gz}^{\mathrm{CFT}}&:= -\frac{2}{\sqrt{\gz}} \frac{\delta S_{\mathrm{CFT}}}{\delta \gz_{ab}} = -2\lim_{r_c\to\infty}e^{2r_c} \Pi^{(2) ab},\\
        \ev{T^{ab}}^{\mathrm{finite}} &:= -e^{2r_c}\frac{2}{\sqrt{h}} \frac{\delta S_{\text{finite}}}{\delta h_{ab}} = -2e^{2r_c} \qty(\Pi^{(2) ab}+\Pi^{(4) ab}).
    \end{split}
\end{equation}
The finite cut-off action can be expanded as  
\begin{equation}
    S_{\text{finite}} = S_{\mathrm{CFT}}[\gz]-2\pi G_N e^{-2r_c}\int d^2 x \sqrt{\gz}\qty(\ev{T^{ab}}_{\gz}^{\mathrm{CFT}}\ev{T_{ab}}_{\gz}^{\mathrm{CFT}}-\qty(\ev{T^a_a}_{\gz}^{\mathrm{CFT}})^2).
\end{equation}
Comparing this to the definition of the $\TTbar$ deformation, we see the relation \eqref{eq:cut-off_coupling} again. Similar results are observed in \cite{Caputa:2020lpa}. We summarized the important points of usual $\TTbar$ deformation,
            \begin{enumerate}
                \item The three times derivatives of the metric is zero,
                \begin{eqnarray}
                    \partial_\mu^3 \gam_{ab}=0.
                \end{eqnarray}
                \item The Einstein equation with Einstein Hilbert action shows that the Fefferman-Graham expansion ends in the order of $\LandauO{e^{-4r}}$ and we can match this expansion and the solutions of the flow equation by which we see the finite cut-off interpretation.
                \item From the holographic renormalization computation shows that the finite cut-off correction  is actually equivalent to the $\TTbar$ deformation.
            \end{enumerate}

\subsection{Flow equations of non-local $\TTbar$ deformation and wormholes}
 Then, we consider the non-local $\TTbar$ deformation,
\begin{equation}
    S_{T\overline{T}}= -\frac{1}{2}\int d^2 x \sqrt{\gam} \ep_{ac}\ep_{bd}T^{ab}F(-\Box_\gamma)T^{cd}.\label{nTTbg}
\end{equation}
The form of $F(x)$ is chosen such that it suppresses the UV contributions. Though $\TTbar$ deformation is an irrelevant deformation, this non-local $\TTbar$ deformation changes only the IR degree of freedom. We postpone the problem of choosing a proper form of $F(x)$ later. We repeat the same procedure as before. Consider the Hubbard-Stratonovich transformation,
\begin{equation}
\begin{split}
    e^{-W_{[\mu+\Delta\mu]}}&=\int [dX]e^{-S_{[\mu]}[X;\gam]}e^{\frac{\Delta\mu}{2}\int d^2 x \sqrt{\gam}\ep_{ab}\ep_{cd}T^{ab}F(-\Box_\gamma)T^{cd}}\\
    &=\int [dX][dh]e^{-S_{[\mu]}[X;\gam]}e^{-\frac{1}{8\Delta\mu}\int d^2 x \sqrt{\gam}\ep^{ab}\ep^{cd}h_{ab}F^{-1}(-\Box_\gamma)h_{cd}+\frac{1}{2}\int d^2 x \sqrt{\gam}T^{ab}h_{ab}}.\nonumber
\end{split}
\end{equation}
The saddle point equation reads
\begin{equation}
    \begin{split}
        -\frac{1}{4\Delta\mu}\ep^{ac}\ep^{bd}F^{-1}(-\Box_\gamma)h^{*}_{cd}+\frac{1}{2}\Hat{T}^{ab}&=0,\\
        h^{*}_{ab}=-2\Delta\mu F(-\Box_\gamma)\hat{T}_{ab}.
    \end{split}
\end{equation}
This gives the flow equation for the metric,
\begin{equation}\label{eq:non_local_TTbar_metric}
    \partial_\mu \gam_{ab}= -2 F(-\Box_\gamma)\qty(\Hat{T}_{[\mu]})_{ab},
\end{equation}
Similarly the flow equation of the effective action is given by
\begin{equation}
    \partial_\mu W_{[\mu]}[\gam]= -\frac{1}{2}\int d^2 x \sqrt{\gam}\ep^{ac}\ep^{bd}(\Hat{T}_{[\mu]})_{ab}F(-\Box_\gamma)(\Hat{T}_{[\mu]})_{cd}.
\end{equation}
Then we consider the variation of the metric $\gam_{ab}$ keeping the flow equation \eqref{eq:non_local_TTbar_metric},
\begin{equation}
    \partial_\mu \delta\gam_{ab}= -2 \delta\qty(F(-\Box_\gamma)\qty(\Hat{T}_{[\mu]})_{ab})=0.
\end{equation}
The right hand side can be written as 
\begin{equation}
    \delta \qty(\Hat{T}_{[\mu]})_{ab}= -F^{-1}(-\Box_\gamma)(\delta F(-\Box_\gamma))\qty(\Hat{T}_{[\mu]})_{ab} = F^{-1}(-\Box_\gamma) F'(-\Box_\gamma)\delta(\Box_\gamma)\qty(\Hat{T}_{[\mu]})_{ab}.\nonumber
\end{equation}
By using this relation and after some tedious algebra we obtain  
\begin{equation}
    \partial_\mu \qty(\Hat{T}_{ab}) = -\frac{1}{2}\qty(\Hat{T}_{ac}\Hat{S}^c_b+\Hat{T}_{bc}\Hat{S}^c_a-3\Hat{T}\Hat{S}_{ab}+3\Hat{T}\Hat{S}_{ab})-N_{ab}
\end{equation}
where $\Hat{S}_{ab}=F(-\Box_\gamma)\Hat{T}_{ab}$ and using the variation formula discussed in the Appendix A of \cite{Biswas:2013cha} we find, 
\begin{equation}
\begin{split}
N^{ab}&:=\Bigg[(\nabla^a \Hat{T}_{cd})(\nabla^b U^{cd}) + (\nabla^b \Hat{T}_{cd})(\nabla^a U^{cd}) - \gamma^{ab}\Big((\Box \Hat{T}_{cd})U^{cd} + (\nabla_e \Hat{T}_{cd})(\nabla^e U^{cd})\Big) \\
&\quad + \Box(\Hat{T}^a_{\phantom{a}c} U^{bc}) + \Box(\Hat{T}^{b}_{\phantom{b}c} U^{ac}) - 2\nabla_e (\Hat{T}^a_{\phantom{a}c} \nabla^e U^{bc}) - 2\nabla_e (\Hat{T}^b_{\phantom{b}c} \nabla^e U^{ac}) \\
&\quad + \nabla_c \nabla^b(\Hat{T}^a_{\phantom{a}d} U^{cd}) + \nabla_c \nabla^a(\Hat{T}^b_{\phantom{b}d} U^{cd}) - 2\nabla_c(\Hat{T}^a_{\phantom{a}d}\nabla^b U^{cd}) - 2\nabla_c(\Hat{T}^b_{\phantom{b}d}\nabla^a U^{cd}) \\
&\quad - \nabla^e\nabla^a(\Hat{T}_{ec}U^{bc}) - \nabla^e\nabla^b(\Hat{T}_{ec}U^{ac}) + 2\nabla^e(\Hat{T}_{ec}\nabla^a U^{bc}) + 2\nabla^e(\Hat{T}_{ec}\nabla^b U^{ac})\Bigg],\nonumber
\end{split}
\end{equation}
where $U_{ab}=F'(-\Box_\gamma)(\Hat{T}_{ab}-\gamma_{ab}\Hat{T})$. In the usual $\TTbar$ deformation, we simply set $S_{ab} = T_{ab}$ and $N_{ab}=0$ and we observe the (magical) cancellation which leads to \eqref{eq:TTbar_three_deriv}. However, in our case of general (non-local) $\TTbar$ deformation (\ref{nTTbg}), we find
 \begin{equation}
     \partial_\m^3 \gam_{ab}\neq 0.
 \end{equation}
\par
\noindent
\textbf{Possible Gravity Dual and Wormhole}
\par 
 From above, we may consider following possibilities for gravity duals of (\ref{nTTbg}):
 \begin{itemize}
     \item We have a finite cut-off holography within Einstein gravity though the finite cut-off surface may not be $r=\const$.

     \item Though we have a finite cut-off holography with $r=\const$, the bulk gravity is no longer the Einstein gravity. The reason for this possibilities is that in non-Einstein gravity, we may have the Fefferman-Graham expansion which is not terminated in the order of $\LandauO{e^{-4r}}$. Then, we can still identify the $\mu$ parameter to the $e^{-r}$.
     \item We do not have any finite cut-off interpretation at all.
 \end{itemize}
The first possibility contradicts with the following argument on symmetries. Suppose we firstly have the un-deformed theory with flat metric $\gz_{ab}=\delta_{ab}$. Then, suppose that the deformed theory is dual to the theory on the cut-off surface $r= \Bar{r}(x^a)$. It is clear that the induced metric on this surface does not have the symmetry of $\delta_{ab}$, namely the two dimensional Poincar\'e symmetry. On the other hand, the non-local $\TTbar$ deformation does not break this symmetry and there is a discrepancy. Also by putting the inhomogeneous cut-off surface in above holographic renormalization group computation, we just observe the $\TTbar$ deformation with inhomogeneous coupling.\par
Thus, we focus on the second possibility. As we willl see, the non-local $\TTbar$ deformation appears as a finite cut-off correction to the following gravitational Hamiltonian:
  \begin{equation}\label{eq:mod_grav}
      \begin{split}
          H &= \int_{\Sigma_r}d^dx \qty(N\mathcal{H}+N_a \mathcal{H}^a),\\
         \frac{1}{\sqrt{h}} \mathcal{H}&=16\pi\GN\qty(\Pi^{a}_bF(-e^{2r}\Box_h)\Pi_a^b-\Pi F(-e^{2r}\Box_h)\Pi)+\frac{1}{16\pi\GN}\qty(R[h]+2),\\ \mathcal{H}^a & = -2D_b \qty(\frac{1}{\sqrt{h}}\Pi^{ab}).
      \end{split}
  \end{equation}
  where $D_a$ is the covariant derivative with respect to $h_{ab}$ and
  \begin{equation}
      F(-e^{2r}\Box_h) = \sum_{k=0}f_k (-e^{2r}\Box_h)^k,   \end{equation}
 The choice of $f_k$ will be discussed later. To discuss the gravity dual of a non-local $\TTbar$ deformation, we need to change the order estimation by including explicit $r$-dependence in the Hamilton constraint. When we include the dependence of $r$-direction in the Hamilton constraint, it is natural to modify the definition of the dilation operator as 
  \begin{equation}
      \delta_D = \int d^2 x \qty(\partial_r + 2h_{ab}\frac{\delta}{\delta h_{ab}}).
  \end{equation}
  Here $\partial_r$ acts only on $r$ which appears explicitly in the Hamilton constraint.
  Then, if we consider the combination of $-e^{2r}\Box_h$, this doesn't change the eigenvalues. Indeed for field $X$ such that $\delta_D X_{ab}= -\lambda X_{ab}$
 \begin{equation}
     \begin{split}
         \delta_D (F(-e^{2r}\Box_h)X_{ab}) &= -\lambda F(-e^{2r}\Box_h)X_{ab},\\
          \delta_D (F(-e^{2r}\Box_h)X) &= -(\lambda+2) F(-e^{2r}\Box_h)X.
     \end{split}
 \end{equation}
Similar to the usual $\TTbar$ deformation we discuss in the previous subsection, we can systematically solve the Hamilton-Jacobi equation order by order. The $1$st order equation is given by 
 \begin{equation}
    0= (16\pi\GN)\qty(\Pie{0}_a^b F(-e^{2r}\Box_h)\Pie{0}^a_b-\Pie{0}F(-e^{2r}\Box_h)\Pie{0})+\frac{1}{8\pi\GN}.
 \end{equation}
 Assuming $\Pie{0}_{ab} = Ph_{ab}$, we obtain
 \begin{equation}
     P= -\frac{1}{\sqrt{f_0}}\frac{1}{16\pi\GN},
 \end{equation}
 where signs are chosen so that the spacetime is asymptotically AdS spacetime.
 The second order equation is given by 
 \begin{equation}
 \begin{split}
   &\quad \Pie{0}_a^b F(-e^{2r}\Box_h)\Pie{2}^a_b+\Pie{2}_a^b F(-e^{2r}\Box_h)\Pie{0}^a_b-\Pie{0}F(-e^{2r}\Box_h)\Pie{2}-\Pie{2}F(-e^{2r}\Box_h)\Pie{0}\\&=-\frac{1}{(16\pi\GN)^2}R[h].\nonumber
 \end{split}
 \end{equation}
 This can be written as 
 \begin{equation}
     \Pie{2} = -\frac{1}{\sqrt{f_0}}\frac{1}{16\pi\GN}Q(-e^{2r}\Box_h)R[h].
 \end{equation}
 where $Q(-e^{2r}\Box_h)$ is the inverse differential operator os $1+\frac{1}{f_0}F(-e^{2r}\Box_h)$ and 
 \begin{equation}
     Q(-e^{2r}\Box_h) = \sum_{k=0}^\infty q_k (-e^{2r}\Box_h)^k,\; q_0 = \frac{1}{2},\;q_1 = -\frac{f_1}{4f_0},\cdots.
 \end{equation}
 Then the second order Lagrangian is topological 
 \begin{equation}
     \int d^2x \sqrt{h} \mtL{2}  =  \frac{1}{32\pi\GN}\frac{1}{\sqrt{f_0}}\int d^2 x \sqrt{h}R[h],
     \end{equation}
    and thus 
    \begin{equation}
        \Tilde{\Pi}^{(2)}_{ab} =0.
    \end{equation}
  The forth order equation reads 
    \begin{equation}
        \begin{split}
            0& = \Pie{0}_a^b F(-e^{2r}\Box_h)\Pie{4}^a_b+\Pie{2}_a^b F(-e^{2r}\Box_h)\Pie{2}^a_b+\Pie{4}_a^b F(-e^{2r}\Box_h)\Pie{0}^a_b\\&-\Pie{0}F(-e^{2r}\Box_h)\Pie{4}-\Pie{2}F(-e^{2r}\Box_h)\Pie{2}-\Pie{4}F(-e^{2r}\Box_h)\Pie{0}.
        \end{split}
    \end{equation}
This is solved as 
    \begin{equation}
        \Pie{4}= -\frac{16\pi\GN}{\sqrt{f_0}}Q(-e^{2r}\Box_h)\qty(\Pie{2}_a^b F(-e^{2r}\Box_h)\Pie{2}_b^a-\Pie{2} F(-e^{2r}\Box_h)\Pie{2}).
    \end{equation}
   Thus the forth order action is computed as 
    \begin{equation}
        \int d^2 x\sqrt{h}\mL_{(4)} = \frac{8\pi\GN}{\sqrt{f_0}}\int d^2 x \sqrt{h}\qty(\Pie{2}_a^b F(-e^{2r}\Box_h)\Pie{2}_b^a-\Pie{2} F(-e^{2r}\Box_h)\Pie{2}).
    \end{equation}
    In terms of the boundary quantities we observe that the effective action of the finite cut-off theory at $r=r_c$ is given by 
    \begin{equation}
    \begin{split}
        S_{\mathrm{finite}} &= \int_{\Sigma_{r_c}} d^2x \sqrt{h} \qty(\mL_{(2)}+\mL_{(4)})\\
        &= S_{\mathrm{CFT}}[\gamma] + \frac{2\pi\GN\; e^{-2r_c}}{\sqrt{f_0}}\int d^2 x\sqrt{\gamma}\ev{T^{ab}}_
    {\gamma}^{\mathrm{CFT}}F(-\Box_{\gamma})\ev{\Hat{T}_{ab}}_{\gamma}^{\mathrm{CFT}}.
    \end{split}
    \end{equation}
    Now we consider the problem of fixing $f_0$. As in the previous section, we would like to regulate the UV degree of freedom in the boundary theories. Also we would like to remove the deformation around the conformal boundary as opposed to the usual $\TTbar$ deformation. Thus, we require $\frac{e^{-2r_c}}{\sqrt{f_0}}F(k^2)\to 0$ for large boundary momentum $k^2$ and large $r_c$. Also we require that the bulk physics only changes in the bulk inside, \ie $r\ll 1$ and the bulk theories around the asymptotic boundary $r\gg 1$ is given by the Einstein gravity. Thus, for a large $r$ we require $F(e^{2r}k_{\mathrm{bulk}}^2)\to 1$. This two conditions are satisfied when we take $f_0 = e^{4r_c}\overline{\mu}^2,\;\overline{\mu}=\LandauO{e^{-0\cdot r_C}}$ and $F(x)\to 1$ for $x\to\infty$, and take $r_c\gg 1$ after the computation. One concrete example for such $F(x)$ is $F(x)= \frac{x/\Lambda^2+e^{4r_c}\overline{\mu}^2}{x/\Lambda^2+1}$ where $\Lambda$ is a UV parameter. With this choice the boundary effective action is now given by 
     \begin{equation}
    \begin{split}
        S_{\mathrm{finite}} 
        &= S_{\mathrm{CFT}}[\gz] - \frac{2\pi\GN\; e^{-2r_c}}{\sqrt{f_0}}\int d^2 x\sqrt{\gz}\ev{T^{ab}}_
    {\gz}^{\mathrm{CFT}}F(-\Box_{\gz})\ev{\Hat{T}_{ab}}_{\gz}^{\mathrm{CFT}}.
    \end{split}
    \end{equation}
    This is nothing but a non-local $\TTbar$ deformation. This means that we obtain the generalized $\TTbar$ deformation including the infinite number derivative as a finite cut off correction of the modified gravity whose bulk action is written as a non-local operators. For large $r_c$ we almost miss the finite cut-off effect and essentially, we observe the deformation changes inside the bulk.Notice that the modified gravitational Hamiltonian \eqref{eq:mod_grav} has explicit $r$-dependence and therefore breaks diffeomorphism invariance involving $r$. From the boundary perspective, the IR fixed point theory no longer possesses conformal symmetry, but instead only Poincaré and scaling invariance. This serves as an interesting loophole in the theorem that global scale invariance in two dimensions implies local conformal invariance under broad conditions \footnote{We thank Yu Nakayama for pointing out this fact.} \cite{Zamolodchikov:1986gt,Polchinski:1987dy,Nakayama:2013is}.\par 
    We may read this bulk non-locality is coming from the effective description of the wormhole contribution. As Coleman discussed that the wormhole affects the low energy or macroscopic physics \cite{Coleman:1988tj,Preskill:1988na,Coleman:1989zu,Hebecker:2018ofv}, if we consider the effective field theory whose scale is larger than the wormhole size, then we have the non-local action,
    \begin{equation}
        \begin{split}
            I_{\eff} &= \frac{1}{2}\int d^{d+1} x \sqrt{g(x)}\int d^{d+1} y \sqrt{g(y)} \sum_{i,j}\Delta_{ij}\Phi_i(x)G(x,y)\Phi_j(y)\\
            &= \sum_{i,j}\int d^{d+1}x \sqrt{g(x)}\Phi_i(x) \frac{\Delta_{ij}}{\Box}\Phi_j(x).
        \end{split}
    \end{equation}
    Here $\Phi_i$ are bulk local operators constructed from the low energy modes and $G(x,y)$ is a Green function related with $\Box$ operator\footnote{In the original Coleman's discussion, they considered the bi-local action with $G(x,y)=\const$. In this sense, our discussion is little bit more general than theirs. }. By thinking conversely, we may think our modified gravitational Hamiltonian \eqref{eq:mod_grav} may appears as an effective description of the some wormholes. The properties of wormholes are reflected to the concrete form of $F(-e^{2r}\Box_h)$. The distance of the wormhole mouths, denoted as $l_{\mathrm{WH}}$, are characterized by the explicit form of $F(x)$. Since $F(e^{2r}k_{\mathrm{bulk}}^2)$ takes non-small value for $e^{2r}k_{\mathrm{bulk}}^2\leq \Lambda$, the bulk probe field with the scale $l_{\mathrm{WH}}\sim k_{\mathrm{bulk}}^{-1}\sim e^{r}\Lambda$ captures the effect of the wormhole. For large $r$,\ie around the asymptotic boundary only the very small momentum captures the wormhole and for $r\to \infty$ only zero momentum can be affected. 
    This is expected from the boundary perspective since the deformed action has support only at the low momenta.  
    The explicit dependence of $r$ in the bulk field may be considered that the wormhole spontaneously breaks the diffeomorphism in the $r$ direction and effective field theory on these wormhole does not have this symmetry. Of course, we may have another mechanism for this non-locality other than the geometric wormhole. 

\subsection{$T_1T_2$ deformation: discussion}\label{subsec:T1T2}
In this section, we come back to our main interest: the possibility to find a CFT dual via a non-local $T_1 T_2$ deformation. Our argument below is quite heuristic and qualitative. Firstly, we start with the local $T_1 T_2$ deformation,
\begin{equation}
    \frac{\partial}{\partial \mu} S_{[\mu]} = \frac{1}{2}\int d^2x \sqrt{\gam}\qty(T^{(1)}_{ab}{T^{(2)}}^{ab}-T^{(1)}T^{(2)}).
\end{equation}
By performing the Hubbard-Stratonovich method, we note that the flow equation is written as (see Appendix \ref{sec:T1T2computation} for derivation)
\begin{equation}\label{eq:T1T2_metric_flow}
    \begin{split}
       \partial_\mu \gam_{ab}^{(1)}&= -2\qty(\Hat{T}^{(2)})_{ab},\\
        \partial_\mu \gam_{ab}^{(2)}&=-2\qty(\Hat{T}^{(1)})_{ab}.
    \end{split}
\end{equation}
To derive the stress tensor flow, we need to notice the difficulty of $T_1T_2$ deformation: after the deformation, we no longer have the two independent metrics dual to the two CFTs. As the interaction breaks a part of diffeomorphism invariance, we have only one stress tensors. To avoid this problem, we focus on the finite temperature case where the CFT$_{(1)}$ and CFT$_{(2)}$ are related by the KMS relation,
\begin{equation}
    {T^{(2)}}^{ab}(t,x) = {T^{(1)}}^{ab}\qty(t+i\frac{\beta}{2},x).
\end{equation}
In this case, we can find the flow equation of the stress tensor 
\begin{equation}
    \partial_{\mu} \Hat{T}^{(1)}_{ab} = -\frac{1}{2} \Hat{T}^{(1)}_{ac} \Hat{T}^{(2)c}_{\phantom{2}b} 
    - \frac{1}{2} \Hat{T}^{(2)}_{ac} \Hat{T}^{(1)c}_{\phantom{1}b} 
    - \frac{3}{2} \Hat{T}^{(1)} \Hat{T}^{(2)}_{ab} 
    + \frac{3}{2} \Hat{T}^{(1)}_{ab} \Hat{T}^{(2)},
\end{equation}
where in the right hand side the everything is written only by the metric  $\gamma^{(1)}_{ab}$. We can also write down the flow equation for ${\Hat{T}^{(2)}}_{ab}$. It is not easy to solve these flow equation. However, if we require $\gamma^{(1)}_{ab}=\gamma^{(2)}_{ab}$ and $\Hat{T}^{(1)}_{ab}=\Hat{T}^{(2)}_{ab}$ at $\mu=0$, then, the flow equations reduce to the one of the usual $\TTbar$ equation and we have the same result as \eqref{eq:finite_cutoff}. In this case the bulk dual will be the usual finite cut-off surface holography which is valid at least at the level of the on-shell action and thermodynamics.  In general cases, we suspect that the two boundaries are related some way around the asymptotic boundaries. One possibilities are gluing two surfaces similar to \cite{Kawamoto:2023wzj, Apolo:2023vnm}. Note that the  paper \cite{Kawamoto:2023wzj} discussed the brane world holography via junction conditions which may be different from the dual of $T_1T_2$ deformation since junction conditions leads the "dynamical" gravity on the brane and we need to assign the Virasoro constraint for the sum of two stress tensors. \par
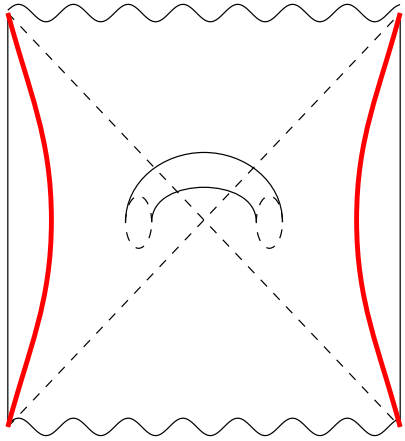
\begin{figure}[ht]
   
   \centering
\resizebox{0.35\textwidth}{!}{%
\begin{circuitikz}
\tikzstyle{every node}=[font=\LARGE]
\draw [short] (8.75,9) -- (8.75,4.25);
\draw [short] (4.25,9) -- (4.25,4.25);
\draw[domain=4.25:8.75,samples=100,smooth] plot (\x,{0.1*sin(10*\x r -10.75 r ) +9});
\draw [dashed] (4.25,4.25) -- (8.75,9);
\draw [dashed] (4.25,9) -- (8.75,4.25);
\draw[domain=4.25:8.75,samples=100,smooth] plot (\x,{0.1*sin(10*\x r -10.75 r ) +4.25});
\draw[red, ultra thick](4.25,9)to [out=-75,in=90](4.75,6.625) to [out = -90, in = 75](4.25,4.25);
\draw[red, ultra thick](8.75,9)to [out=180+75,in=90](8.25,6.625) to [out = -90, in = 180-75](8.75,4.25);
\draw [black, dashed] (5.75,6.6) ellipse (0.15cm and 0.3cm);
\draw [black, dashed] (7.25,6.6) ellipse (0.15cm and 0.3cm);
\draw[black](5.6,6.6)to [out=90,in=180] (6.5,7.4)to [out=0,in=90](7.4,6.6);
\draw[black](5.9,6.6)to [out=90,in=180] (6.5,7)to [out=0,in=90](7.1,6.6);
\end{circuitikz}
}%
\caption{Sketch of possible wormhole corresponding to non-local $T_1T_2$ deformation. The red curves corresponds to the cut-off surface where the boundary theories live for finite $r_c$. }
\label{fig:non-localT1T2}
\end{figure}

 Now we extend the above argument to  the non-local $T_1T_2$ deformations to see the effect in the deep inside of the bulk. That is, we consider the deformation,
 \begin{equation}
  \frac{\partial}{\partial \mu} S_{[\mu]}  =\frac{1}{2}\int d^2 x\sqrt{-\gam}\;\ep_{ab}\ep_{cd}(\Tmu^{(1)})^{ab}F(-\Box_\gamma)(\Tmu^{(2)})^{cd}.
\end{equation}
Though it is not easy to analyze these deformation in an analytic manner, we will find the similar properties to the non-local $\TTbar$ deformations. To discuss the bulk dual, we may consider the trick that we focus on the finite temperature case and identify
\begin{equation}
    T^{(2)}_{ab}(t,x)  = T^{(1)}_{ab}\qty(t+i\frac{\beta}{2},x) = e^{i \frac{\beta}{2}\partial_t}T^{(1)}_{ab}\qty(t,x).
\end{equation}
Then, we may think the differential operation $ e^{i \frac{\beta}{2}\partial_t}$ as an another segment of the non-locality $F(-\Box_\gamma)$. Similar to the non-local case, we may think that the bulk dual has non-local gravitational theory but with additional $e^{i e^{2r}\frac{\beta}{2}\partial_t}$ factors. This factors can be thought to imply wormholes connecting the Rindler wedges as sketched in Fig.\ref{fig:non-localT1T2}. This claim is also supported from our previous analysis of the deformation of boundary condition (\ref{deformb}) in section \ref{sec:deformb} by simply extending the scalar operator to the energy stress tensor. This shows that even if we start from two disconnected AdS geometries, the double trace deformation creates the connection between two AdS geometries in the IR region. If we apply this to an eternal AdS black hole as we discussed here, we indeed expect to have wormhole-like connections are created in the near horizon region. We would like to leave more examinations of this speculative argument for a future problem.

\section{Conclusions and discussions}
\label{sec:conclude}

In this paper, we studied the holographic duality for traversable AdS wormholes. We started with a pair of CFTs, denoted by CFT$_{(1)}$ and CFT$_{(2)}$. They are originally decoupled or are in a thermofield double (TFD) state, which are dual to two disconnected AdS geometries or an eternal AdS black hole, respectively. We argued that we can deform it into a traversable AdS wormhole by adding (i) Janus deformation (model A) or (ii) double trace deformation (model B), between the two CFTs.

To understand basic properties of traversable AdS wormholes, in section \ref{sec:AdSWH}, we examined a simple example by gluing two AdS geometries. We computed holographic two point functions of scalar operator $\mathcal{O}_{i}$ in the CFT$_{(i)}$ for $i=1,2$.
In the UV limit, the Euclidean two point function  $\ev{\mathcal{O}_1(k)\mathcal{O}_2(-k)}$, is exponentially dumped, while $\ev{\mathcal{O}_1(k)\mathcal{O}_1(-k)}$ coincides with that in the CFT vacuum. 
Thus, we understand two CFTs are decoupled in this limit. On the other hand, in the IR region, both $\ev{\mathcal{O}_1(k)\mathcal{O}_2(-k)}$ and $\ev{\mathcal{O}_1(k)\mathcal{O}_2(-k)}$ are non-trivial and becomes the same order. This shows that the two CFTs are strongly interacting at low energy and is due to the presence of wormhole.

Moreover, in the Lorentzian signature, the two point function $\ev{\mathcal{O}_1(t,x)\mathcal{O}_2(t,x')}$ gets divergent exactly when the two points are connected by a null geodesic in the wormhole geometry. Thus this divergence in the CFTs is another manifestation of the presence of traversable wormhole in the gravity dual. A similar property can also be found in our second example of traversable wormhole constructed by gluing the BTZ black hole geometry. 
Through calculations of geodesic lengths and two point functions between left and right boundaries, we confirmed that they are causally connected and that they have the same divergences of the two point functions. 

Another probe which is useful to study the wormhole holography, is the holographic entanglement entropy, which can be found from the extremal surface areas in the AdS wormhole geometry. We compute this in the above wormhole and found rich structures of phase transitions. However, we have to be careful here because due to the Janus or double trace deformations in the dual CFTs, we expect that the density matrix for the total system is not hermitian for generic global time slices in the AdS wormhole as emphasized in section \ref{sec:twosetups}. We can still define so called pseudo entropy by using the expression of von-Neumann entropy and argue that it is computed from the extremal surface area. We noted that the pseudo entropy  gets complex valued when the time difference between the two CFTs gets larger than a critical time.
This critical time is exactly when the extremal surface between the two boundaries becomes light-like and this is characteristic to the traversable wormhole.

As a new class of CFT duals of traversable wormholes, we raised Janus deformations of TFD states in the paired CFTs. The Janus deformation introduces an exactly marginal perturbation, proportional to a parameter $\gamma$ of the two CFTs in an asymmetric way. We pointed out that AdS$_3$ Janus solutions at imaginary values of $\gamma$ become traversable wormholes in section \ref{sec:Janus}. We see that this deformation has a real spacetime metric with a traversable wormhole but with a complex valued dilaton.
We calculated the geodesic length between two points on different boundaries and confirmed that it becomes vanishing in finite time.  The time evolution of the length is quantitatively similar to ones in the glued BTZ case. When $\gamma$ is imaginary, we find that Janus solutions are dual to the setups where the initial state and final state are different and this is the reason why we can have a traversable wormhole which makes the causal propagation from CFT$_{(1)}$ and CFT$_{(2)}$ possible. We also studied a free scalar CFT example of a Janus CFT and showed that it qualitatively shares similar properties expected from the AdS wormhole. This model is successful to explain the fact that the pseudo entropy $S_A$ increases as the imaginary Janus deformation is added, exceeding the entropy of the (un-deformed) AdS black hole. However, we found that the two point functions get divergent at infinite time but not finite time. This may suggest that to reproduce the null geodesic at finite time we need strongly coupled and large $N$ CFTs so that they are dual to classical gravity. This clearly deserves future studies. Moreover, it would be an intriguing future problem to extend this construction of Janus traversable wormhole to higher dimensions and more general deformations such as non-marginal perturbations. 

Another CFT dual of a traversable wormhole is the double trace deformation, as studied in section \ref{sec:DT}. Even though similar constructions have been already studied by many authors, our analysis covered many aspects which have not been discussed as summarized as follows. We analyzed the case where we turn on the deformation to two decoupled CFTs and we also clarified how the quantum states look like from a quantum information theoretic viewpoint. In order to perform the double trace deformations for not only relevant but also irrelevant perturbations to have a macroscopic wormhole, we consider non-local version of double trace deformations. We confirmed that we can reproduce the expected holographic two point functions computed in section \ref{sec:AdSWH} by a suitable choice of the double trance deformations. This analysis reveals that in general the throat region of the wormhole is described by highly quantum gravity, though the asymptotically AdS region is well treated by the classical gravity. 

Notably, due to the double trace interactions, we found that the quantum state in the combined system of two CFTs at a generic time slice is not standard in that its density matrix is not hermitian. Moreover, it is not pure. This is because two CFTs are interacting and a time slice in each CFT is not space-like separated from the one in the other CFT. This is clearly shown by our result that the pseudo entropy for the total state is non-vanishing, confirmed in a free scalar CFT example. This fact that the global time slice of the wormhole corresponds to a mixed state is the special feature only for the double trance deformation and is missing for the Janus deformation. We argued that in the gravity side, these suggest that the presence of many microscopic wormholes due to the quantum effect as in Fig.\ref{fig:MWH}.  Further investigations of this interpretation and implication to quantum gravity are left for a future problem.

As another important class of the double trace deformation, we discussed the non-local generalization of the $\TTbar$ deformation in section \ref{sec:TTbar}.
We concretely wrote down the flow equation of the non-local $\TTbar$ deformation and found the third derivative of the metric is non-zero. This is different from the usual $\TTbar$ deformation and the bulk dual should deviate from the finite cut-off proposal with pure Einstein gravity. We proposed that the possible gravity dual is a non-local gravitational theory which may appears as a large scale effective description of fracton like wormhole geometry. Since the non-locality is suppressed around the conformal boundaries, we see that the deformation acts in the IR region of the boundary theory similar to the scalar double trace deformation. Also we discuss the $T_1T_2$ deformation with non-locality. In this case, we heuristically discuss that there are wormhole which connects the two Rindler patches. This may be the traversable wormhole which carries information from one side to the other. More detail discussion for $T_1T_2$ deformation will be the interesting future work.

\section*{Acknowledgments}


We are grateful to Kristan Jensen, Hiroki Kanda, Rene Meyer and Robert Myers for useful discussions and comments. TT would like to thank "OIST-ExU workshop: Quantum Extreme Universe: Matter, Information, and Gravity" held at OIST and "Gauge/Gravity Duality 2024" held at TSIMF, where a part of this work was presented and completed. This work is supported by MEXT KAKENHI Grant-in-Aid for Transformative Research Areas (A) through the ``Extreme Universe'' collaboration: Grant Number 21H05187. TT is also supported by Inamori Research Institute for Science, and by JSPS Grant-in-Aid for Scientific Research (A) No. 21H04469 and JSPS Grant-in-Aid for Scientific Research (B) No. 25K01000. TK is supported by Grant-in-Aid for JSPS Fellows No. 23KJ1315.

\appendix


\section{Calculation of geodesic lengths in BTZ wormholes}
\label{ap:BTZ}
We provide details of geodesic lengths in glued BTZ wormhole in section \ref{sec:BTZ}.

\subsection{Geodesic in BTZ black hole}
In the BTZ \eqref{BTZnull}, the geodesic distance $D_{12}$ between $(u_1,v_1,x_1)$ and $(u_2,v_2,x_2)$ reads
\ba
\cosh D_{12}=\frac{2(u_1v_2+u_2v_1)+(1-u_1v_1)(1-u_2v_2)\cosh\frac{x_1-x_2}{a}}
{(1+u_1v_1)(1+u_2v_2)}.  \label{disbtz}
\ea

First let us compute the geodesic distance between $(t_1,\ep,x_1)$ in the left boundary (CFT$_{(1)}$) and $(t_2,\ep,x_2)$ in the right boundary (CFT$_{(2)}$) 
in the full standard BTZ (=gluing with $T=0$) (\ref{BTZm}), where $\ep$ is the UV cut off.
At both boundaries we have 
\ba
&& u_1\simeq -\left(1-\frac{\ep}{a}\right)e^{-\frac{t_1}{a}},\ \ \ v_1\simeq \left(1-\frac{\ep}{a}\right)e^{\frac{t_1}{a}},\no
&& u_2\simeq \left(1-\frac{\ep}{a}\right)e^{\frac{t_2}{a}},\ \ \  v_2\simeq -\left(1-\frac{\ep}{a}\right)e^{-\frac{t_2}{a}}.
\ea

In this setup the geodesic distance reads from (\ref{disbtz}):
\ba
\cosh D_{12}=\frac{a^2}{\ep^2}\left(\cosh\frac{t_1+t_2}{a}+\cosh\frac{x_1-x_2}{a}\right).
\ea
Thus we find
\ba
D_{12}\simeq \log\left[\frac{2a^2}{\ep^2}\left(\cosh\frac{t_1+t_2}{a}+\cosh\frac{x_1-x_2}{a}\right)\right].
\ea

\subsection{Geodesic distance in Lorentzian wormhole}

We compute $D(t_1,t_2)$ in the setup in Fig.\ref{fig:BTZglue}. 
Denote the end point on EOW brane as $(u,v) = (\zeta, \eta)$ in the left region and $(u,v) = (\eta, \zeta)$ in the right one. 
They satisfy \eqref{EOWa}:
\begin{equation}
  \eta = \frac{\zeta - \lambda}{\lambda \zeta + 1} 
  ~~ \Leftrightarrow ~~ 1 + \zeta \eta = \frac{1}{\lambda} \qty(\zeta - \eta).
\end{equation}
Using \eqref{disbtz}, the distance between the left boundary point $(u_1, v_1)$ and end point on EOW brane $(\zeta, \eta)$ is 
\begin{align}
  \cosh D_{2} &= \frac{a}{\epsilon} \frac{\eta e^{t_2 / a} - \zeta e^{-t_2 / a} + (1 - \zeta \eta)}{1 + \zeta \eta}. \\
  D_2 &\simeq \log \qty (\frac{2a}{\epsilon} \frac{\eta e^{t_2 / a} - \zeta e^{-t_2 / a} + (1 - \zeta \eta)}{1 + \zeta \eta}). \label{geolen}
\end{align}
The distance between right boundary point $(u_2, v_2)$ and end point of EOW $(\eta, \zeta)$ is also written in the form \eqref{geolen}, but replacing $t_1$ to $t_2$. 
Therefore, the total distance is
\begin{align}
  D_{12} &= D_1 + D_2 \notag \\
  &= \log \qty[\frac{4 a^2}{\epsilon^2} \frac{1}{(1 + \zeta \eta)^2} \qty{\eta e^{t_1 / a} - \zeta e^{-t_1 / a} + (1 - \zeta \eta)} \qty{\eta e^{t_2 / a} - \zeta e^{-t_2 / a} + (1 - \zeta \eta)}] .\label{d12}
\end{align}

\subsubsection{$t_1 = -t_2$ case}
In the case of $t_1 = -t_2$, the maximum geodesic distance is given when
\begin{equation}
  (\zeta, \eta) = \qty(\frac{-1 + \sqrt{1+ \lambda^2}}{\lambda}, \frac{1 - \sqrt{1+ \lambda^2}}{\lambda}).
\end{equation}
Inserting this to eq.\eqref{d12}, we gain
\begin{equation}
  D_{12} = 2 \log \qty[\frac{2a}{\epsilon} \qty(- \lambda \cosh \qty(\frac{t_1}{a}) + \sqrt{1 + \lambda^2})]. \label{t1nt2}
\end{equation}

\subsubsection{$t_1 = t_2$ case}
In the case of $t_1 = t_2$, \eqref{d12} is simplified to
\begin{equation}
  \begin{aligned}
    D_{12} &= 2 \log \qty[\frac{2a}{\epsilon} \frac{1}{1 + \zeta \eta} \qty(\eta e^{t_1 / a} - \zeta e^{-t_1 / a} + (1 - \zeta \eta))]=: 2 \log \qty[\frac{2a}{\epsilon} f(\zeta)] .\label{d12_2}
  \end{aligned}
\end{equation}
To maximize the distance, we compute the maximum point of $f(\zeta)$;
\begin{equation}
  f'(\zeta) = 0 ~~
  \Leftrightarrow ~~ 2 \sinh(\frac{t_1}{a}) (\zeta \eta ' - \eta) - 2 (1 - \eta ') = 0.
\end{equation}
Here we denote the derivative of $\zeta$ as '. The solution reads
\begin{align}
  \zeta &= \frac{\lambda \sinh(t_1/a) - 1 + \sqrt{\lambda ^2 + 1} \cosh(t_1/a)}{\sinh(t_1/a) + \lambda}, \\
  \eta &= \frac{- \lambda \sinh(t_1/a) - 1 - \sqrt{\lambda ^2 + 1} \cosh(t_1/a)}{\sinh(t_1/a) - \lambda}.
\end{align}
Plugging this into eq.\eqref{d12_2}, we get the geodesic distance.
\begin{equation}
  D_{12} = 2 \log \qty[\frac{2a}{\epsilon} \cosh(\frac{t_1}{a}) \qty(-\lambda + \sqrt{\lambda^2  + 1})].\label{t1t2}
\end{equation}


\section{Calculation of two point functions in BTZ wormholes} \label{app2pt}
We show the detailed calculations of two point function in BTZ wormholes.

\subsection{Coordinate change} \label{corcha}
We will give the relations among some kinds of coordinates of BTZ black hole geometry. 
We start from eq.\eqref{BTZm}. 
\begin{equation}
  ds^2=-\frac{a^2-z^2}{z^2}dt^2+\frac{a^2 dz^2}{z^2 \left(1-\frac{z^2}{a^2}\right)}+\frac{a^2 dx^2}{z^2}. 
\end{equation}
We define $\displaystyle r := a^2 / z$, and redefine $t \rightarrow t / a$ and $x \rightarrow x / a$ so that they becomes dimensionless. Then the metric becomes
\begin{equation}
  ds^2 = - \qty(r^2 - a^2) d t^2 + \frac{1}{(r^2 / a^2) - 1} dr^2 + r^2 dx^2. 
\end{equation}
which is the standard BTZ coordinate \eqref{BTZ_sta}. We perform more coordinate changes;
\begin{align}
  &\begin{cases}
    \displaystyle
    \tau &:= e^x \sinh t  \sqrt{1 - \frac{a^2}{r^2}}, \\
    \chi &:= e^x \cosh t \sqrt{1 - \frac{a^2}{r^2}}, \\    
    \zeta &:= e^x \frac{a}{r},
  \end{cases} \\
  &\Rightarrow\; ds^2 = \frac{1}{\zeta^2} \qty(- d \tau^2 + d \zeta^2 + d \chi^2). 
\end{align}
\begin{align}
  &\begin{cases}
    \displaystyle
    \frac{y}{\cosh (\rho/a)} &:= \zeta, \\
    y \tanh(\rho / a) &:= \chi, 
  \end{cases} \\
  &\Rightarrow\; ds^2 = d \rho^2 + a^2 \cosh^2 \qty(\frac{\rho}{a}) \qty(\frac{- d\tau^2 + dy^2}{y^2}). 
\end{align}
The final equation is the form we used in section \ref{2ptBTZ}, eq.\eqref{BTZ_glo_Poi}. 

\subsection{Solutions of the wave equation} \label{soluti}
We solved the wave equation on BTZ wormhole geometry \eqref{EOM_BTZ};
\begin{equation}
  \qty[\pdv[2]{\rho} + \frac{2}{a} \tanh(\frac{\rho}{a}) \pdv{\rho} + \frac{y^2}{a^2 \cosh(\rho / a)} \qty(- \pdv[2]{\tau} + \pdv[2]{y}) - m^2] \phi(\rho, \tau, y) = 0 . 
\end{equation}
We separate valuables as $\phi(\rho, \tau, y) = \tilde{\phi}(\rho) \varphi (\tau, y)$, then impose $\varphi (\tau, y)$ to satisfy the following equation \cite{Akal:2020wfl};
\begin{equation}
  \qty[y^2 \qty(-\pdv[2]{\tau} + \pdv[2]{y}) - M^2] \varphi_M (t, y) = 0. 
\end{equation}
Here we introduced a constant $M$. This is just the Klein-Gordon equations in AdS$_2$ geometry with mass $M$, so we are familiar with its solution. 
\begin{equation}
  \varphi_\nu (\tau, y) \int \frac{\dd{\omega}}{2 \pi} e^{-i \omega \tau} \qty[C_1 \sqrt{y} J_{\nu} (\omega y) + C_2 \sqrt{y} Y_{\nu} (\omega y) ], \qquad \nu := \sqrt{\frac{1}{4} + M^2}. \label{BTZ_sol2}
\end{equation}
By substituting this, we can write the equation of $\tilde{\phi}(\rho)$. 
\begin{equation}
  \qty[\pdv[2]{\rho} + \frac{2}{a} \tanh(\frac{\rho}{a}) \pdv{\rho} + \frac{M^2}{a^2 \cosh(\rho / a)} - m^2] \tilde{\phi}(\rho) = 0. \label{BTZ_eom}
\end{equation}
The solutions are written by the hypergeometric function as following;
\begin{equation}
  \phi_{\nu} = 
  \left\{
    \begin{aligned}
      \qty(e^{\frac{2 \rho}{a}})^{\frac{1}{2}(1 + \beta)} &\qty(1 + e^{\frac{2 \rho}{a}})^{\frac{1}{2}(-1 + 2 \nu)}  \\
      &\tensor[_2]{F}{_1}\qty(\frac{1}{2}(1 + 2 \nu),~ \frac{1}{2}(1 + 2 \nu + 2\beta);~ 1 + \beta;~ -e^{\frac{2 \rho}{a}}), \\
      \qty(e^{\frac{2 \rho}{a}})^{\frac{1}{2}(1 - \beta)} &\qty(1 + e^{\frac{2 \rho}{a}})^{\frac{1}{2}(-1 + 2 \nu)}  \\
      &\tensor[_2]{F}{_1}\qty(\frac{1}{2}(1 + 2 \nu),~ \frac{1}{2}(1 + 2 \nu - 2\beta);~ 1 - \beta;~ -e^{\frac{2 \rho}{a}}) .
    \end{aligned} \label{BTZ_sol1}
  \right.
\end{equation}
Here we defined $\beta := \sqrt{1 + a^2 m^2}$. 
We get the full solutions of wave equation by multiplying eq.\eqref{BTZ_sol1} and \eqref{BTZ_sol2}, and taking linear combinations for each $M$. 
If we set $m^2 = - 3 / 4 a^2$, eq.\eqref{BTZ_sol1} reduces to eq.\eqref{sincos}.

\subsection{Two point function in BTZ wormhole} \label{2ptwor}
We got two point function of the dual CFT of glued BTZ wormhole;
\begin{align}
  \ev{\mathcal{O}_1 (t, x) \mathcal{O}_1 (t', x')} &= \frac{1}{4 \pi a^3 (\cosh t \cosh t')^{\frac{3}{2}}} \sum_{n=0}^{\infty} \qty(\nu(n))^2 Q_{\nu(n) - \frac{1}{2}} \qty(\xi), \\
  \ev{\mathcal{O}_2 (t, x) \mathcal{O}_1 (t', x')} &= \frac{1}{4 \pi a^3 (\cosh t \cosh t')^{\frac{3}{2}}} \sum_{n=0}^{\infty} (-1)^{n-1} \qty(\nu(n))^2 Q_{\nu(n) - \frac{1}{2}} \qty(\xi), \\
  \xi &= \frac{\cosh(x-x') \pm \sinh t \sinh t'}{\cosh t \cosh t'}.
\end{align} 

\subsubsection{BTZ black hole case}
We will check that without the EOW brane, previous result reduces to usual finite temperature two point function. 
We use the integral representation of Legendre function;
\begin{equation}
  Q_\nu (z) = \int_0^{z - \sqrt{z^2 - 1}} \frac{t^\nu}{\sqrt{t^2 - 2 tz + 1}} \dd{t}. 
\end{equation}
Substituting this, we can perform the sum of $n$ easily, and we can represent two point function in integral form
\begin{align}
  \ev{\mathcal{O}_1 (t, x) \mathcal{O}_1 (t', x')} &= \frac{1}{4 \pi a^3 (\cosh t \cosh t')^{\frac{3}{2}}} \notag \\
  &\hspace{40pt} \int_0^{\xi - \sqrt{\xi^2 - 1}} \frac{N^2 t^{N-1} (1 + t^N)}{(1 - t^N)^3} \sqrt{\frac{t}{t^2 - 2 t \xi + 1}} \dd{t}, \\
  \ev{\mathcal{O}_2 (t, x) \mathcal{O}_1 (t', x')} &= \frac{1}{4 \pi a^3 (\cosh t \cosh t')^{\frac{3}{2}}} \notag \\
  &\hspace{40pt} \int_0^{\xi - \sqrt{\xi^2 - 1}} \frac{- N^2 t^{N-1} (-1 + t^N)}{(1 + t^N)^3} \sqrt{\frac{t}{t^2 - 2 t \xi + 1}} \dd{t}. 
\end{align}
In the ordinary BTZ black hole case, $N = 1$, we can perform the integral explicitly; 
\begin{align}
  \ev{\mathcal{O}_1 (t, x) \mathcal{O}_1 (t', x')} 
  &= \frac{1}{8 a^3} \frac{1}{(\sinh(t-t'+x-x') \sinh(t-t'-x+x'))^{\frac{3}{2}}}, \\
  \ev{\mathcal{O}_2 (t, x) \mathcal{O}_1 (t', x')} 
  &= \frac{1}{8 a^3} \frac{1}{(\cosh(t-t'+x-x') \cosh(t-t'-x+x'))^{\frac{3}{2}}}.
\end{align}
These are exactly well-known CFT results (up to constant factor). 

\subsubsection{Wormhole case}
We can also perform the integral for some special cases. 
Set $N = 2$, which corresponds to placing the EOW brane in $\rho_0 = a \log(\cot(\pi/8))$. In this case, 
\begin{align}
  \ev{\mathcal{O}_2 (t, x) \mathcal{O}_1 (t', x')} &= \frac{3}{32 \sqrt{2} a^3 (\cosh t \cosh t')^{\frac{3}{2}}} \xi^{\frac{3}{2}} \\
  &= \frac{3}{32 \sqrt{2} a^3} \frac{1}{(\cosh(x-x') - \sinh t \sinh t')^{\frac{3}{2}}}. \notag 
\end{align}
Considering $x = x' = 0$ and $t' = t$ case, then this function diverge at 
\begin{equation}
  \sinh t = 1  ~ \Leftrightarrow ~ t = \log (1 + \sqrt{2}).
\end{equation}
This is exactly the same time that the geodesic becomes null, which is derived from eq.\eqref{ttt2} and \eqref{EOWloc};
\begin{equation}
  \cosh t = \coth (\rho_0 / a). \label{nulltime}
\end{equation}
Set $N = 3$, which corresponds to putting EOW brane at $\rho_0 = a \log(\cot(\pi/12))$. In this case, 
\begin{align}
  \ev{\mathcal{O}_2 (t, x) \mathcal{O}_1 (t', x')} &= \frac{1}{96 a^3 (\cosh t \cosh t')^{\frac{3}{2}}} \qty(\frac{\sqrt{2}}{(1+\xi)^{\frac{3}{2}}} + \frac{18}{(-1 + 2 \xi)^{\frac{5}{2}}} - \frac{4}{(-1 + 2 \xi)^\frac{3}{2}}).
\end{align}
Considering $x = x' = 0$ and $t' = t$ case, then this function diverge at 
\begin{equation}
  \xi = \frac{1}{2} ~ \Leftrightarrow ~ t = \frac{1}{2} \log 3.
\end{equation}
This is again consistent with eq.\eqref{nulltime}.


\section{Calculation of coupled harmonic oscillators for the model B}
\label{HarmoB}

We present details of calculations of the coupled harmonic oscillators in section \ref{tchpb}.

\subsection{Evaluation of $\rho_{AB}$}
To evaluate (\ref{rabb}), it is useful to consider the following quantity 
\ba
F&\equiv & \la 0|e^{xa+yb}e^{-iHT}e^{za^\dagger+wb^\dagger}|0\lb\no
&=&\sum_{m,n,p,q=0}^\infty \frac{1}{\s{m!n!p!q!}}x^n y^m z^q w^p
\la n|_A\la m|_B e^{-iHT} |q\lb_A |p\lb_B,
\ea
from which we can find $\la n|_A\la m|_B e^{-iHT}|q\lb_A|p\lb_B$ by its series expansion.

By a direct computation, after some algebras,  we obtain
\ba
F=\frac{1}{\cosh^2\theta-\sinh^2\theta e^{-2iT}}\cdot e^{A(xy+zw)+B(xz+yw)},\label{Ffor}
\ea
where 
\ba
A=\frac{\tanh\theta(e^{-2iT}-1)}{1-\tanh\theta^2 e^{-2iT}},
\ \ \ \ B=\frac{e^{-iT}}{\cosh^2\theta-\sinh^2\theta e^{-2iT}}.
\ea

For example we find 
\ba
\la 0|_A\la 1|_B e^{-iHT}|0\lb_A|1\lb_B
=\la 1|_A\la 0|_B e^{-iHT}|1\lb_A|0\lb_B
=\frac{e^{-iT}}{(\cosh^2\theta-\sinh^2\theta e^{-2iT})^2}.
\ea
This shows that the transition matrix is not hermitian when 
$T\neq \pi {\mathbb{Z}}$. We can also confirm that when $T=0$ we have $A=0$ and $B=1$. 

To derive (\ref{Ffor}) we rewrite $a,b$ in terms of $\ti{a},\ti{b}$ using (\ref{bgla}) and 
\ba
|0\lb_{AB}=\frac{1}{\cosh\theta}e^{\tanh\theta \ti{a}^\dagger \ti{b}^\dagger}|\ti{0}\lb_{AB}.
\ea
We can rewrite (we set $c=\cosh\theta$, $s=\sinh\theta$ and $t=\tanh\theta$)
\ba
F&=&\frac{1}{c^2}\la \ti{0}|e^{t\ti{a}\ti{b}}e^{x(c\ti{a}-s\ti{b}^\dagger)}
e^{y(-s\ti{a}^\dagger +c\ti{b})e^{z(c\ti{a}^\dagger e^{-iT}}-s\ti{b}e^{iT})}
e^{w(-se^{iT}\ti{a}+ce^{-iT}\ti{b}^\dagger)}e^{t\ti{a}^\dagger \ti{b}^\dagger e^{-2iT}}|\ti{0}\lb \no
&=&\frac{1}{c^2}\la \ti{0}|e^{t\ti{a}\ti{b}}e^{(xc-swe^{iT})\ti{a}}
e^{(cy-sze^{iT})\ti{b}}e^{(zce^{-iT}-ys)\ti{a}^\dagger}
e^{(wce^{-iT}-xs)\ti{b}^\dagger}e^{t\ti{a}^\dagger \ti{b}^\dagger e^{-2iT}}|\ti{0}\lb\no
&& \ \ \times e^{(xy+zw)sc-(yw+xz)s^2e^{iT}}\no
&=&\frac{1}{c^2}\frac{1}{1-t^2 e^{-2iT}}\cdot e^{sc(xy+wz)-s^2e^{iT}(xz+yw)}
\no
&& \ \ \times
\exp\left[\frac{1}{1-t^2e^{-2iT}}\left((xz+yw)(c^2 e^{-iT}+s^2e^{iT}-2s^2e^{-iT})+(xy+zw)(s^2t+sc e^{-2iT}-2sc)\right)\right],\nonumber
\ea
where we employed (\ref{FPD}). This leads to (\ref{Ffor}).

\subsection{Computation of $\rho_A$}

The reduced transition matrix $\rho_A$ is computed as 
(\ref{rhoa}). To derive this, we introduce 
\ba
G&\equiv& \int \frac{dzd\bar{z}}{\pi\tanh\theta}e^{\frac{z\bar{z}}{\tanh\theta}}
\la 0|e^{x(a+b)}e^{-iHT}e^{z a^\dagger+\bar{z}b^\dagger}|0\lb \no
&=&\sum_{p=0}^\infty (-\tanh\theta)^p \la 0|e^{x(a+b)}e^{-iHT}|p\lb_A|p\lb_B\no
&=&\sum_{m=0}^\infty \frac{x^{2m}}{m!}\cdot\left[\sum_{p=0}^\infty
(-\tanh\theta)^p\la m|_A\la m|_B e^{-iHT}|p\lb_A|p\lb_B\right]. \label{Ga}
\ea
On the other hand, by using (\ref{FPD}) we can directly evaluate the integral, leading to 
\ba
G=e^{-\tanh\theta x^2}.  \label{Gb}
\ea
By comparing the coefficient of $x^{2m}$ for (\ref{Ga}) and (\ref{Gb}), we obtain
\ba
\sum_{p=0}^\infty
(-\tanh\theta)^p\la m|_A\la m|_B e^{-iHT}|p\lb_A|p\lb_B
=(-\tanh\theta)^m.
\ea
Thus we obtain (\ref{rhoa}).

\subsection{Pseudo entropy}
\par 
Now let us compute the second Renyi pseudo entropy for $\rho_{AB}$. We can derive the expression (\ref{PEHOQ}) from the following evaluation:
\ba
&& \mbox{Tr}[\rho_{AB}^2]\no
&=&\frac{1}{\cosh^4\theta(\cosh^2\theta-\sinh^2\theta e^{-2iT})^2}\int \frac{dxd\bar{x}dyd\bar{y}dzd\bar{z}dwd\bar{w}}{\pi^2}\no
&&\times
e^{-|x|^2-|y|^2-|z|^2-|w|^2}e^{-\tanh\theta A(xy+\bar{x}\bar{y}+zw+\bar{z}\bar{w})}e^{-\tanh\theta B(xz+\bar{x}\bar{z}+yw+\bar{y}\bar{w})}\no
&=&\frac{2}{1+e^{-2iT}+(1-e^{-2iT})\cosh 4\theta}.
\ea

\subsection{Two point functions}
\par 
Finally we compute the two point function $\la x_A(T) x_B(0)\lb$.
To derive (\ref{TPFH}), we can calculate this two point function as follows:
\ba
&& \la x_A(T)x_B(0)\lb \no
&=&\frac{1}{2}\cdot \left(-\frac{\tanh\theta}{\cosh^2\theta}\right)
\cdot \sum_{n,q=0}^\infty (-\tanh\theta)^{n+q}\s{q+1}\s{n+1}\no
&& \times \left[\la n|\la n+1|e^{-iHT}|q\lb|q+1\lb +\la n+1|\la n|e^{-iHT}
|q+1\lb|q\lb\right] \no
&&=\left(-\frac{\tanh\theta}{\cosh^2\theta}\right)\int\frac{dxd\bar{x}dzd\bar{z}}{\pi^2}e^{-|x|^2-|z|^2}
\la 0|b e^{xa-\tanh\theta\bar{x}b}e^{-iHT}e^{-\tanh\theta za^\dagger+\bar{z}b^\dagger}b^\dagger|0\lb\no
&&=\left(-\frac{\tanh\theta}{\cosh^2\theta}\right)\int\frac{dxd\bar{x}dzd\bar{z}}{\pi^2}e^{-|x|^2-|z|^2}\Biggl[-\tanh\theta Be^{-\tanh\theta A(|x|^2+|z|^2)}e^{-\tanh\theta
B(xz+\bar{z}\bar{z})}\no
&&\ \ \ \ \ +\tanh^2\theta(Ax+Bz)(A\bar{z}+B\bar{x})e^{-\tanh\theta A(|x|^2+|z|^2)}e^{-
\tanh\theta B(xz+\bar{x}\bar{z})}\Biggr]\no
&&=-e^{-iT}\sinh\theta\cosh\theta,
\ea
where we introduced $\eta=1+\tanh\theta A$.


\section{Details of free scalar Janus CFT}
\label{ap:januscft}
Consider the interface CFT of a compactified free scalar where the radius of a massless free scalar changes from $R_1$ to $R_2$ when we cross the interface. We employ the boundary state formulation \cite{Cardy:2004hm,DiVecchia:1999mal} via the doubling trick as described in \cite{Bachas:2001vj,Sakai:2008tt} (see  Fig.\ref{fig:Fold}).

To describe our TFD setup, we take the Euclidean time to be an interval: $0\leq \tau\leq \frac{\beta}{2}$ with a space direction $0\leq \sigma\leq 2\pi$. The mode expansions of two massless scalar fields $\phi^{(1)}$ and $\phi^{(2)}$ look like
\ba
&&\phi^{(i)}_L=x^{(i)}_L-ip^{(i)}_L(\tau-i\sigma)+i\sum_{m\in Z}
\frac{\ap^{(i)}_m}{m}e^{-m(\tau-i\sigma)},\no
&&\phi^{(i)}_R=x^{(1)}_R-ip^{(1)}_R(\tau+i\sigma)+i\sum_{m\in Z}
\frac{\tilde{\ap}^{(i)}_m}{m}e^{-m(\tau+i\sigma)},\ \  (i=1,2) \label{modee}\nonumber
\ea
where the oscillator operators satisfy
\ba
[\ap^{(i)}_m,\ap^{(j)}_n]=m\delta_{ij}\delta_{m+n,0},\ \ \ [x^{(i)}_{L,R},p^{(j)}_{L,R}]=i\delta_{ij}.
\ea
By compactifying $\phi^{(i)}$ on a circle with the radius $R_i$, the momenta are quantized as:
\ba
P^{(i)}_L=\frac{n_i}{R_i}+\frac{w_iR_i}{2},\ \ \ P^{(i)}_R=\frac{n_i}{R_i}-\frac{w_iR_i}{2},\label{nw}
\ea
where $n_i$ (momenta) and $w_i$ (winding number) take integer values. 

\subsection{Boundary condition}

At $\tau=0$ and $\tau=\beta/2$, we impose the mixed boundary condition \cite{Bachas:2001vj,Sakai:2008tt}:
\ba
&& \de_\tau(\cos\theta \phi^{(1)}-\sin\theta \phi^{(2)})=0,\no
&& \de_\sigma(\sin\theta \phi^{(1)}+\cos\theta \phi^{(2)})=0, \label{bca}
\ea
at any $\sigma$, where we set $\tan\theta=\frac{R_2}{R_1}$.
Note that at $R_1=R_2$ i.e. $\theta=\frac{\pi}{4}$, the interface becomes transparent and the time coordinate just describes a circle 
$0\leq \tau\leq \beta$ by regarding $\phi^{(2)}$ live in $\frac{\beta}{2}\leq \tau\leq \beta$. $\theta\neq \frac{\pi}{4}$ corresponds to the Janus deformation. At $\theta=0$, the field $\phi^{(1)}$ obeys the Neumann boundary condition and $\phi^{(2)}$ does the Dirichlet boundary condition, thus they decouple completely. At $\theta=\frac{\pi}{2}$, we have the opposite boundary condition and they decouple again.

These conditions (\ref{bca}) can be rewritten as 
\ba
&& \left[(\sin\theta\ap^{(1)}_m+\cos\theta\ap^{(2)}_m)-(\sin\theta\ti{\ap}^{(1)}_{-m}+\cos\theta\ti{\ap}^{(2)}_{-m})\right]|B\lb=0.\no
&& \left[(\cos\theta\ap^{(1)}_m-\sin\theta\ap^{(2)}_m)-(\cos\theta\ti{\ap}^{(1)}_{-m}-\sin\theta\ti{\ap}^{(2)}_{-m})\right]|B\lb=0.
\ea
This is solved as follows
\ba
|B\lb=e^{\sum_{m=1}^\infty\frac{1}{m}\left[\cos2\theta(-\ap^{(1)}_{-m}\ti{\ap}^{(1)}_{-m}+\ap^{(2)}_{-m}\ti{\ap}^{(2)}_{-m})+\sin2\theta(\ap^{(1)}_{-m}\ti{\ap}^{(2)}_{-m}+\ti{\ap}^{(1)}_{-m}\ap^{(2)}_{-m})\right]}|0\lb.  \label{BST}
\ea

Now we introduce the creation and annihilation operators as 
\ba
&& a_n=\frac{1}{\s{n}}(\cos\theta\ap^{(1)}_m-\sin\theta\ap^{(2)}_m),
\ \ \ a^\dagger_n=\frac{1}{\s{n}}(\cos\theta\ap^{(1)}_{-m}-\sin\theta\ap^{(2)}_{-m}),\no
&& \ti{a}_n=\frac{1}{\s{n}}(\cos\theta\ti{\ap}^{(1)}_m-\sin\theta\ti{\ap}^{(2)}_m),
\ \ \ \ti{a}^\dagger_n=\frac{1}{\s{n}}(\cos\theta\ti{\ap}^{(1)}_{-m}-\sin\theta\ti{\ap}^{(2)}_{-m}),\no
&& b_n=\frac{1}{\s{n}}(\sin\theta\ap^{(1)}_m+\cos\theta\ap^{(2)}_m),
\ \ \ b^\dagger_n=\frac{1}{\s{n}}(\sin\theta\ap^{(1)}_{-m}+\cos\theta\ap^{(2)}_{-m}),\no
&& \ti{b}_n=\frac{1}{\s{n}}(\sin\theta\ti{\ap}^{(1)}_m+\cos\theta\ti{\ap}^{(2)}_m),
\ \ \ \ti{b}^\dagger_n=\frac{1}{\s{n}}(\sin\theta\ti{\ap}^{(1)}_{-m}+\cos\theta\ti{\ap}^{(2)}_{-m}).  \label{relaq}
\ea
The boundary states (\ref{BST}) can be simply rewritten as
\ba
|B\lb=e^{\sum_{m=1}^\infty\left(-a^\dagger_n \ti{a}^\dagger_n+b^\dagger_n \ti{b}^\dagger_n\right)}|0\lb.
\ea

The total Hamiltonian with the zero energy $H=L_0+\ti{L}_0-\frac{1}{6}$ reads
\ba
H=\sum_{n=1}^\infty
n(a^\dagger_n a_n+\ti{a}^\dagger_n \ti{a}_n+b^\dagger_n b_n+\ti{b}^\dagger_n \ti{b}_n)
+\frac{n_1^2}{R^2_1}+\frac{w^2_1R^2_1}{4}+\frac{n_2^2}{R^2_2}+\frac{w^2_2R^2_2}{4}-\frac{1}{6}.
\ea
It is useful to note that the boundary conditions (\ref{bca}) for zero modes (\ref{nw}) lead to 
\ba
w_1+w_2=0,\ \ \ \ n_1=n_2.  \label{qqq}
\ea

\subsection{Two point functions}

Here we explicitly evaluate the two point function $\la V_1V_2\lb$
for the vertex operators (\ref{vextoa}).
The two point function in our TFD geometry with the Janus deformation, is given by 
\ba
&& \la V_1V_2\lb_U=\la B|e^{-\frac{\beta}{2} H}V_1(\tau_1)V_2(\tau_2)|B\lb \no
&&=\la 0|e^{\sum_{n}(-a_n\ti{a}_n+b_n\ti{b}_n)e^{-\beta n}}
e^{i\lambda_+\phi^{(1)}_L(\tau_1)+i\lambda_-\phi^{(1)}_R(\tau_1)}
e^{i\mu_+\phi^{(2)}_L(\tau_2)+i\mu_-\phi^{(2)}_R(\tau_2)}
e^{\sum_{m=1}^\infty\left(-a^\dagger_n \ti{a}^\dagger_n+b^\dagger_n \ti{b}^\dagger_n\right)}|0\lb,\nonumber
\ea
where the subscript $U$ in $ \la V_1V_2\lb_U$ means that this is a unnormalized two point function. The standard normalized two point function  denoted simply by $\la V_1V_2\lb$ is obtained by dividing the unnormalized one by the vacuum partition function $\la 1\lb_U$.  

We can express $\phi^{(i)}_{L,R}$ given by (\ref{modee}) in terms of 
$a_n, \ti{a}_n, b_n$ and $\ti{b}_n$  by (\ref{relaq}), namely
\ba
\ap^{(1)}_n=\s{n}(\cos\theta a_n+\sin\theta b_n),
\ \ \ \ \ap^{(2)}_n=\s{n}(-\sin\theta a_n+\cos\theta b_n),\ \ \  \mbox{etc.}
\ea

We can also use the formula \cite{Takayanagi:2010wp}
\ba
\la 0|e^{\mp a\ti{a} e^{-\beta} }e^{a_La+a_R\ti{a}}e^{b_La^\dagger+b_R \ti{a}^\dagger}e^{\mp a^\dagger \ti{a}^\dagger}|0\lb
=\frac{1}{1-e^{-\beta}}\cdot e^{\frac{1}{1-e^{-\beta}}\left(a_Lb_L+a_Rb_R\mp a_La_R\mp e^{-\beta n}b_Lb_R\right)}.\no \label{FPD}
\ea

After some calculations we find
\ba
&&\la V_1V_2\lb_U=Z_0\cdot \frac{e^{\frac{\beta}{2}}}{\prod_{n=1}^\infty(1-e^{-\beta n})^2}
\cdot e^{\sum_{m=1}^{\infty}\frac{K_m}{m(1-e^{-\beta m})}}=\frac{1}{|\eta\left(\frac{i\beta}{2\pi}\right)|^2}\cdot e^{\sum_{m=1}^{\infty}\frac{K_m}{m(1-e^{-\beta m})}},\no 
&&K_m\equiv -(\lambda^+)^2-(\lambda^-)^2-(\mu^+)^2-(\mu^-)^2
-\lambda^+\lambda^-\cos2\theta e^{-2m\tau_1}+\mu^+\mu^-\cos2\theta e^{-2m\tau_2}\no
&&\ +(\mu^+\lambda^-+\mu^{-}\lambda^{+})\sin2\theta e^{-m(\tau_1+\tau_2)}+e^{-\beta m}\left[-\lambda_+\lambda_-\cos2\theta e^{2m\tau_1}+\mu_+\mu_-\cos2\theta e^{2m\tau_2}\right]\no
&&+e^{-\beta m}\left[(\lambda_+\mu_-+\lambda_{-}\mu_+)\sin2\theta e^{m(\tau_1+\tau_2)}\right].\label{VVA}
\ea
It is useful to note 
\ba
\cos2\theta=\frac{R_1^2-R_2^2}{R_1^2+R_2^2},\ \ \ \ 
\sin 2\theta=\frac{2R_1R_2}{R_1^2+R_2^2}.
\ea

By using the formula
\ba
\log (1-x)=-\sum_{m=1}^{\infty}\frac{x^m}{m},
\ea
we obtain 
\ba
\sum_{m=1}^\infty\frac{K_m}{m(1-e^{-\beta m})}
&=&\sum_{m=1}^\infty\sum_{n=0}^\infty\frac{1}{m}e^{-\beta nm}K_m\no
&=&\sum_{n=0}^\infty\Biggl[((\lambda^+)^2+(\lambda^-)^2+(\mu^+)^2+(\mu^-)^2)\log(1-e^{-\beta n})\no
&&\ \ \ \ \ +\lambda^+\lambda^-\cos2\theta\log(1-e^{-2\tau_1}e^{-\beta n})-\mu^+\mu^{-}\cos2\theta\log(1-e^{-2\tau_2}e^{-\beta n})\no
&&
\ \ \ -(\mu^+\lambda^-+\mu^-\lambda^+)\sin 2\theta
\log(1-e^{-\tau_1-\tau_2}e^{-\beta n})\no
&& \ \ \  +\lambda^+\lambda^- \cos2\theta \log(1-e^{2\tau_1}e^{-\beta(n+1)})-\mu^+\mu^-\cos2\theta \log(1-e^{2\tau_2}e^{-\beta(n+1)})\no
&&\ \ \ -(\lambda^+\mu^-+\lambda^-\mu^+)\sin 2\theta
\log(1-e^{\tau_1+\tau_2}e^{-\beta(n+1)})
\Biggr].
\ea

The zero mode part $Z_0$ looks like ($H_0$ is the zero mode part of Hamiltonian $H$):
\ba
&& Z_0\no
&&={\ca{N}^2}\sum_{\ti{w},\ti{n}}\la \ti{n},\!\ti{n},\!\ti{w},\!-\ti{w}|e^{-\frac{\beta H_0}{2}}
e^{\tau_1H_0}V_1^{(w,0,n,0)}e^{-\tau_1 H_0}e^{\tau_2 H_0}V_2^{(0,-w,0,n)}e^{-\tau_2 H_0}|\ti{n}\!-\!n,\ti{n}\!-\!n,\ti{w}\!-\!w,\!-\ti{w}+w\lb,\no
&&={\ca{N}^2}\cdot e^{-\Delta_1\tau_1-\Delta_2\tau_2}\sum_{\ti{w},\ti{n}\in Z} e^{-\frac{\beta}{2}\left[\frac{(w+\ti{w})^2(R_1^2+R_2^2)}{4}+(n+\ti{n})^2\left(\frac{1}{R_1^2}+\frac{1}{R_2^2}\right)\right]}\cdot e^{\left(\frac{2n\ti{n}}{R_1^2}+\frac{w\ti{w}R_1^2}{2}\right)\tau_1+\left(\frac{2n\ti{n}}{R_2^2}+\frac{w\ti{w}R_2^2}{2}\right)\tau_2},\no
&& \Delta_i\equiv\frac{n_i^2}{R^2_i}+\frac{w_i^2R_i^2}{4},\ \ \ \ (i=1,2).
\ea
In the end we found the following result of the two point function:
\ba
&&\la V_1(\tau_1)V_2(\tau_2)\lb_U  \no
&&{\cal N}^2\cdot \sum_{\ti{w},\ti{n}\in Z} e^{-\frac{\beta}{2}\left[\frac{\ti{w}^2(R_1^2+R_2^2)}{4}+\ti{n}^2\left(\frac{1}{R_1^2}+\frac{1}{R_2^2}\right)\right]}\cdot e^{\left(\frac{2n\ti{n}}{R_1^2}+\frac{w\ti{w}R_1^2}{2}\right)\tau_1+\left(\frac{2n\ti{n}}{R_2^2}+\frac{w\ti{w}R_2^2}{2}\right)\tau_2}\cdot \frac{1}{|\eta\left(\frac{i\beta}{2\pi}\right)|^2} \no
&& \cdot \left[\frac{\theta_1\left(\frac{i\tau_1}{\pi}\Bigr|\frac{i\beta}{2\pi}\right)}{\eta\left(\frac{i\beta}{2\pi}\right)^3}
\right]^{\left[\left(\frac{n}{R_1}\right)^2-\left(\frac{wR_1}{2}\right)^2\right]\cos2\theta}\cdot 
 \left[\frac{\theta_1\left(\frac{i\tau_2}{\pi}\Bigr|\frac{i\beta}{2\pi}\right)}{\eta\left(\frac{i\beta}{2\pi}\right)^3}
\right]^{\left[-\left(\frac{n}{R_2}\right)^2+\left(\frac{wR_2}{2}\right)^2\right]\cos2\theta}\no
&&  \cdot\left[\frac{\theta_1\left(\frac{i(\tau_1+\tau_2)}{2\pi}\Bigr|\frac{i\beta}{2\pi}\right)}{\eta\left(\frac{i\beta}{2\pi}\right)^3}
\right]^{-2\left[\frac{n^2}{R_1R_2}+\frac{w^2 R_1R_2}{4}\right]\sin2\theta}. \label{VVa}
\ea
Note that this is not a normalized two point function. To have a normalized one we need to divide it by the vacuum partition function 
\ba
Z=\la 1\lb_U ={\cal N}^2\cdot \sum_{\ti{w},\ti{n}\in Z} e^{-\frac{\beta}{2}\left[\frac{\ti{w}^2(R_1^2+R_2^2)}{4}+\ti{n}^2\left(\frac{1}{R_1^2}+\frac{1}{R_2^2}\right)\right]}\cdot\frac{1}{|\eta\left(\frac{i\beta}{2\pi}\right)|^2}.
\ea
This leads to the final expression (\ref{VV}) of the normalized two point function.

\subsection{High temperature limit}

To compare with the BTZ Janus, it is useful to take the high temperature limit $\beta\to 0$. For this, we perform the Poisson resummation formula of the zero mode part 
\ba
&&F_0=\sum_{\ti{w},\ti{n}\in Z} e^{-\frac{\beta}{2}\left[\frac{\ti{w}^2(R_1^2+R_2^2)}{4}+\ti{n}^2\left(\frac{1}{R_1^2}+\frac{1}{R_2^2}\right)\right]+\left(\frac{2n\ti{n}}{R_1^2}+\frac{w\ti{w}R_1^2}{2}\right)\tau_1+\left(\frac{2n\ti{n}}{R_2^2}+\frac{w\ti{w}R_2^2}{2}\right)\tau_2}\no
&&=\frac{2\pi}{\beta}\cdot \frac{R_1R_2}{R_1^2+R_2^2}
\sum_{p,q\in Z}e^{-\frac{2\pi^2}{\beta}\cdot\frac{R_1^2R_2^2}{R_1^2+R_2^2}\left[p+\frac{in}{\pi}\left(\frac{\tau_1}{R_1^2}+\frac{\tau_2}{R_2^2}\right)\right]^2}\cdot e^{-\frac{8\pi^2}{\beta}\cdot\frac{1}{R_1^2+R_2^2}\left[q+\frac{iw}{4\pi}\left(R_1^2\tau_1+R_2^2\tau_2\right)\right]^2},\no
\ea

Thus, in the high temperature limit $\beta\to 0$, by setting $p=q=0$ we obtain
\ba
&&\frac{\sum_{\ti{w},\ti{n}\in Z} e^{-\frac{\beta}{2}\left[\frac{\ti{w}^2(R_1^2+R_2^2)}{4}+\ti{n}^2\left(\frac{1}{R_1^2}+\frac{1}{R_2^2}\right)\right]+\left(\frac{2n\ti{n}}{R_1^2}+\frac{w\ti{w}R_1^2}{2}\right)\tau_1+\left(\frac{2n\ti{n}}{R_2^2}+\frac{w\ti{w}R_2^2}{2}\right)\tau_2}}{ \sum_{\ti{w},\ti{n}\in Z} e^{-\frac{\beta}{2}\left[\frac{\ti{w}^2(R_1^2+R_2^2)}{4}+\ti{n}^2\left(\frac{1}{R_1^2}+\frac{1}{R_2^2}\right)\right]}}\no
&&\simeq \exp\Biggl[\left(\frac{2n^2R_2^2}{R_1^2(R_1^2+R_2^2)}+\frac{w^2 R_1^4}{2\beta(R_1^2+R_2^2)}\right)\tau^2_1+\left(\frac{2n^2R_1^2}{R_2^2(R_1^2+R_2^2)}+\frac{w^2 R_2^4}{2\beta(R_1^2+R_2^2)}\right)\tau^2_2  \no
&&\ \ \ \  \ \ \  +\left(\frac{4n^2}{\beta(R_1^2+R_2^2)}+\frac{w^2R_1^2R_2^2}{\beta(R_1^2+R_2^2)}\right)\tau_1\tau_2\Biggr].  \label{zeroex}
\ea

For the oscillator part, we perform the modular transformation:
\ba
\frac{\theta_1\left(\frac{i\tau}{\pi}\Bigr|\frac{i\beta}{2\pi}\right)}{\eta\left(\frac{i\beta}{2\pi}\right)^3}=\frac{\beta}{2\pi}\cdot e^{\frac{2\tau^2}{\beta}}\cdot \frac{\theta_1\left(\frac{2\tau}{\beta}\Bigr|\frac{2\pi i}{\beta}\right)}{\eta\left(\frac{2\pi i}{\beta}\right)^3}.
\ea
This leads to the high temperature limit $\beta\to 0$:
\ba
\frac{\theta_1\left(\frac{i\tau}{\pi}\Bigr|\frac{i\beta}{2\pi}\right)}{\eta\left(\frac{i\beta}{2\pi}\right)^3}\simeq \frac{\beta}{\pi}\cdot e^{\frac{2\tau^2}{\beta}}\cdot \sin\left(\frac{2\pi \tau}{\beta}\right).
\ea

In the total two point function (\ref{VV}), the exponential factor in the zero mode (\ref{zeroex}) is precisely canceled by the $e^{\frac{2\tau^2}{\beta}}$ factors from theta functions after the modular transformation. In this way, finally we obtain the expression (\ref{hightempqx}).

\section{Computation in $T_1T_2$ deformation}\label{sec:T1T2computation}
Here we give a elaborate computation discussed in the section \ref{subsec:T1T2}.
Let us define the inner product on the space of pair of the two rank symmetric tensor filed by
\begin{equation}
     \ev{\mathrm{T},\mathrm{S}} := \sum_{q=1,2}\frac{1}{2}\int d^2 x \sqrt{\gamma} \ep_{ac}\ep_{bd}\; {T^{(q)}}^{ab}(x) {S^{(q)}}^{cd}(x).
 \end{equation}
 where $\mathrm{T}=(T^{(1)}_{ab}(x),T^{(2)}_{ab}(x)),\mathrm{S}=(S^{(1)}_{ab}(x),S^{(2)}_{ab}(x))$. For finite temperature case we notice that $\gam^{(2)}(t,x)=\gam^{(1)}\qty(t+i\frac{\beta}{2},x)$ and denote $\gam^{(1)}_{ab}=\gamma_{ab}$. For notational simplicity we omit $\mu$ dependence unless we need it.
We also introduce a matrix on the space of pair of the two rank symmetric tensor filed, where each component is a matrix function $\mathrm{A}=(A_{pq}(x,y))$. Especially if we consider 
\begin{equation}
    A_{pq}(x,y)= \begin{pmatrix}
        0 & \delta_{x,y}\\
        \delta_{x,y} & 0
    \end{pmatrix}=\begin{pmatrix}
        0 & 1\\1 & 0
    \end{pmatrix}\otimes\delta_{x,y}
\end{equation}
then, the $T_1 T_2$ deformation is expressed as 
\begin{equation}
    S_{T_1T_2} = \Delta\mu\ev{\mathrm{T},\mathrm{A}\mathrm{T}}
\end{equation}
Then, the Hubbard-Stratonivich transformation is given by 
\begin{equation}
\begin{split}
    e^{\Delta\mu\ev{\mathrm{T},\mathrm{A}\mathrm{T}}}&=\int [d\mathrm{h}] e^{-\frac{1}{8\Delta\mu}\ev{\mathrm{h},\mathrm{A}^{-1}\mathrm{h}}+\ev{\mathrm{T},\mathrm{h}}}
\end{split}
\end{equation}
By rewriting $\Hat{(h_1)}_{ab}\to\frac{1}{2}(h_{1})_{ab},\Hat{(h_2)}_{ab}\to\frac{1}{2}(h_{2})_{ab}$,
\begin{equation}
\begin{split}
     e^{\Delta \mu S_{T_1T_2}}= \int [dh_1][dh_2] \mathrm{exp}&\left(-\frac{1}{4\Delta \mu}\int d^2 x \sqrt{\gamma} \ep^{ac}\ep^{bd} (h_1)_{ab}(h_2)_{cd}\right.\\ &\left.-\int d^2 x \sqrt{\gamma} \frac{1}{2}{T^{(1)}}^{ab} (h_1)_{ab}-\int d^2 x \sqrt{\gamma} \frac{1}{2}{T^{(2)}}^{ab} (h_2)_{ab}\right)
\end{split}
\end{equation}
By doing the saddle point approximation in $\Delta\mu$ and identify $\gamma_{[\mu+\Delta\mu]}^{(q)}=\gamma_{[\mu]}^{(q)}+h^{(q)}$, we obtain \eqref{eq:T1T2_metric_flow}. 
Similarly, the flow equation for the effective action can be read off from the saddle point computation for $h$ integral, 
\begin{equation}
\begin{split}
    W_{[\mu+\Delta\mu]}[\gamma_{[\mu+\Delta\mu]}]= W_{[\mu]}-\frac{\Delta\mu}{2}\int d^2 x\sqrt{-\gam}\ep^{ab}\ep^{cd}{(\Hat{T}^{(1)})}_{ab}{(\Hat{T}^{(2)})}_{cd}
\end{split}
\end{equation}

Then, as usual in the $T\Bar{T}$ deformation above, we deform the metric $\gamma \to \gamma + \delta\gamma$ while keeping the flow equation,
\begin{equation}\label{eq:flow_eq_KMS_metric}
    \partial_\mu \delta\gamma_{ab}= -2\delta {(\Hat{T}^{(1)})}_{ab}=0.
\end{equation}
We demand that for all complex times $t$, the variation of the hat stress tensor vanishes, i.e.,
\begin{equation}
    \delta{(\Hat{T}^{(1)})}_{ab}=0,\;\delta{(\Hat{T}^{(2)})}_{ab}=0.
\end{equation}
Then, we find the following equation,
\begin{equation}
    \partial_\mu \qty(\sqrt{-\gamma}{(\Hat{T}^{(1)})}^{ab})\delta\gamma_{ab}= \delta\qty(\sqrt{-\gamma}\epsilon^{ac}\epsilon^{bd}){(\Hat{T}^{(1)})}_{ab}{(\Hat{T}^{(2)})}_{cd}.
\end{equation}
By doing a short algebra for the variation of the metric, we find
\begin{equation}
    \begin{split}
        \partial_\mu\qty(\sqrt{-\gamma}{(\Hat{T}^{(1)})}^{ab})&=\frac{1}{2}\qty({(\Hat{T}^{(1)})}^a_c {(\Hat{T}^{(2)})}^{bc}+{(\Hat{T}^{(2)})}^a_c {(\Hat{T}^{(1)})}^{bc}-{(\Hat{T}^{(1)})}^{ab}{(\Hat{T}^{(2)})}-{(\Hat{T}^{(1)})}{(\Hat{T}^{(2)})}^{ab})
    \end{split}
\end{equation}

 Next, we consider solving the flow equation \eqref{eq:flow_eq_KMS_metric}. 
 Similar to \cite{Guica:2019nzm}, we consider the flow equation for $(\Hat{T})_{ab}$. By using the identity ${(\Hat{T}^{(1)})}_{ab}=-\epsilon_{ac}\epsilon_{bd}{(\Hat{T}^{(1)})}^{cd}$, we obtain
\begin{equation}
    \partial_{\mu}{(\Hat{T}^{(1)})}_{ab}=-\frac{1}{2}{(\Hat{T}^{(1)})}_{ac}{(\Hat{T}^{(2)})}^c_b-\frac{1}{2}{(\Hat{T}^{(2)})}_{ac}{(\Hat{T}^{(1)})}^c_b-\frac{3}{2}{(\Hat{T}^{(1)})}{(\Hat{T}^{(2)})}_{ab}+\frac{3}{2}{(\Hat{T}^{(1)})}_{ab}{(\Hat{T}^{(2)})}.
\end{equation}

\bibliographystyle{JHEP}
\bibliography{ConnectAdS}


\end{document}